# Multiscale Dynamic Dependence Estimation over Networks

Cristian F. Jiménez-Varón
Department of Mathematics, University of York
and
Marina I. Knight
Department of Mathematics, University of York
and
Matthew A. Nunes*
Department of Mathematical Sciences, University of Bath

August 16, 2026

**Abstract**

In many settings, observed multivariate time series are often nonstationary in nature, i.e., their second order properties vary over time. An additional feature is that their cross-channel dependencies are structured by an underlying network. Together, they give rise to complex interactions between temporal dynamics and network topology. We propose Locally Stationary Wavelet processes on Networks (Net-LSW), a new framework for modelling multiscale, time-varying dependencies that explicitly incorporates the network structure. Unlike traditional multivariate approaches, the Net-LSW process encodes the graph directly in the covariance structure of its random increments. We introduce the concept of local partial correlation graph, mathematically connecting absent edges to zero entries in the time-scale dependent inverse wavelet spectral structure. For inference on the local cross-nodal (partial) dependence, we develop a novel subprocess-based estimation scheme and establish its consistency properties. This *new pipeline for network-based nonstationary process modelling, complete with estimation and simulation capabilities* that extend outside time-varying vector autoregressive models, is shown to accurately recover evolving dependence structures whilst respecting the underlying graph topology. The analysis of daily stock price volatilities across a global bank network captures multiscale, highly nonstationary dependencies and identifies time-varying systemic shifts during major financial shocks, including Brexit and the COVID-19 pandemic.

*Keywords:* graphs; nonstationary time series; spectral estimation; wavelets

*Authors gratefully acknowledge support from the EPSRC NeST Programme Grant EP/X002195/1 and the Viking high performance computing facility use provided by the University of York, UK.

# 1 Introduction

Modern datasets frequently comprise multivariate time series which are nonstationary in nature, i.e., whose second order (autocovariance and spectrum) properties vary over time. Such series are common in a wide range of scientific applications, from health sciences (Wilson et al. 2021, Knight et al. 2024, Wu et al. 2026), geophysics (Wilson et al. 2019), engineering (Wen et al. 2023), environmental science (Polanco-Martínez et al. 2020, Chebana & Ouarda 2021), and economics (Asai & McAleer 2009, Hunter et al. 2017).

In many settings, such data are structured by an underlying, known network topology, which gives rise to complex interactions between the time-evolving dynamics of the process and the network configuration. The work in this article is particularly motivated by the analysis of volatility in financial networks, a topic of wider focus in recent literature (Karpman et al. 2023, Cen et al. 2026). An example network of banking institutions is shown in Figure 1. In this setting, data such as equity prices are observed for a set of interconnected financial institutions and are nonstationary in nature, as is well-established (Jiménez-Varón & Knight 2026, Killick et al. 2025). It is important in this context to estimate the time-dependent cross-institution equity dynamics, for example for examining interbank spillover effects, as well as market liquidity and risk (Demirer et al. 2018, Muñoz Mendoza et al. 2024), whilst incorporating the known cross-nodal connectivity (edges).

This type of data also arises in a range of other applications, including structural connections across localized brain regions (Achard & Bullmore 2007) and geographical connectivity in climate networks (Nunes et al. 2015). Since the underlying dynamics evolve over time, modelling such data requires robust statistical frameworks capable of simultaneously capturing the interactions between the temporal dynamics at each node, the multiscale network structure, and the spatial constraints imposed by the underlying network topology.

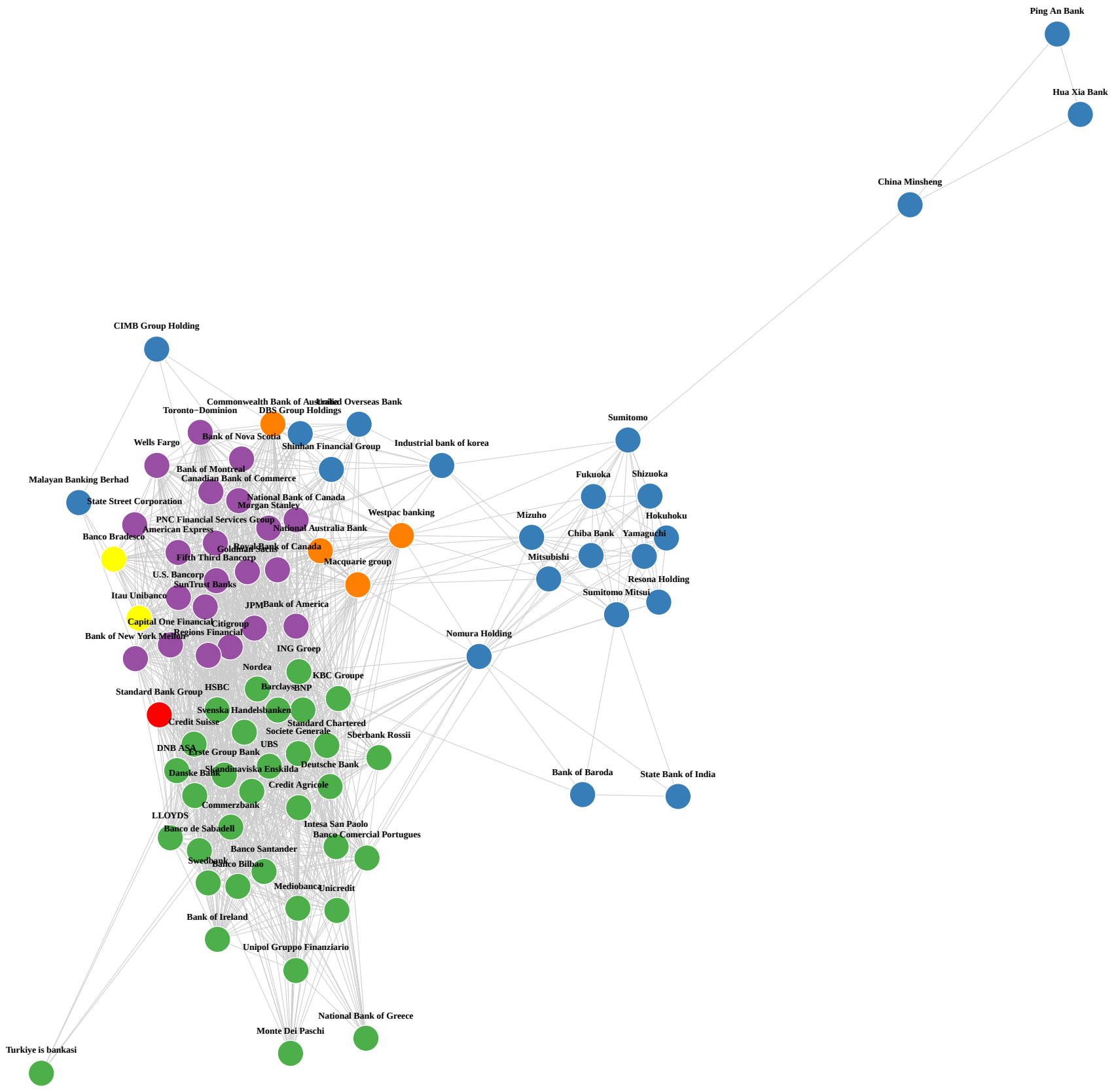

Figure 1: Global banking connectedness network (Demirer et al. 2018). Each node represents a financial institution, with edges reflecting institutional links. Node colours indicate regional origin: ● North America, ● Europe, ● Asia, ● Oceania, ● South America, and ● Africa.

To model nonstationary time series, the statistical literature has traditionally relied on the foundational concept of evolutionary spectra introduced by Priestley (1965). This paradigm was subsequently revolutionized by Dahlhaus (1997) through the mathematical formalization of *locally stationary processes*, which model time series whose statistical properties change smoothly over time, allowing them to be approximated locally by stationary processes. While these classical models operate in the Fourier domain, wavelet-based models have emerged as an invaluable tool for extracting highly localized features simultaneously across both time and frequency. The foundation for time- and scale-varying spectral modelling is the univariate Locally Stationary Wavelet (LSW) process introduced by Nason et al. (2000).

This was extended to model localized linear dependencies between pairs of nonstationary series by Sanderson et al. (2010), later generalized to larger, unconstrained systems via the Multivariate Locally Stationary Wavelet (MvLSW) framework by Park et al. (2014), and to continuous-time processes by Palasciano et al. (2025).

To represent and analyze multivariate time series structured by a known graph topology, an increasingly popular approach is via so-called *network time series* models. In particular, parsimonious time-domain models—most notably Generalized Network Autoregressive (GNAR) processes (Knight et al. 2016, 2020)—model high-dimensional interactions via node neighbourhood parametrizations, a paradigm recently extended to accommodate time-varying parameters and latent groupings (Chen et al. 2025, Li et al. 2026). However, existing network time series models are unsuitable for representing nonstationary time series. Models which explicitly, and practically, capture time-varying, scale-dependent cross dependencies while honouring the underlying graph topology are distinctly lacking in the current literature, despite such a modelling framework being vital to correctly extract insight from data such as the volatility series considered in this work.

This article fills this gap, introducing Locally Stationary Wavelet processes on Networks (**Net-LSW**), a new framework for modelling multiscale, time-varying dependencies that explicitly incorporates the network structure. Unlike traditional unconstrained multivariate approaches, the Net-LSW process restricts the node-specific amplitudes to a diagonal transfer function matrix and encodes the graph architecture into the covariance structure of the process's random innovation increments. By embedding the network topology into these internal innovation dynamics, our framework perfectly aligns the time-varying, scale-dependent spectral properties of the data with the network structure.

A first theoretical contribution of this article is the formal connection between these network-structured innovations and the foundational time series graphical modelling literature established by Dahlhaus (2000*a*). By shifting from the classical stationary Fourier paradigm to our localized wavelet domain model, we introduce a generalized notion of a *local partial correlation graph*. This multiscale network formulation complements recent process-wide conditional specifications designed for stationary series (Bhattacharya et al. 2025), as well as infinite-dimensional operator representations used to study nonstationary graph structures for time-varying vector autoregressive models (Basu & Subba Rao 2023, Krampe & Subba Rao 2024), but which stop short of developing graphical model estimation in this setting. Crucially, we establish a fundamental structural equivalence under the Net-LSW framework: the absence of an edge between nodes in the underlying physical network exactly corresponds to a structural zero entry in the associated localized network precision matrix across all wavelet scales and rescaled time points. This direct mathematical relationship allows the inverse evolutionary wavelet spectral domain to uniquely reflect the graph structure.

The second theoretical contribution of our work is at the level of the estimation of the local cross-nodal dependence quantities under the proposed model, where our approach differs fundamentally from traditional time series graphical models and conventional multivariate wavelet estimation. Traditional estimation methods for (unconstrained) nonstationary wavelet models apply a joint linear matrix correction across all scales simultaneously (Park et al. 2014). However, this multiscale mixing contaminates the network's topology constraints, meaning that enforced structural zeros in the true parameters would not map to structural zeros in the final inverse evolutionary spectrum. Drawing on Wu et al. (2026), who introduced a subprocess-based approach in the context of multiscale wavelet coherence estimation, we resolve these unsatisfactory properties of current methods by developing a novel network-constrained *scale-by-scale estimation* procedure. This approach enables us to formulate a finite-dimensional, convex, network-penalized optimisation problem for each individual scale. We establish the mean-square consistency of our local smoothed subprocess spectral estimators and prove that optimization via graphical covariance selection (Hastie et al. 2009) yields unique, consistent spectral estimators faithful to the network topology. Finally, since latent subprocesses are unobservable in practice, we introduce a robust projection method to reconstruct them from the aggregate observed coefficients, proving that (asymptotically) their underlying covariance structures match.

To the best of our knowledge, Net-LSW provides the first unified pipeline for modelling and feasible estimation of nonstationary network time series co-dependencies. We illustrate its utility by analyzing the time-scale dependency structures of daily stock price volatilities across a high-dimensional network of major international banking institutions (see Figure 1). We show that direct cross-bank dependencies shift dynamically across both time windows and wavelet scales, capturing clear structural changes during the Brexit referendum, the COVID-19 pandemic, and global macro-financial shocks.

The remainder of this article is organized as follows. In Section 2, we establish notation and review the necessary background material. Section 3 details the mathematical formulation of the novel Net-LSW process and its essential properties. In Section 4, we propose the wavelet-domain Local Partial Correlation Graph and prove the structural equivalence between missing network edges and zeros in the localized precision matrix. Section 5 presents our subprocess-based inference framework and establishes its asymptotic consistency theory. Section 6 demonstrates the application of the proposed framework to the motivating data. The Supplementary Material contains all technical proofs and auxiliary derivations, as well as an extensive simulation study that reinforces the theoretical properties.

## 2 Background

In this section we introduce the pertinent theoretical background for the techniques proposed in this article, and start with an overview of locally stationary wavelet modelling.

## 2.1 Network and Data Notation

In what follows, we assume that the data are observed on a network, here understood as a connected graph. We denote the node set as $\mathcal{N} = \{1, \ldots, n\}$ with $n \in \mathbb{N}$ nodes connected by a set of $m$ edges, denoted by $\mathcal{E}$. Let the neighbourhood or connection set of node $p$ be denoted by $\text{ne}(p) = \{q \in \mathcal{N} : (p, q) \in \mathcal{E}\}$. The entire network is formally characterized by the graph $\mathcal{G} = (\mathcal{N}, \mathcal{E})$.

The primary scope of this work is to build a framework for the modelling and subsequent analysis of second order nonstationary processes recorded through time at each node of this known, fixed network $\mathcal{G}$. We denote the observed time series as $\{X_t^{(p)}\}_{t=0}^{T-1}$, where $p \in \mathcal{N}$ indexes the network nodes and $t = 0, \ldots, T-1$ represents discrete time.

The complete collection of these process observations across the network nodes is written as $\{\mathbf{X}_t^{\mathcal{N}}\}_{t=0}^{T-1}$, where the vector $\mathbf{X}_t^{\mathcal{N}}$ collects the observed values across all nodes $\mathcal{N}$ at a specific time index $t$. We drop the superscript $(^{\mathcal{N}})$ when the time series are treated simply as a standard multivariate process and no explicit network structure is assumed.

## 2.2 Brief Review of Locally Stationary Wavelet (LSW) Processes

The foundation for multivariate time-varying dependence modelling using wavelets lies in the multivariate locally stationary wavelet process, introduced by Park et al. (2014) (see also Sanderson et al. (2010) for the bivariate case). This model extends the univariate LSW framework of Nason et al. (2000) to model a collection of $n$ nonstationary signals, $\{\mathbf{X}_t\}_{t=0}^{T-1}$, where $\mathbf{X}_t$ is an $n \times 1$ vector of observations, for the purpose of estimating time- and scale-dependent linear dependence between the series. The MvLSW process $\{\mathbf{X}_t\}$ has a representation

$$\mathbf{X}_t = \sum_{j \geq 1} \sum_{k \in \mathbb{Z}} \mathbf{V}_j \left( \frac{k}{T} \right) \psi_{j,k}(t) \boldsymbol{\xi}_{j,k},$$

where $\mathbf{V}_j(\frac{k}{T})$ is a lower-triangular $n \times n$ transfer function matrix at each scale $j$ and location $k$; $\{\boldsymbol{\xi}_{j,k}\}_{j,k}$ is a multivariate vector of zero-mean, unit variance, uncorrelated random increments, and $\{\psi_{j,k}\}_{j,k}$ is a set of discrete non-decimated wavelets (Nason et al. 2000).

The MvLSW second order structure is governed by the collection across scales $j$ and rescaled times $z \in (0,1)$ of $n \times n$ Local Wavelet Spectral (LWS) matrices, $\mathbf{S}_j(z)$, defined as

$$\mathbf{S}_j(z) = \mathbf{V}_j(z)\, \mathbf{V}_j(z)^\top, \tag{1}$$

and whose diagonal elements, $S_j^{(p,p)}(\cdot)$, are the evolutionary wavelet spectra (EWS) for channel $p$, and the off-diagonal elements, $S_j^{(p,q)}(\cdot)$, are the evolutionary wavelet cross-spectra between channels $p$ and $q$. The spectral-domain quantities are in a one-to-one correspondence with their time-domain counterparts, namely the localized process covariance. The localized (cross-)covariance between channels $p$ and $q$ at rescaled time $z$ and lag $\tau$ is

$$c^{(p,q)}(z,\tau) = \sum_{j \geq 1} S_j^{(p,q)}(z)\, \Psi_j(\tau),\ \forall z,\ \tau, \tag{2}$$

where $\Psi_{j\ell}(\cdot) = \sum_{n \in \mathbb{Z}} \psi_j(n)\psi_\ell(n+\cdot)$ is the cross-scale autocorrelation wavelet, here expressed at single scale $j = \ell$ (see Killick et al. (2020) for their detailed properties). When $p = q$, equation (2) gives the channel $p$ localized autocovariance.

A key dependence measure between channels $p$ and $q$ in the MvLSW model is the locally stationary wavelet coherence at scale $j$ and rescaled time $z$, first introduced in the bivariate setting (Sanderson et al. 2010) as

$$\rho_j^{(p,q)}(z) = \frac{S_j^{(p,q)}(z)}{\left\{S_j^{(p)}(z)\, S_j^{(q)}(z)\right\}^{1/2}}.$$

This measure provides the wavelet domain correlation between the innovations of the two processes at a given scale $j$ and rescaled time $z$, taking values in $[-1,1]$ and having the usual linear dependence interpretation, including indirect associations. We adopt the notation $\{\boldsymbol{\rho}_j^{(p,q)}(z)\}_{j,k}$ for the scale- and time-dependent wavelet coherence matrix function over all pairs of channels $p, q \in \{1, \ldots, n\}$. Park et al. (2014) additionally introduced the concept of the locally stationary wavelet *partial* coherence for the MvLSW process, $\{\boldsymbol{\eta}_j^{(p,q)}(z)\}_{j,k}$: this measure quantifies the direct linear association between two channels, $p$ and $q$, by removing the linear effects of all other observed signals.

Unlike the bivariate LSW framework of Sanderson et al. (2010), which partitions the dependence structure between the transfer functions and the process innovations, the MvLSW framework of Park et al. (2014) encapsulates the entire spectral structure (including cross-channel dependence) within the transfer function matrices of process "amplitudes", $\{\mathbf{V}_j(z)\}_{j,k}$. This key distinction permits the straightforward estimation of wavelet partial coherence and leads to a computationally efficient estimation scheme; our model formulation below is motivated by this feature of the MvLSW framework.

# 3 Novel Nonstationary Process over the Nodes of a Network (Net-LSW)

We now give details of our proposed modelling approach for the nonstationary process observed on a network. The proofs of the results presented in this section can be found in the Supplementary Material, Section A.

**Definition 3.1 (Net-LSW process).** *A sequence of stochastic processes $\{\mathbf{X}_t^{\mathcal{N}}\}_{t=0}^{T-1}$ is a locally stationary wavelet process over the node set $\mathcal{N}$ of a network $\mathcal{G} = (\mathcal{N}, \mathcal{E})$ with known edge set $\mathcal{E}$ (Net-LSW), if it admits the following representation in the mean-squared sense*

$$\mathbf{X}_t^{\mathcal{N}} = \sum_{j\geq 1} \sum_{k\in\mathbb{Z}} \mathbf{W}_j^{\mathcal{N}} \left(\frac{k}{T}\right) \psi_{j,k}(t) \boldsymbol{\xi}_{j,k}^{\mathcal{N}}, \tag{3}$$

*where*

- $\{\psi_{j,k}(t)\}_{j,k}$ *is a discrete non-decimated wavelet family across scales $j$ and times $k$,*

- $\mathbf{W}_j^{\mathcal{N}}(z) = \left[diag\left(W_j^{(p,p)}(z)\right)\right]_{p\in\mathcal{N}}$ *is an* $n \times n$ *diagonal matrix of individual-node amplitude (transfer) functions at scale* $j$ *and rescaled time* $z = \frac{k}{T}$*. The transfer functions are assumed to be Lipschitz continuous in* $z$*, with Lipschitz constants* $\{L_j^{(p,p)}\}_j$ *collectively admitting* $\sum_{j\geq 1} 2^j L_j^{(p,p)} < \infty$ *for each node* $p$*.*

- $\{\boldsymbol{\xi}_{j,k}^{\mathcal{N}}\}_{j,k}$ *is a jointly multivariate normal random sequence of* $n \times 1$ *(column) zero-mean innovation vectors satisfying Properties 3.1 and 3.2 below.*

**Property 3.1.** *As the proposed model seeks to concomitantly reflect the nonstationarity in the process at each node and the multiscale cross-correlation between signals recorded at different nodes, we define the covariance structure of the increments* $\{\boldsymbol{\xi}_{j,k}^{\mathcal{N}}\}_{j,k}$ *to be for any nodes* $p$*,* $q \in \mathcal{N}$ *and time* $k$*,*

$$\begin{aligned}\operatorname{cov}(\xi_{j,k}^{(p)}, \xi_{j',k'}^{(p)}) &= \delta_{j,j'}\delta_{k,k'}, \ \textit{for any scales } j,\, j', \\ \operatorname{cov}(\xi_{j,k}^{(p)}, \xi_{j',k'}^{(q)}) &= \delta_{j,j'}\delta_{k,k'}\lambda_j^{(p,q)}\left(\frac{k}{T}\right),\end{aligned}$$

*where* $\lambda_j^{(p,q)}(\cdot)$ *are assumed to be Lipschitz continuous through time, and* $|\lambda_j^{(p,q)}(\cdot)| \leq 1$ *with* $\lambda_j^{(p,p)}(\cdot) = 1$ *ensuring unit nodal increment variance, across all scales* $j$*.*

*At each scale* $j$ *and rescaled time* $z$*, we define the* $n \times n$ *process increment covariance matrix* $\boldsymbol{\Lambda}_j^{\mathcal{N}}(z) = \left[\lambda_j^{(p,q)}(z)\right]_{p,q\in\mathcal{N}}$*, whose diagonal elements are by convention* $\lambda_j^{(p,p)}(z) = 1, \forall p \in \mathcal{N}$*. We denote the* $n \times n$ *increment concentration matrix by* $\boldsymbol{\Gamma}_j^{\mathcal{N}}(z) = \left(\boldsymbol{\Lambda}_j^{\mathcal{N}}(z)\right)^{-1}$*.*

The following property formally links the network topology to the concentration matrix of the process increments, providing a multiscale characterization of the graph.

**Property 3.2.** *The process structure reflects the underlying network* $\mathcal{G}$ *such that the absence of an edge,* $(p,q) \notin \mathcal{E}$*, corresponds to a zero entry in the increment concentration matrix,*

$$\left[\boldsymbol{\Gamma}_j^{\mathcal{N}}(z)\right]_{p,q} = 0, \quad \textit{for all } j, z. \tag{4}$$

We now introduce a time–scale specific edge decomposition of process dependence in our network time series, to which we will refer as the *network cross-spectrum.*

**Definition 3.2.** *We define the evolutionary Net-LSW cross-spectrum at scale $j$ and rescaled time $z$ as the $n \times n$ symmetric, positive semi-definite matrix $\mathbf{S}_j^{\mathcal{N}}(z) = \left[S_j^{(p,q)}(z)\right]_{p,q \in \mathcal{N}}$, as*

$$\mathbf{S}_j^{\mathcal{N}}(z) = \mathbf{W}_j^{\mathcal{N}}(z)\,\mathbf{\Lambda}_j^{\mathcal{N}}(z)\,\mathbf{W}_j^{\mathcal{N}}(z)\,. \tag{5}$$

Equivalently, using the diagonal structure of the amplitude matrix, we explicitly obtain the Net-LSW cross-spectrum associated with nodes $p$ and $q$,

$$S_j^{(p,q)}(z) = W_j^{(p,p)}(z)\,\lambda_j^{(p,q)}(z)\,W_j^{(q,q)}(z)\,. \tag{6}$$

Our methodological development in Section 5 considers a transformed version of the spectrum, in order to exploit its theoretical properties. We introduce this rephrased spectrum next.

**Definition 3.3.** *We define the evolutionary Net-LSW $\boldsymbol{\beta}$-spectrum at scale $j$ and rescaled time $z = \frac{k}{T}$, as the $n \times n$ matrix $\boldsymbol{\beta}_j^{\mathcal{N}}(z) = \left[\beta_j^{(p,q)}(z)\right]_{p,q \in \mathcal{N}}$, given by*

$$\boldsymbol{\beta}_j^{\mathcal{N}}(z) = \sum_{\ell=1}^{J} A_{j,\ell}\mathbf{S}_\ell^{\mathcal{N}}(z)\,, \tag{7}$$

where $J = \log_2(T)$ and the matrix $\mathbf{A}$ is the discrete autocorrelation wavelet inner product matrix defined in Nason et al. (2000).

In a similar fashion to (2) for the standard MvLSW model, we can quantify the dependence between series at two nodes in a network as follows.

**Definition 3.4.** *We define the localized cross-covariance between series at nodes $p$ and $q$ at rescaled time $z$ and lag $\tau$, as*

$$c^{(p,q)}(z,\tau) = \sum_{j \geq 1} S_j^{(p,q)}(z)\,\Psi_j(\tau).$$

**Proposition 3.1.** *For a Net-LSW process $\{\mathbf{X}_t^{\mathcal{N}}\}_{t=0}^{T-1}$ as in Definition 3.1, the localized cross-covariance is connected to the process cross-covariance for nodes $p$ and $q$ through*

$$c^{(p,q)}(z,\tau) = \operatorname{cov}\left(X_{\lfloor zT \rfloor,T}^{(p)}, X_{\lfloor zT \rfloor+\tau,T}^{(q)}\right) + \mathcal{O}(T^{-1}).$$

The proof makes use of similar techniques to those in the proof of Proposition 1 in Sanderson et al. (2010), and is thus omitted here. This relationship captures node dependence not only across time, but also across the network nodes.

Our focus is the (wavelet-domain) time-varying normalized dependence across the network, namely the local Net-LSW cross-coherence and its partial version, which we next define.

**Definition 3.5.** *We define the localized Net-LSW cross-coherence at scale $j$ and rescaled time $z$, as the $n \times n$ matrix*

$$\boldsymbol{\rho}_j^{\mathcal{N}}(z) = diag\left(\mathbf{S}_j^{\mathcal{N}}(z)\right)^{-1/2} \mathbf{S}_j^{\mathcal{N}}(z)\, diag\left(\mathbf{S}_j^{\mathcal{N}}(z)\right)^{-1/2}.$$

**Property 3.3.** *For a Net-LSW process $\{\mathbf{X}_t^{\mathcal{N}}\}_{t=0}^{T-1}$ as in Definition 3.1, the process coherence and increment covariance matrices coincide across all scales and times, specifically*

$$\boldsymbol{\rho}_j^{\mathcal{N}}(\cdot) = \boldsymbol{\Lambda}_j^{\mathcal{N}}(\cdot)\,, \forall j. \tag{8}$$

**Definition 3.6.** *Let $\boldsymbol{\Theta}_j^{\mathcal{N}}(z) = \left(\mathbf{S}_j^{\mathcal{N}}(z)\right)^{-1}$ be the precision matrix and $\mathbf{H}_j(z)$ be a diagonal matrix whose pth diagonal entry is $\left(\Theta_j^{(p,p)}(z)\right)^{-1/2}$. The localized Net-LSW partial coherence at scale $j$ and rescaled time $z$ is defined as the $n \times n$ matrix,*

$$\boldsymbol{\eta}_j^{\mathcal{N}}(z) = -\mathbf{H}_j(z)\,\boldsymbol{\Theta}_j^{\mathcal{N}}(z)\,\mathbf{H}_j(z)\,. \tag{9}$$

The off-diagonal elements of this matrix, $[\boldsymbol{\eta}_j^{\mathcal{N}}(z)]_{p,q}$, $p \neq q$, correspond to the localized (at rescaled time $z$) scale-dependent wavelet partial coherence between the process at nodes $p$ and node $q$. By the definition of the partial coherence, this is the coherence between the given nodes after the linear effects of all other nodes have been removed.

**Property 3.4.** *For a Net-LSW process $\{\mathbf{X}_t^{\mathcal{N}}\}_{t=0}^{T-1}$ as in Definition 3.1 and denoting $\widetilde{\mathbf{H}}_j(z) = \mathbf{H}_j(z)\left(\mathbf{W}_j^{\mathcal{N}}(z)\right)^{-1/2}$, the process partial coherence and increment concentration matrices are connected as follows at all scales $j$ and rescaled times $z$,*

$$\boldsymbol{\eta}_j^{\mathcal{N}}(z) = -\widetilde{\mathbf{H}}_j(z)\,\boldsymbol{\Gamma}_j^{\mathcal{N}}(z)\,\widetilde{\mathbf{H}}_j(z)\,. \tag{10}$$

**Remark 1 (Implication for inverse spectrum).** *Property 3.4 establishes a direct, entry-wise link between the process partial coherence and the increment concentration matrix, revealing that*

$$\left[\boldsymbol{\Gamma}_j^{\mathcal{N}}(z)\right]_{p,q} = 0 \iff \left[\boldsymbol{\eta}_j^{\mathcal{N}}(z)\right]_{p,q} = 0, \quad \textit{for all } j,\ z.$$

To formalize how structural constraints or the absence of network dependencies map directly onto these zero-valued matrix entries, we next introduce our proposed graphical model.

To avoid notational clutter, in the remainder of the paper we drop the $\mathcal{N}$ superscript for Net-LSW process-derived quantities, whenever there is no potential confusion with the multivariate counterpart.

# 4 Proposed Local Partial Correlation Graph

We next propose a generalized notion for the partial correlation graph of Dahlhaus (2000*a*) suitable for cases when (i) the multivariate process observed over the network is no longer stationary; and we characterize (ii) cross-nodal dependencies in the *wavelet domain* rather than in the Fourier domain. The proofs of the results in this section can be found in Supplementary Material, Section B.

**Remark 2 (Notation convention).** *Throughout this development, we shall follow the locally stationary paradigm that enhances the process quantities by localising them at rescaled times. As in Section 3, for any Net-LSW process* $\{\mathbf{X}_t^{\mathcal{N}}\}$*, we shall use notation such as* $c^{(p,q)}(z,\tau)$ *and* $\{S_j^{(p,q)}(z)\}_j$ *for the local (cross-) covariance and spectrum between any pair of nodes* $p$ *and* $q$*, respectively, where* $z$ *denotes rescaled time,* $\tau$ *denotes lag, and* $j$ *is the wavelet scale (see* (6) *for their one-to-one connection).*

*To emphasize the graphical aspect of the data and model, we also establish a matrix-based notation. For any fixed nodal pair* $(a,b)$ *that may, or may not, correspond to an*

*edge in the network, we* block-partition *the* $n \times n$ *process localized covariance matrix* $\mathbf{C}(z,\tau) = \left[c^{(p,q)}(z,\tau)\right]_{p,q \in \mathcal{N}}$ *as*

$$\mathbf{C}(z,\tau) = \left[\begin{array}{c|c} \mathbf{A}(z,\tau) & \mathbf{B}(z,\tau) \\ \hline \mathbf{B}(z,\tau)^\top & \mathbf{D}(z,\tau) \end{array}\right], \tag{11}$$

*where*

*(i) the upper-left block is the* $2 \times 2$ *matrix* $\mathbf{A}(z,\tau) = \left[c^{(p,q)}(z,\tau)\right]_{p,q \in \{a,b\}}$ *containing the localized auto- and cross-covariances between the specific nodes* $a$ *and* $b$*,*

*(ii) the upper-right block is the* $2 \times (n-2)$ *localized cross-covariance matrix* $\mathbf{B}(z,\tau) = \left[c^{(a,p)}(z,\tau), c^{(b,q)}(z,\tau)\right]_{p \in \{a,b\}, q \notin \{a,b\}}$*,*

*(iii) the lower-right block* $\mathbf{D}(z,\tau) = \left[c^{(p,q)}(z,\tau)\right]_{p,q \notin \{a,b\}}$ *encompasses the localized cross-covariance information at all nodes other than* $a$ *and* $b$*.*

**Definition 4.1.** For a Net-LSW process $\{\mathbf{X}_t^{\mathcal{N}}\}$ observed over a network node set and for fixed nodes $a$ and $b$, we define the bivariate residual locally stationary process $\{\boldsymbol{\epsilon}_t\}_{t=0}^{T-1}$ as

$$\boldsymbol{\epsilon}_t := \left[\epsilon_t^{(p)} = X_t^{(p)} - \text{Proj}_{\overline{\text{sp}}\{\mathbf{X}_t^{-ab}\}}(X_t^{(p)})\right]_{p \in \{a,b\}} \in \mathbb{R}^2,$$

with the residual process obtained by removing the linear dependencies attributed to the remaining $(n-2)$ nodes in the network, $\{\mathbf{X}_t^{-ab}\}$, via the usual projection operator subject to minimizing the corresponding mean squared error (Brockwell & Davis 2016).

To ensure clarity of exposition, whenever necessary, we enhance the Remark 2 notation by introducing a superscript to indicate which process the quantity refers to, e.g., $\mathbf{C}^X(z,\tau)$ for the original process local covariance $n \times n$ matrix and $\mathbf{C}^\epsilon(z,\tau)$ for the residual local covariance $2 \times 2$ matrix.

**Definition 4.1 (Local partial correlation graph).** Let $\{\mathbf{X}_t^{\mathcal{N}} = (X_t^{(p)})_{p \in \mathcal{N}}\}_{t=0}^{T-1}$ be an $n$-dimensional locally stationary process (Net-LSW) as in Definition 3.1, observed over the nodes $p \in \mathcal{N}$ of a network $\mathcal{G} = (\mathcal{N}, \mathcal{E})$. We define the graph $G = (\mathcal{N}, E)$ to be the local partial correlation graph of $\{\mathbf{X}_t^{\mathcal{N}}\}$ if the absence of an edge between two specific nodes, say $(a, b) \notin E$, is equivalent to the (symmetrical) local residual covariance matrix $\mathbf{C}^{\epsilon}(z, \tau)$ being diagonal, i.e., $[\mathbf{C}^{\epsilon}(z, \tau)]_{a,b} = 0$, $\forall z$, $\tau$.

**Remark 3** **(Local stationarity equivalence).** *As in the classical stationary setup, the condition that specific nodes a and b are not connected via an edge in the graph G is equivalent to the conditional covariance of the processes* $\{X_t^{(a)}\}$ *and* $\{X_t^{(b)}\}$ *given the information at the remaining* $(n-2)$ *nodes,* $\{\mathbf{X}_t^{-ab}\}$*, being zero for all times t and lags* $\tau$*. The use of the word "local" in Definition 4.1 reflects the time-localization inherent to the Net-LSW framework – recall that as shown in Proposition 3.1, a unique localized covariance structure asymptotically characterizes the process covariance.*

**Remark 4** **(Covariance connection to Schur complement).** *From Definition 4.1 coupled with Remark 3, the local process conditional covariance is given by the localized counterpart of the Schur complement of the* $(n-2) \times (n-2)$ *covariance matrix,* $\mathrm{cov}(\mathbf{X}_t^{-ab}, \mathbf{X}_{t+\tau}^{-ab})$*, of the process that excludes all information from nodes a, b, in the* $n \times n$ *full process covariance matrix,* $\mathrm{cov}(\mathbf{X}_t^{\mathcal{N}}, \mathbf{X}_{t+\tau}^{\mathcal{N}})$*, namely*

$$\mathbf{C}^{\epsilon}(z, \tau) := \mathbf{C}^{X}(z, \tau)/\mathbf{D}(z, \tau) = \mathbf{A}(z, \tau) - \mathbf{B}(z, \tau)\left(\mathbf{D}(z, \tau)\right)^{-1}\mathbf{B}(z, \tau)^{\top}.$$

*It is this connection that we shall follow through into the wavelet spectral domain (see also Dahlhaus 2000a, for related Fourier-based developments).*

We next establish the expression for the spectral structure of the residual process, and show that node $(a, b)$-unconnectedness in $G$ is reflected by the presence of a zero-valued entry in the corresponding inverse spectral matrix, $[\mathbf{S}_j^{\epsilon}(z)]_{a,b}^{-1}$ for all scales $j$ and rescaled times $z$.

**Proposition 4.1.** *The spectral matrices* $\{\mathbf{S}_j^{\epsilon}(z)\}_j$ *of the residual process* $\{\boldsymbol{\epsilon}_t\}$ *in Defini-* *tion 4.1 have the following structure inherited from the original process* $\{\mathbf{X}_t^{\mathcal{N}}\}$*:*

$$\mathbf{S}_j^{\epsilon}(z) = \mathbf{A}_j^S(z) - \mathbf{B}_j^S(z)\left(\mathbf{D}_j^S(z)\right)^{-1}\mathbf{B}_j^S(z)^\top, \tag{12}$$

$$= \mathbf{S}_j^X(z)/\mathbf{D}_j^S(z),\ \forall j,\ z, \tag{13}$$

*where the scale- and time-dependent matrices* $\mathbf{A}_j^S(z)$*,* $\mathbf{B}_j^S(z)$ *and* $\mathbf{D}_j^S(z)$ *denote respectively the* $2 \times 2$*,* $2 \times (n-2)$ *and* $(n-2) \times (n-2)$ *dimensional blocks that partition the spectral matrix of the process* $\{\mathbf{X}_t^{\mathcal{N}}\}$ *as*

$$\mathbf{S}_j^X(z) = \left[\begin{array}{c|c} \mathbf{A}_j^S(z) & \mathbf{B}_j^S(z) \\ \hline \mathbf{B}_j^S(z)^\top & \mathbf{D}_j^S(z) \end{array}\right]. \tag{14}$$

The proof exploits the properties of Schur decompositions in conjunction with the locally stationary covariance and spectral wavelet matrix connections. As an aside, let $\mathbf{B}_{1,j}(z)$ and $\mathbf{B}_{2,j}(z)$ denote the first and second rows of the $2 \times (n-2)$ cross-spectral block $\mathbf{B}_j^S(z)$ corresponding to nodes $a$ and $b$, respectively. A diagonal structure of the Schur complement of $\mathbf{D}_j^S(z)$ in $\mathbf{S}_j^X(z)$ implies that the original cross-spectral entry satisfies

$$S_j^{(a,b)}(z) = \mathbf{B}_{1,j}(z)\left(\mathbf{D}_j^S(z)\right)^{-1}\left(\mathbf{B}_{2,j}(z)\right)^\top, \quad \forall j,\ z. \tag{15}$$

The residual spectral equation (12) is the localized wavelet counterpart of the Fourier result in Brockwell & Davis (2016) that forms the essence of the edge identifiability proposal in Dahlhaus (2000*b*, equation (2.3)), as well as in our framework, as described next.

**Proposition 4.2.** *For a Net-LSW process as in Definition 3.1 observed over a graph* $\mathcal{G}$*, its local partial correlation graph* $G$ *coincides with* $\mathcal{G}$*. That is, an edge* $(a, b)$ *is absent from* $\mathcal{E}$ *if and only if* $[\mathbf{S}_j^X(z)]_{a,b}^{-1} = 0$ *for all scales* $j$ *and locations* $z$*. Equivalently,* $(a, b) \notin \mathcal{E}$ *if and only if the local partial spectral coherence satisfies* $\boldsymbol{\eta}_j^X(z) = 0$ *for all* $j, z$*.*

This explicit linkage between the zero pattern of the increment concentration matrix $\boldsymbol{\Gamma}_j(z)$ (see also Remark 1) and the missing edges of the local partial correlation graph $G$ aligns perfectly with classic time series graphical model theory (Dahlhaus 2000*a*, Basu & Subba Rao 2023), extending the equivalence of conditional uncorrelatedness and innovation precision structure directly into the nonstationary wavelet spectrum. In other words, the graph $\mathcal{G}$ over which the process is observed is in fact its associated local partial correlation graph.

Section 5 next introduces a new framework for the time-localized estimation of multiscale (cross) nodal dependencies intrinsic to a Net-LSW process and formally characterizes their desirable asymptotic behaviour.

# 5 Net-LSW (Partial) Coherence Estimation

To establish an estimation methodology for the Net-LSW process, we exploit the fact that the network structure is encoded in the inverse spectral matrix $\boldsymbol{\Theta} = \mathbf{S}^{-1}$, as established in Section 4.1. Since the network $\mathcal{G}$ is known, the locations of the zero entries in $\boldsymbol{\Theta}$ corresponding to the complement of the edge set $\mathcal{E}$ are known *a priori*. This information can, in principle, be incorporated into the estimation procedure via appropriate regularization.

A natural approach to *network-agnostic* estimation amounts to relying on the associated wavelet coefficients $\{\mathbf{d}_{j,k}\}_{j,k}$ of the observed *multivariate* process $\{\mathbf{X}_t^{\mathcal{N}}\}_{t=0}^{T-1}$. Under a multivariate MvLSW framework with Gaussian innovations and uniformly absolutely summable covariance functions, these coefficients satisfy asymptotically in $T \to \infty$ (Knight et al. 2024)

$$\mathbf{d}_{j,k} \dot{\sim} N_n(\mathbf{0}, \boldsymbol{\beta}_j(k/T)) \,, \tag{16}$$

where $\{\boldsymbol{\beta}_j(k/T)\}_j$ denotes the scale-dependent rephrased wavelet spectra, obtained as in equation (7) and (uniquely) connected to the wavelet spectra $\{\mathbf{S}_j(k/T)\}_j$ through

$$\mathbf{S}_j\,(k/T) = \sum_{\ell=1}^{J} A^{-1}_{j,\ell} \boldsymbol{\beta}_\ell\,(k/T)\,,\ \forall j,\ k. \tag{17}$$

Here, $A^{-1}_{j,\ell}$ is the $(j, \ell)$ entry of the inverse inner product operator (Nason et al. 2000).

Stemming from the distribution of the wavelet coefficients in (16), one might envisage directly using a typical likelihood-based constrained optimization formulation with the aim of enforcing the known edge set pattern $\mathcal{E}$ (see Hastie et al. (2009) for the a multivariate version lacking the time series aspect; Jiménez-Varón & Knight (2026) for a Fourier spectral-domain version for stationary time series). However, such an approach would instead yield the known zeros in the rephrased inverse spectrum, $\boldsymbol{\beta}_j^{-1}(k/T)$, and (17) highlights that these enforced zero entries would not map to zero entries in the target inverse spectral matrix, $\mathbf{S}_j^{-1}(k/T)$, as needed and indicated by the theoretical evidence in Proposition 4.2.

Rather than operating on the full process, we propose a new approach that focuses on **scale-specific subprocesses**, substantially simplifying the relationship between their associated $\boldsymbol{\beta}$- and $\mathbf{S}$-spectra. As a result, the Gaussian likelihood (Friedman et al. 2007) remains convex when augmented with network-informed penalties on $\mathbf{S}_j^{-1}(k/T)$, yielding a tractable and stable estimation procedure. Hence we next propose a subprocess-based estimation strategy which offers a spectrum $\mathbf{S}$-targeted approach. The proofs of the results in this section appear in Supplementary Material, Section C.

## 5.1 Scale-dependent Net-LSW Subprocesses

The subprocess construction has been introduced by Wu et al. (2026) in the context of a locally stationary multivariate process augmented with cross-scale interactions. From a multivariate viewpoint, such cross-scale interactions are absent in our setup here, but the interactions between series in our modelling setup is network-informed. For our model, we define the subprocesses and their corresponding wavelet coefficients as follows.

**Definition 5.1.** *Let $\{\mathbf{X}_t^{\mathcal{N}}\}$ be a Net-LSW process as in Definition 3.1. The* **scale-$j$ Net-LSW subprocess** $\{\mathbf{X}_{j,t}\}$ *is the component of the signal that exists exclusively at resolution*

*$j$, and is defined by*

$$\mathbf{X}_{j,t} = \sum_{k\in\mathbb{Z}} \mathbf{W}_j^{\mathcal{N}}(k/T)\,\psi_{j,k}(t)\,\boldsymbol{\xi}_{j,k}^{\mathcal{N}}.$$

Viewed as a process itself, the scale-$j$ subprocess $\{\mathbf{X}_{j,t}\}_t$ has a local covariance function that is defined as $c_j^{(p,q)}(z,\tau) = S_j^{(p,q)}(z)\Psi_j(\tau)$, and akin to Wu et al. (2026), it can be shown to approximate the process covariance, $\left|\mathrm{cov}(X_{j,[zT]}^{(p)}, X_{j,[zT]+\tau}^{(q)}) - c_j^{(p,q)}(z,\tau)\right| = \mathcal{O}(2^{-j}T^{-1})$.

**Definition 5.2.** *The* **scale $j$-subprocess wavelet coefficient** $\{\mathbf{d}_{\ell,k}^j\}_{\ell,k}$ *is the vector obtained by applying the non-decimated wavelet transform at scales $\ell = 1,\ldots,J$ and time-location $k \in \mathbb{Z}$ to the isolated subprocess $\{\mathbf{X}_{j,t}\}$, namely*

$$\mathbf{d}_{\ell,k}^j = \sum_t \mathbf{X}_{j,t}\,\psi_{\ell,k}(t),\ \forall \ell,\ k.$$

**Remark 5.** *The superscript $j$ identifies the subprocess corresponding to resolution $j$, while the subscripts $(\ell,k)$ denote the analysis scale and location indices of the wavelet transform.*

**Remark 6.** *Note that $\mathbf{X}_t^{\mathcal{N}} = \sum_{j\geq 1}\mathbf{X}_{j,t}$ follows naturally from the Net-LSW process and subprocess definitions. Hence the non-decimated wavelet coefficient of the full process at scale $\ell$ and location $k$ satisfies $\mathbf{d}_{\ell,k} = \sum_{j\geq 1}\mathbf{d}_{\ell,k}^j$ due to the linearity of the wavelet transform.*

### 5.1.1 Distribution of Subprocess Wavelet Coefficients

To establish a tractable likelihood for the Net-LSW process, we derive the distribution of the subprocess wavelet coefficients for each scale $j$. Combining the scale-dependent subprocess in Definition 5.1 with the formal Definition 5.2, we obtain the wavelet coefficient of scale-$j$ subprocess at an arbitrary analysis scale $\ell$ and location $k$, can be re-written as

$$\mathbf{d}_{\ell,k}^j = \sum_{k'} \left(\Psi_{j,\ell}(k'-k)\mathbf{W}_j^{\mathcal{N}}(k'/T)\right)\boldsymbol{\xi}_{j,k'}^{\mathcal{N}}. \tag{18}$$

The statistical properties and asymptotic distribution of these coefficients are formalized in the following proposition.

**Proposition 5.1.** *Let* $\{\mathbf{X}_{j,t}\}$ *be the scale-j subprocess of a Net-LSW process satisfying Definition 3.1. Then, the subprocess wavelet coefficients* $\mathbf{d}^j_{\ell,k}$ *have mean zero,*

$$\mathbb{E}\left[\mathbf{d}^j_{\ell,k}\right] = \mathbf{0}, \quad \forall \ell, k, \tag{19}$$

*and a covariance structure given by*

$$Cov(\mathbf{d}^j_{\ell,k}, \mathbf{d}^j_{\ell,k+\nu}) = A^{(\nu)}_{j,\ell}\mathbf{S}_j(k/T) + \mathcal{O}(T^{-1}), \tag{20}$$

*where* $A^{(\nu)}_{j,\ell} = \sum_\tau \Psi_j(\tau)\Psi_\ell(\tau+\nu)$ *is the* $(j,\ell)$*-element of the shifted wavelet inner product matrix* $\mathbf{A}^{(\nu)}$ *(Sanderson et al. 2010, Killick et al. 2020), and* $\mathbf{S}_j(\cdot)$ *is the wavelet spectral matrix of the original process* $\{\mathbf{X}^{\mathcal{N}}_t\}$*, at scale* $j$*.*

*The contemporaneous subprocess wavelet coefficients at the matching scale* $(\ell = j,\ \nu = 0)$ *asymptotically follow the multivariate normal distribution as* $T \to \infty$ *for all* $(j,k)$ *pairs*

$$\mathbf{d}^j_{j,k} \dot{\sim} N_n\left(\mathbf{0}, A_{j,j}\mathbf{S}_j(k/T)\right). \tag{21}$$

The proof manipulates the scale-$j$ subprocess definition coupled with the random vector increment properties. This result highlights that the spectral matrix $\mathbf{S}_j(k/T)$ can now be viewed as a (scaled) covariance matrix, a key property we shall use in Section 5.4.

## 5.2 Subprocess-based Wavelet Periodogram

For a given scale $j$, we define the raw subprocess wavelet periodogram at time $k$ as

$$\mathbf{I}_{j,k} = \mathbf{d}^j_{j,k}\left(\mathbf{d}^j_{j,k}\right)^\top, \tag{22}$$

where we note that since the scale-$j$ subprocess is active at resolution level $j$, the corresponding wavelet coefficients have been chosen to match that level.

## 5.3 Smoothing and Consistency

To obtain a consistent estimator for the local wavelet spectrum $\mathbf{S}_j(k/T)$, we reduce the variance of the raw periodogram in (22) by smoothing over a local temporal neighbourhood. We define the smoothed subprocess periodogram at scale $j$ and rescaled time $z = k/T$ as

$$\widehat{\mathbf{S}}_j(k/T) = \frac{1}{A_{j,j}(2M_T+1)} \sum_{m=-M_T}^{M_T} \mathbf{I}_{j,k+m}, \tag{23}$$

where $M_T$ is a smoothing bandwidth chosen such that $M_T \to \infty$ and $M_T T^{-1} \to 0$ as $T \to \infty$. Denoting the scale-$j$ subprocess wavelet coefficient entry corresponding to node $p$ and scale-time pair $(j,\, k')$ by $d_{j,k'}^{j,(p)}$, the cross-nodal spectral activity for any two nodes $p, q \in \mathcal{N}$ is given by the $(p,q)$-element of this matrix, explicitly written as

$$\widehat{S}_j^{(p,q)}(k/T) = \frac{1}{A_{j,j}(2M_T+1)} \sum_{m=-M_T}^{M_T} d_{j,k+m}^{j,(p)} d_{j,k+m}^{j,(q)}.$$

**Proposition 5.2 (Consistency and Boundedness in Probability).** *Let $\{\mathbf{X}_t^{\mathcal{N}}\}$ be a Net-LSW process as in Definition 3.1. Under the further assumption of local subprocess (auto- and cross-) covariance summability, specifically $\sup_{z\in(0,1)} \sum_{\tau} |c_j^{(p,q)}(z,\tau)| = \mathcal{O}(1)$ for each scale $j$ and node pair $(p,q)$, the estimator in (23) is consistent for the true spectrum, i.e,*

$$\begin{aligned} \widehat{S}_j^{(p,q)}(z) - S_j^{(p,q)}(z) &= o_P(1) \qquad \text{and} \\ \widehat{S}_j^{(p,q)}(z) - S_j^{(p,q)}(z) &= \mathcal{O}_P\left(M_T^{-1/2} \vee (M_T T^{-1})^{1/2}\right), \ \forall j,\ z, \end{aligned}$$

*for a smoothing bandwidth $M_T$ such that $M_T \to \infty$ and $M_T T^{-1} \to 0$ as $T \to \infty$.*

## 5.4 Constrained Likelihood and Node-Wise Estimation

Akin to Hastie et al. (2009), the Gaussian log-likelihood for the scale-$j$, location-$z$ subprocess precision is (up to a proportionality constant)

$$\mathcal{L}(\boldsymbol{\Theta}_j(z)) = \log\det \boldsymbol{\Theta}_j(z) - \mathrm{Tr}(\widehat{\mathbf{S}}_j(z)\boldsymbol{\Theta}_j(z)).$$

Under the Net-LSW framework, the sparsity of $\boldsymbol{\Theta}_j(z) = \mathbf{S}_j^{-1}(z)$ follows the edge set $\mathcal{E}$. Thus, to enforce the network structure, we define the constrained objective function that employs the spectral estimator in (23) at each $(j, z)$, as

$$\mathcal{L}(\boldsymbol{\Theta}_j(z)) = \log\det \boldsymbol{\Theta}_j(z) - \mathrm{Tr}(\widehat{\mathbf{S}}_j(z)\boldsymbol{\Theta}_j(z)) - \sum_{(p,q)\notin\mathcal{E}} l_j^{(p,q)} \Theta_j^{(p,q)}(z), \tag{24}$$

where $l_j^{(p,q)}$ are the Lagrange multipliers restricting $\Theta_j^{(p,q)}(z) = 0$ for non-adjacent node pairs. This objective will be optimized using the paradigm proposed by Hastie et al. (2009) leading to the final network-constrained estimators, $\widehat{\boldsymbol{\Theta}}_j^*(z)$ and $\widehat{\mathbf{S}}_j^*(z) = \left(\widehat{\boldsymbol{\Theta}}_j^*(z)\right)^{-1}$.

**Proposition 5.3.** *Let $\{\mathbf{X}_t^{\mathcal{N}}\}_{t=0}^{T-1}$ be a Net-LSW process as in Definition 3.1. Let $\widehat{\mathbf{S}}_j(z)$ be the initial symmetric, positive-definite spectral matrix estimator at scale $j$ and rescaled time $z$, given in (23). Then the network-constrained precision matrix estimator $\widehat{\boldsymbol{\Theta}}_j^*(z)$ that maximizes (24) exists, is unique, and is consistent for the true localized precision structure for all scales $j \geq 1$ and rescaled time points $z \in (0, 1)$ as $T \to \infty$, i.e.,*

$$\widehat{\boldsymbol{\Theta}}_j^*(z) - \boldsymbol{\Theta}_j(z) = o_P(1), \quad \textit{hence} \quad \widehat{\mathbf{S}}_j^*(z) - \mathbf{S}_j(z) = o_P(1). \tag{25}$$

## 5.5 Subprocess Approximation via Wavelet projection

In practice, the latent subprocesses $\{\mathbf{X}_{j,t}\}$ introduced in Definition 5.1 are not directly observable. Instead, the available data consists of the multivariate process $\{\mathbf{X}_t^{\mathcal{N}}\}_{t=0}^{T-1}$ which, under the Net-LSW structural assumption, is connected to its scale-specific components through $\mathbf{X}_t^{\mathcal{N}} = \sum_{j\geq 1} \mathbf{X}_{j,t}$. Consequently, performing scale-dependent network inference requires a robust approximation of these latent components from the observed signal.

To isolate the subprocess in practice, an approximation $\widetilde{\mathbf{X}}_{j,t}$ is constructed via a synthesis of raw analysis coefficients $\mathbf{d}_{j,m} = \sum_s \mathbf{X}_s^{\mathcal{N}} \psi_{j,m}(s)$. Specifically,

$$\widetilde{\mathbf{X}}_{j,t} = \sum_m 2^{-j} \mathbf{d}_{j,m} \psi_{j,m}(t).$$

By reordering the summations, the internal term collapses into the autocorrelation wavelet,

$$\widetilde{\mathbf{X}}_{j,t} = \sum_s 2^{-j} \mathbf{X}_s^{\mathcal{N}} \left( \sum_m \psi_{j,m}(s)\psi_{j,m}(t) \right) = \sum_s 2^{-j} \mathbf{X}_s^{\mathcal{N}} \Psi_j(t-s). \tag{26}$$

The property of additively recovering the process from the true subprocesses also holds for the approximated subprocesses, as $\sum_j 2^{-j}\Psi_j(\tau) = \delta_{\tau,0}$ (Nason et al. 2000) yields

$$\sum_j \widetilde{\mathbf{X}}_{j,t} = \sum_s \mathbf{X}_s^{\mathcal{N}} \sum_j \left( 2^{-j}\Psi_j(t-s) \right) = \sum_s \mathbf{X}_s^{\mathcal{N}} \delta_{t-s,0} = \mathbf{X}_t^{\mathcal{N}}.$$

We subsequently obtain the wavelet coefficients of the approximated subprocess, denoted $\widetilde{\mathbf{d}}_{\ell,k}^{j}$, by re-analysing the approximated subprocess at analysis scale $\ell$

$$\widetilde{\mathbf{d}}_{\ell,k}^{j} = \sum_{t=0}^{T-1} \widetilde{\mathbf{X}}_{j,t} \psi_{\ell,k}(t). \tag{27}$$

**Proposition 5.4.** *Let* $\{\mathbf{X}_t^{\mathcal{N}}\}$ *be a Net-LSW process as in Definition 3.1, whose subprocess approximation and corresponding wavelet coefficients are constructed at every scale as in equations (26) and (27), respectively. Under the assumptions of uniform local process and subprocess (auto- and cross-) covariance absolute summability, and with* $T \to \infty$, *we have*

$$\widetilde{\mathbf{d}}_{j,k}^{j} \dot{\sim} N_n \left( \mathbf{0}, A_{j,j} \mathbf{S}_j(k/T) \right). \tag{28}$$

The property in (28) suggests the use of the smoothed approximated subprocess periodogram estimator at scale $j$ and rescaled time $z = k/T$, namely

$$\widetilde{\mathbf{S}}_j(k/T) = \frac{1}{A_{j,j}(2M_T+1)} \sum_{m=-M_T}^{M_T} \widetilde{\mathbf{d}}_{j,k+m}^{j} \left( \widetilde{\mathbf{d}}_{j,k+m}^{j} \right)^{\top}, \tag{29}$$

with desirable limiting behaviour, as we next show.

**Proposition 5.5 (Consistency and boundedness in probability using approximated subprocesses).** *Let* $\{\mathbf{X}_t^{\mathcal{N}}\}$ *be a Net-LSW process as in Definition 3.1. Under the further assumptions detailed in Proposition 5.4, for each scale* $j$, *rescaled time* $z$ *and node pair*

*$(p,q)$, the estimator in (29) is consistent for the true spectrum in the sense that*

$$\begin{aligned}\widetilde{S}_j^{(p,q)}(z) - S_j^{(p,q)}(z) &= o_P(1) \qquad \text{and} \\ \widetilde{S}_j^{(p,q)}(z) - S_j^{(p,q)}(z) &= \mathcal{O}_P\left(M_T^{-1/2} \vee \left(2^{-j} M_T T^{-1/2}\right)^{1/2}\right),\end{aligned}$$

*for a smoothing bandwidth $M_T$ such that $M_T \to \infty$ and $M_T T^{-1/2} \to 0$ as $T \to \infty$.*

Note that the use of approximated subprocesses results in a weaker convergence rate when compared to that obtained in Proposition 5.2 that uses true subprocesses. Consequently, our choice of smoothing bandwidth departs from the usual choice in the multivariate local stationary environment ($M_T = \lfloor \sqrt{T} \rfloor$ of Park et al. (2014)), and using empirical guidance we propose to use $M_T = \lfloor \log_2^2(T)/2 \rfloor$.

This convergence result provides the theoretical justification for utilizing the approximated estimator in (29) within the constrained optimization framework detailed in Section 5.4 at each $(j, z)$, yielding the network-constrained estimators for the precision matrix, $\widetilde{\mathbf{\Theta}}_j^*(z)$, and its corresponding wavelet spectral matrix, $\widetilde{\mathbf{S}}_j^*(z) = \left(\widetilde{\mathbf{\Theta}}_j^*(z)\right)^{-1}$. The complete estimation procedure is summarized in Algorithm 1.

Further scale-specific dependence measures, such as coherence and partial coherence, can then be directly derived by plugging these estimators into Definitions 3.5 and 3.6.

**Corollary 5.1 (Consistency of network dependence measures).** *Let $\widetilde{\mathbf{S}}_j^*(z)$ and $\widetilde{\mathbf{\Theta}}_j^*(z)$ be the network-constrained spectral and precision estimators obtained as detailed in Algorithm 1. For any pair of nodes $(p,q)$, let the plug-in estimators for the localized coherence $\rho_j^{(p,q)}(z)$ and partial coherence $\eta_j^{(p,q)}(z)$ be defined as*

$$\widehat{\rho}_j^{(p,q)}(z) = \frac{\left[\widetilde{\mathbf{S}}_j^*(z)\right]_{p,q}}{\sqrt{\left[\widetilde{\mathbf{S}}_j^*(z)\right]_{p,p}}\sqrt{\left[\widetilde{\mathbf{S}}_j^*(z)\right]_{q,q}}} \tag{30}$$

*and*

$$\widehat{\eta}_j^{(p,q)}(z) = -\frac{\left[\widetilde{\mathbf{\Theta}}_j^*(z)\right]_{p,q}}{\sqrt{\left[\widetilde{\mathbf{\Theta}}_j^*(z)\right]_{p,p}}\sqrt{\left[\widetilde{\mathbf{\Theta}}_j^*(z)\right]_{q,q}}}. \tag{31}$$

*Then, these estimators are consistent such that*

$$\widehat{\rho}_j^{(p,q)}(z) - \rho_j^{(p,q)}(z) = o_P(1) \quad \textit{and} \quad \widehat{\eta}_j^{(p,q)}(z) - \eta_j^{(p,q)}(z) = o_P(1), \tag{32}$$

*for all scales $j \geq 1$ and rescaled time points $z \in (0, 1)$ as $T \to \infty$.*

### 5.6 Simulation Findings

A detailed simulation study is provided in the Supplementary Material, Section D. A naive approach to estimation would be to apply the constrained optimization step (24) to the MvLSW network-agnostic estimates. However, due to the cross-scale bias correction inherent to this framework, their estimates in practice often suffer from non-positive definiteness (although theoretically this should hold), rendering this approach computationally infeasible in the context of enforcing the network structure. In contrast, the simple form of our proposed estimator that does not require correction ensures positive definiteness.

In the simulation study, we therefore compare our approach to unconstrained MvLSW estimation and benchmark it with an oracle estimator using the true subprocess. Here, we highlight our key findings. Our estimation method provides improvements of 59–92% over MvLSW, displaying higher stability (lower uncertainty) as network sparsity decreases. As the series length, $T$, increases, our method reflects the proven asymptotic properties, whereas MvLSW strikingly does not. In addition, the estimation performance of our methodology is robust to increasing the number of nodes, $n$.

## 6 High-Dimensional Global Banking Dynamics

In this section, we study the localized dependence structure of global banks under the Net-LSW framework introduced in Section 3.

**Algorithm 1:** Subprocess-based network spectral estimation via wavelet projection.

**Input :** Observed multivariate process $\{\mathbf{X}_t\}_{t=0}^{T-1}$; network $\mathcal{G} = (\mathcal{N}, \mathcal{E})$ where $\text{ne}(p) = \{q \in \mathcal{N} : (p, q) \in \mathcal{E}\}$ is the neighbourhood of node $p$; smoothing window $M_T$.

**Output:** Estimated network-constrained spectral matrices $\{\widetilde{\mathbf{S}}_j^*(k/T)\}$ and network-constrained precision matrices $\{\widetilde{\mathbf{\Theta}}_j^*(k/T)\}$.

**Assume** $\{\mathbf{X}_t\}_{t=0}^{T-1}$ *is a Net-LSW process with* $T = 2^J$*.*

```
// Stage 1:  Subprocess Approximation via Wavelet Projection
```

**for** $j = 1$ **to** $J$ **do**

  Extract observed coefficients: $\mathbf{d}_{j,k} = \sum_{t=0}^{T-1} \mathbf{X}_t \psi_{j,k}(t)$

  Project to time domain: $\widetilde{\mathbf{X}}_{j,t} = \sum_k 2^{-j} \mathbf{d}_{j,k} \psi_{j,k}(t)$

  Compute approximated coefficients: $\widetilde{\mathbf{d}}_{\ell,k}^j = \sum_{t=0}^{T-1} \widetilde{\mathbf{X}}_{j,t} \psi_{\ell,k}(t)$

**end**

```
// Stage 2:  Initial Local Spectral Estimation
```

**for** $j = 1$ **to** $J$ **do**

  $\widetilde{\mathbf{S}}_j(k/T) = \frac{1}{A_{j,j}(2M_T+1)} \sum_{m=-M_T}^{M_T} \widetilde{\mathbf{d}}_{j,k+m}^j \left(\widetilde{\mathbf{d}}_{j,k+m}^j\right)^\top$

**end**

```
// Stage 3:  Node-wise Optimization (Hastie et al. 2009)
```

**for** *scale* $j \in \{1, \ldots, J\}$ *and time* $k \in \{0, \ldots, T-1\}$ **do**

  Initialize working covariance matrix: $\mathbf{W} = \widetilde{\mathbf{S}}_j(k/T)$

  **repeat**

    **for** *node* $p = 1$ **to** $n$ **do**

      Partition $\mathbf{W}$ into $\mathbf{W}_{11}$ (all rows/cols except node $p$) and $\mathbf{w}_p$ (working col for node $p$, excluding row $p$)

      Isolate target cross-covariance vector for node $p$ from initial matrix $\widetilde{\mathbf{S}}_j(k/T)$, denoted $\widetilde{\mathbf{s}}_p = \left[\widetilde{\mathbf{S}}_j(k/T)\right]_{-p,p}$

```
      // Solve neighbourhood-constrained system on sub-matrix W11
```

      Solve for active neighbour weights $\hat{\boldsymbol{\ell}}_{\text{ne}(p)}^*$: $[\mathbf{W}_{11}]_{\text{ne}(p),\text{ne}(p)} \hat{\boldsymbol{\ell}}_{\text{ne}(p)}^* - \widetilde{\mathbf{s}}_{p,\text{ne}(p)} = 0$

```
      // Enforce network structure constraints
```

      Construct full vector $\hat{\boldsymbol{\ell}} \in \mathbb{R}^{n-1}$ by setting active positions to $\hat{\boldsymbol{\ell}}_{\text{ne}(p)}^*$ and $\hat{\ell}_q = 0$ for non-neighbors $q \notin \text{ne}(p)$

```
      // Update Part 2 of working matrix
```

      Update working row and column for node $p$: $\mathbf{w}_p = \mathbf{W}_{11} \hat{\boldsymbol{\ell}}$

```
      // Extract sparse precision elements during final cycle
```

      **if** *final iteration cycle* **then**

        $\left[\widetilde{\mathbf{\Theta}}_j^*(k/T)\right]_{p,p} = \left(\left[\widetilde{\mathbf{S}}_j(k/T)\right]_{p,p} - \mathbf{w}_p^\top \hat{\boldsymbol{\ell}}\right)^{-1}$

        $\left[\widetilde{\mathbf{\Theta}}_j^*(k/T)\right]_{-p,p} = -\hat{\boldsymbol{\ell}} \left[\widetilde{\mathbf{\Theta}}_j^*(k/T)\right]_{p,p}$

      **end**

    **end**

  **until** *convergence of* $\mathbf{W}$

  Invert optimized precision matrix to obtain final constrained spectrum:

  $\widetilde{\mathbf{S}}_j^*(k/T) = \left(\widetilde{\mathbf{\Theta}}_j^*(k/T)\right)^{-1}$

**end**

## 6.1 Data

The data we analyze consists of daily equity prices for $n = 83$ major global banking institutions observed from August 13, 2015, through December 29, 2023, with a sample size of $T = 2048$ observations. For each institution $i$ and time index $t$, volatility is proxied using the range-based estimator of Garman & Klass (1980),

$$\widetilde{\sigma}_{it}^2 = 0.511(H_{it}-L_{it})^2-0.019\Big[(C_{it}-O_{it})(H_{it}+L_{it}-2O_{it})-2(H_{it}-O_{it})(L_{it}-O_{it})\Big]-0.383(C_{it}-O_{it})^2,$$

with $H_{it}, L_{it}, O_{it}$, and $C_{it}$ the (log) high, low, open, and close prices (Demirer et al. 2018).

The multivariate time series $\mathbf{X}_t = (X_t^{(1)}, \dots, X_t^{(n)})^\top$ is constructed by setting each element $X_t^{(i)} = \log \widetilde{\sigma}_{it}^2$. As our framework takes an underlying network topology as a given input, we incorporate a pre-specified banking connectedness graph directly into the estimation of local spectral quantities, using the variance spillover framework of Demirer et al. (2018).

The graph $\mathcal{G} = (\mathcal{N}, \mathcal{E})$ in Figure 1 prescribes the structural network constraints over whose nodes we model the data $\{\mathbf{X}_t\}$ as a Net-LSW process. The model choice is driven by the stationarity test of Nason (2013) via the `locits` `R` package (Nason 2025), showing that 60 out of 83 nodes reject the null hypothesis of second order stationarity across the nodal volatility series. We will examine how direct cross-bank dependencies evolve jointly across rescaled time and wavelet scales while strictly preserving the network's structural properties.

## 6.2 Localized Time-Scale Dependence Structures

To evaluate the time-varying dependencies across the banking network, we highlight selected institutional pairs that represent distinct systemic, regional, and cross-border transmission channels. We compare network-informed coherence with network-informed partial coherence to isolate direct, bilateral dependence from broad, market-wide co-movement. Throughout this empirical analysis, we report and plot the absolute values of both metrics.

### Dependence Dynamics of US Money-Center Banks

We first examine the dependence structure between *JPMorgan Chase* and *Citigroup*, which serve as central nodes in global dollar funding markets. Figure 2 displays the network-informed coherence (left) and the network-informed partial coherence (right panel) across scales. Comparing these two profiles reveals a distinct structural decoupling once network-wide dependencies are conditioned out.

As shown in Figure 2(a), the unconditioned network-informed coherence indicates a persistent, high-intensity co-movement dominating the intermediate and coarse scales ($j \geq 4$) across the entire temporal sample. This broad-band synchronization reflects common exposure to global macroeconomic factors, systemic liquidity conditions, and market sentiment.

Conversely, the network-informed partial coherence in Figure 2(b) controls for these confounding global factors. Once conditioned on the remaining network nodes, the long-horizon scales ($j \geq 5$) become negligible, falling to near-zero values. The remaining dependence is highly localized and confined strictly to the finer scales ($j = 1$ to 4) and the direct connection is intermittent. Specifically, we observe localized spikes in partial coherence, such as a sharp coupling at scale $j = 2$ in late 2018 and at scale $j = 3$ in late 2020. The direct linkage between these two money-center institutions operates primarily at high frequencies, driven by transient liquidity adjustments rather than permanent, long-term structural shifts.

### Cross-Border and Domestic UK Banking Dynamics

We next evaluate the localized dependence profiles within European and UK banking channels, focusing on *HSBC*, *Société Général*, *Lloyds*, and *Standard Chartered*. Figure 3 presents the network-informed partial coherence for both institutional pairs, isolating their direct interbank channels by filtering out broad market trends.

For the cross-border *HSBC–Société Général* pair in Figure 3(a), the partial coherence demonstrates that direct dependence is restricted to fine horizons ($j \leq 4$), whereas coarser scales ($j \geq 5$) display no significant co-movement. The dependence is primarily driven by scale $j = 3$, exhibiting recurrent periods of high intensity, most markedly during Brexit and 2020–2021, suggesting a specialized, intermediate-frequency liquidity or settlement mechanism that activates regularly between these institutions independently of long-term macroeconomic trends, whilst reacting to global shocks.

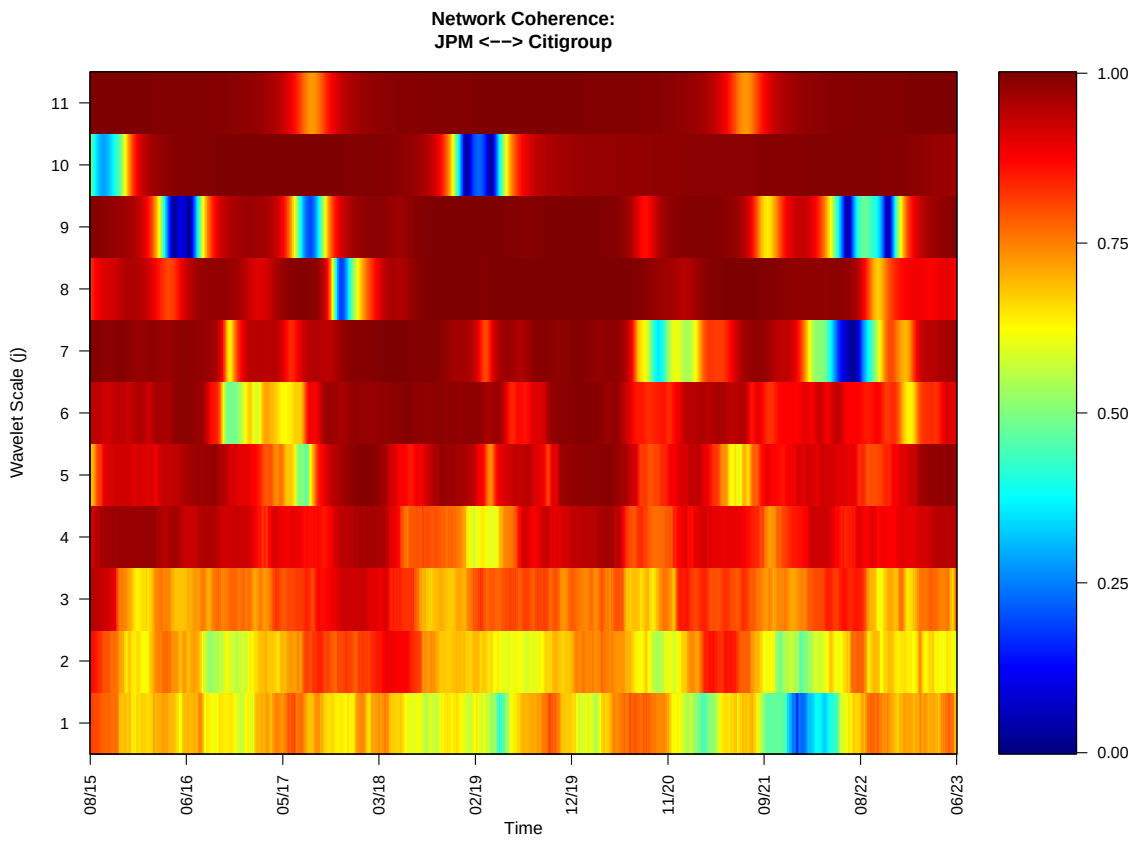

(a) Network-Informed Coherence

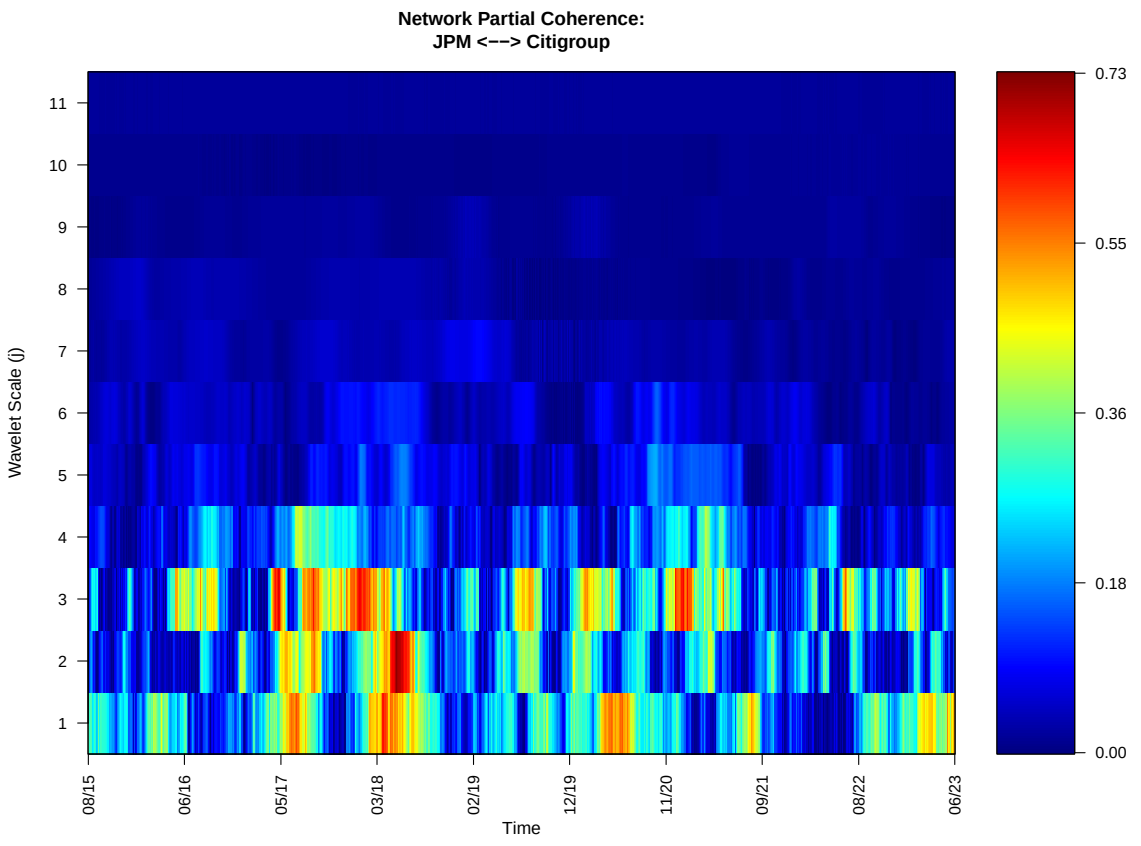

(b) Network-Informed Partial Coherence

Figure 2: Localized dependency profiles for the JPMorgan Chase–Citigroup pair, contrasting network-informed coherence with network-informed partial coherence.

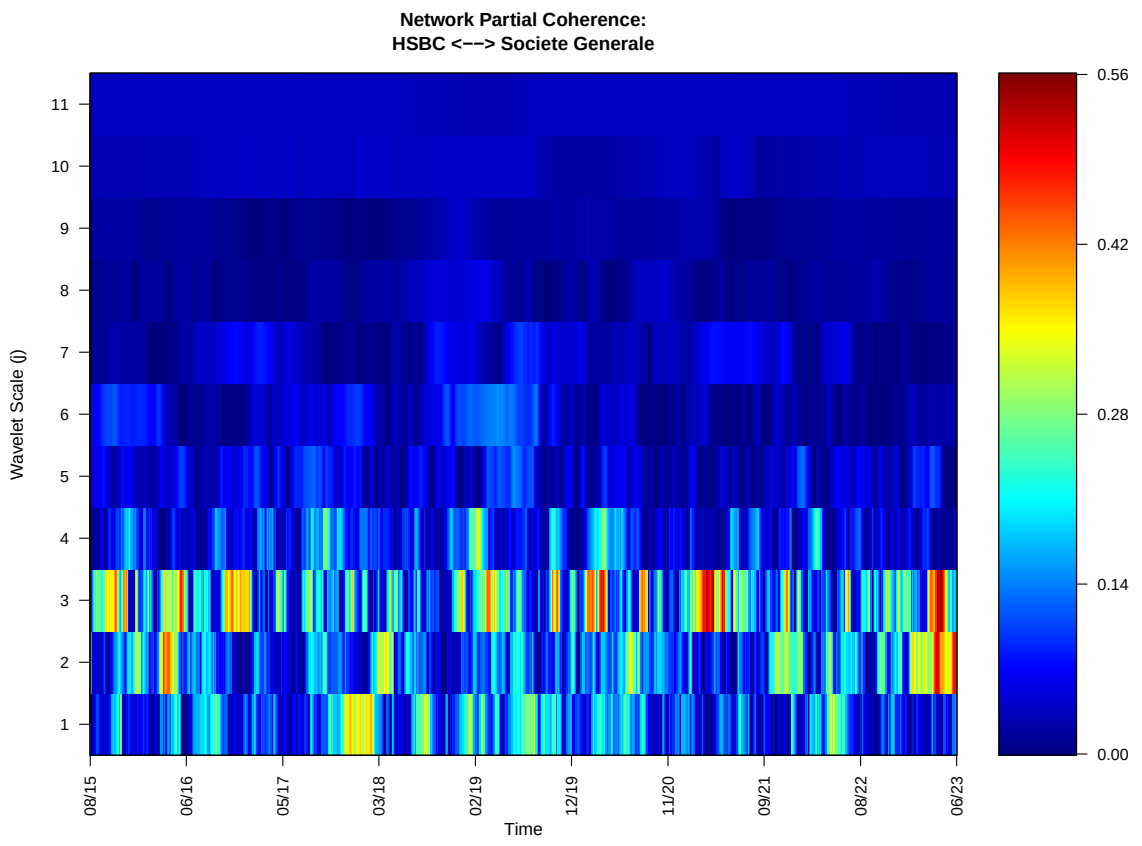

(a) HSBC–Société Général

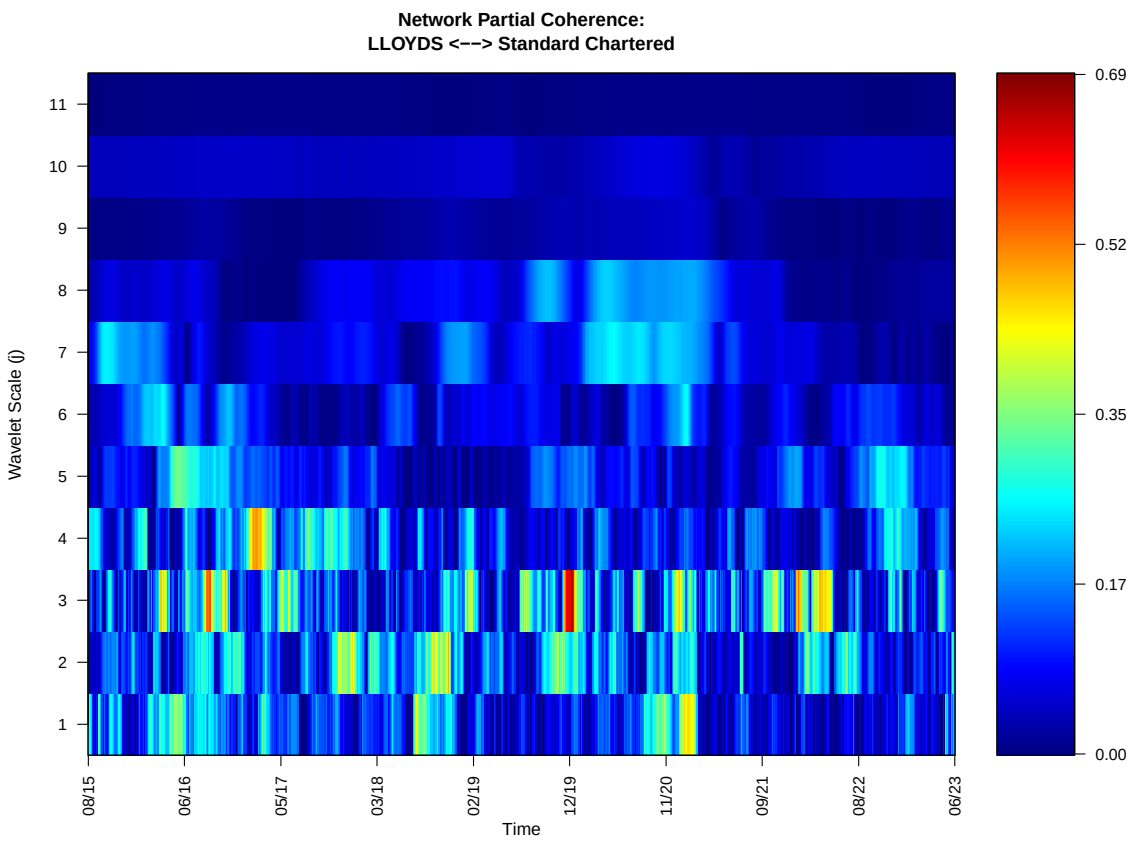

(b) Lloyds–Standard Chartered

Figure 3: Network-informed partial coherence across European and UK banking channels: strictly localized fine-scale corridor (left); extended scale dependency footprint (right).

The *Lloyds–Standard Chartered* pair in Figure 3(b) exhibits a wider, multiscale frequency footprint, punctuated by Brexit and the start of the pandemic. At medium-finer scales ($j = 1$ to 4), the direct relationship is characterized by volatile, intermittent spikes throughout the timeline. Crucially, unlike the HSBC–Société Général pair, this pair shows a pronounced cluster of intermediate-scale activity localized between medium-coarser scales $j = 6$ and 8. This mid-to-long horizon dependence manifests as elevated partial coherence during 2020 before tapering off at the coarser boundary ($j \geq 9$). Comparing the two panels demonstrates that the proposed framework uncovers highly individualized, pair-specific dependency structures—ranging from isolated fine-scale links to broader, multiscale transmission bands—that are otherwise masked by aggregate market measures.

### Asian Regional Banking Corridor

To analyze regional transmission dynamics in East Asia, we examine the Japanese banking corridor formed by *Mitsubishi* and *Sumitomo Mitsui*. Figure 4 presents the network-informed coherence (left panel) alongside the network-informed partial coherence (right panel) across scales. Comparing these profiles illustrates the structural adjustment that occurs when conditioning out systemic trends.

Figure 4(a) shows the network-informed coherence reveals a continuous block of high co-movement dominating the long-horizon scales ($j \geq 6$) across the entire sample period. This broad-band synchronization indicates that at coarser horizons, these two dominant regional nodes are structurally bound by common macroeconomic factors and domestic monetary conditions. At finer scales ($j = 1$ to 3), this unconditioned co-movement is much weaker, fluctuating across the timeline.

However, when conditioning on the broader financial network to isolate their direct relation-ship, the network-informed partial coherence in Figure 4(b) uncovers a different localized structure. The long-horizon co-movement at coarser scales ($j \geq 8$) is filtered out and instead, an intermediate-scale behaviour emerges at scales $j = 4$ and 5. Within this intermediate band, the direct connection manifests as a series of recurring high intensity periods during 2016–2019, followed by high coupling at the beginning of COVID-19 and re-emerging in the second half of 2021 toward the end of the sample. This contrast demonstrates that while the two institutions appear synchronized due to common market factors, their direct relationship operates through an independent, cyclical intermediate-scale channel.

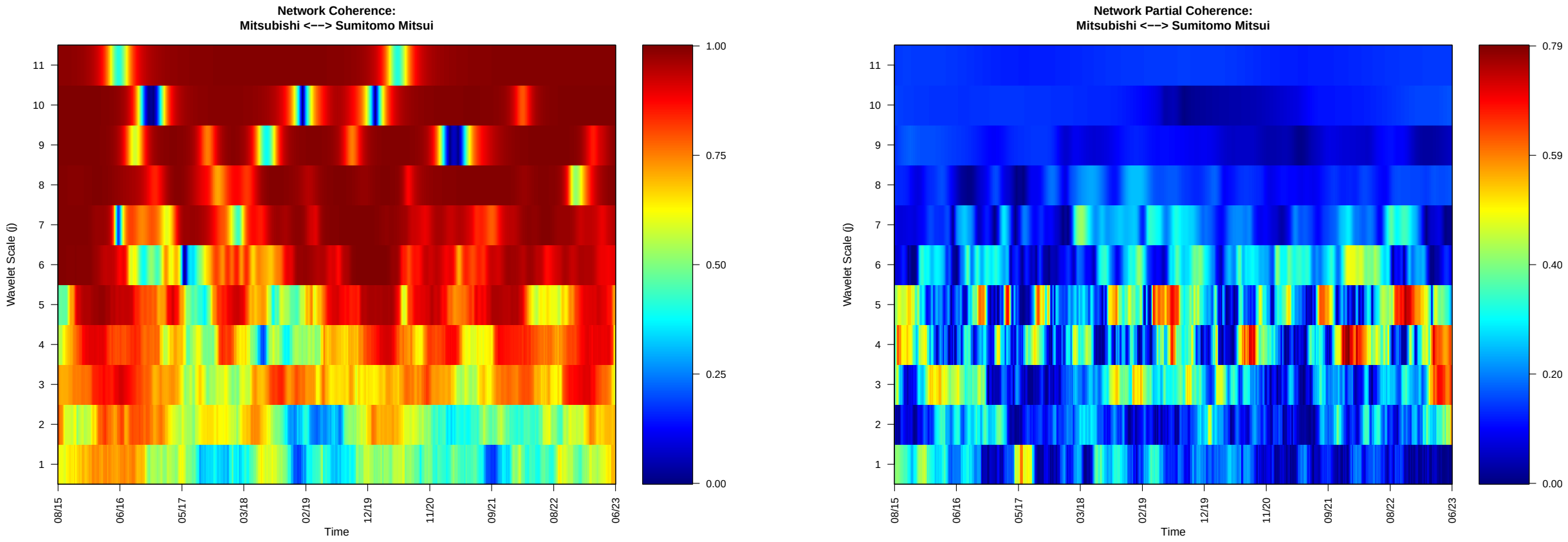

(a) Network-Informed Coherence

(b) Network-Informed Partial Coherence

Figure 4: Profiles for the Mitsubishi–Sumitomo Mitsui case, contrasting broad-band network-informed coherence with frequency-specific network-informed partial coherence.

# 7 Discussion

In this article, we addressed the challenge of capturing multiscale, time-varying linear cross-dependencies in nonstationary multivariate time series, structured by an underlying network topology. While traditional frameworks allow for unconstrained dependence estimation, they are network-agnostic and computationally expensive—estimating all pairwise dependencies as if the system were fully connected—thus leading to severe over-parametrisation in high dimensions. In contrast, our proposed Locally Stationary Wavelet process on Networks (Net-LSW) framework accommodates the network-induced topological constraints by directly encoding the existing edge connections into the model formulation. To the best of our knowledge, this is the first complete development to tackle network nonstationary process modelling alongside estimation and simulation capabilities, whilst addressing complex data scenarios, outside of the classical time-varying vector autoregressive modelling setting.

# Disclosure statement

The authors report there are no competing interests to declare.

# Data Availability Statement

Deidentified data have been made available at the following URL.

## Supplementary Material

This provides additional theoretical details, mathematical proofs, and a complete simulation study. Section A contains technical proofs for the basic properties of the Net-LSW process. Section B presents proofs for the Local Partial Correlation Graph results. Section C contains the proofs and asymptotic consistency results for the subprocess-based estimation framework. Section D presents a comprehensive simulation study evaluating the finite-sample performance, robustness, and scalability of the proposed framework.

The complete `R` code for simulation and estimation will be made publicly available upon publishing.

# Supplementary Material for "Multiscale Dynamic Dependence Estimation over Networks"

Cristian F. Jiménez-Varón
Department of Mathematics, University of York
and
Marina I. Knight
Department of Mathematics, University of York
and
Matthew A. Nunes
Department of Mathematical Sciences, University of Bath

August 16, 2026

This Supplementary Material provides additional theoretical details, mathematical proofs, and a complete simulation study accompanying the main manuscript. Section A contains technical proofs for the basic properties of the Net-LSW process (Section 3). Section B presents proofs for the Local Partial Correlation Graph results (Section 4). Section C provides the mathematical proofs and asymptotic consistency results for the subprocess-based estimation framework (Section 5). Finally, Section D presents a comprehensive simulation study evaluating the finite-sample performance, robustness, and scalability of the proposed framework (briefly summarized in Section 5.6 of the main manuscript).

## A Technical proofs for Section 3 (Net-LSW)

*Proof of Property 3.3.* The element corresponding to a pair of nodes $p$ and $q$ of the wavelet coherence matrix $\boldsymbol{\rho}_j^{\mathcal{N}}(\cdot)$ at rescaled time $z = \frac{k}{T}$, is written as

$$\rho_j^{(p,q)}(z) = \frac{S_j^{(p,q)}(z)}{\sqrt{S_j^{(p,p)}(z)}\sqrt{S_j^{(q,q)}(z)}},$$

and simple manipulations show

$$\begin{aligned} \rho_j^{(p,q)}(z) &= \frac{W_j^{(p,p)}(z)\lambda_j^{(p,q)}(z)W_j^{(q,q)}(z)}{W_j^{(p,p)}(z)W_j^{(q,q)}(z)}, \\ &= \lambda_j^{(p,q)}(z) = [\mathbf{\Lambda}_j(z)]_{p,q}, \end{aligned}$$

where we have used the definition of the Net-LSW spectrum in equation (6) and process specifications in Definition 3.1. □

*Proof of Property 3.4.* Similar to the (cross-)coherence, the partial coherence turns out to be connected to the increment inverse covariance $\mathbf{\Gamma}_j^{\mathcal{N}}(z)$, since

$$\begin{aligned} \boldsymbol{\eta}_j^{\mathcal{N}}(z) &= -\mathbf{H}_j(z)\,\mathbf{\Theta}_j^{\mathcal{N}}(z)\,\mathbf{H}_j(z), \\ &= -\mathbf{H}_j(z)\left(\mathbf{W}^{\mathcal{N}}(z)\right)^{-1/2}\left(\mathbf{\Lambda}^{\mathcal{N}}(z)\right)^{-1}\left(\mathbf{W}^{\mathcal{N}}(z)\right)^{-1/2}\mathbf{H}_j(z), \end{aligned}$$

where we have used the definition of the Net-LSW cross-spectrum, and the $p$th diagonal element of the $\widetilde{\mathbf{H}}_j(z)$ diagonal matrix is $\widetilde{H}_j^{(p,p)}(z) = H_j^{(p,p)}(z)\left(W^{(p,p)}(z)\right)^{-1/2}$. □

# B Technical proofs for Section 4 (Local Partial Correlation Graph)

**Preliminary notation.** Suppose we have a $(p+q)\times(p+q)$ matrix $\mathbf{M}$ assumed to be (semi-) positive definite, which can be represented as a block partitioned matrix as

$$\mathbf{M} = \left[\begin{array}{c|c} \mathbf{E} & \mathbf{F} \\ \hline \mathbf{F}^\top & \mathbf{H} \end{array}\right], \tag{1}$$

where $\mathbf{E}$, $\mathbf{F}$, and $\mathbf{H}$ are $p\times p$, $p\times q$, and $q\times q$ matrices, respectively. Recall that if $\mathbf{H}$ is invertible and symmetric, then the Schur complement of $\mathbf{H}$ with respect to $\mathbf{M}$ is given by

$$\mathbf{M}/\mathbf{H} = \mathbf{E} - \mathbf{F}\mathbf{H}^{-1}\mathbf{F}^\top. \tag{2}$$

Then we have

$$\mathbf{y}^\top \mathbf{M} \mathbf{y} = \mathbf{x}^\top \left[\begin{array}{c|c} \mathbf{M}/\mathbf{H} & \mathbf{0} \\ \hline \mathbf{0} & \mathbf{H} \end{array}\right] \mathbf{x}, \tag{3}$$

where

$$\begin{aligned} \mathbf{y} &= \left[\begin{array}{c|c} \mathbf{I} & \mathbf{0} \\ \hline -\mathbf{H}^{-1}\mathbf{F}^\top & \mathbf{I} \end{array}\right] \mathbf{x} \\ &= \left[\begin{array}{c|c} \mathbf{I} & \mathbf{0} \\ \hline -\mathbf{H}^{-1}\mathbf{F}^\top & \mathbf{I} \end{array}\right] (\mathbf{x}_l^\top, \mathbf{x}_r^\top)^\top \\ &= (\mathbf{x}_l^\top,\ (\mathbf{x}_r - \mathbf{H}^{-1}\mathbf{F}^\top \mathbf{x}_l)^\top)^\top, \end{aligned}$$

for any $\mathbf{x} \in \mathbb{R}^{p+q}$, $\mathbf{x}_l \in \mathbb{R}^p$, and $\mathbf{x}_r \in \mathbb{R}^q$. Equivalently,

$$\mathbf{x} = \left[\begin{array}{c|c} \mathbf{I} & \mathbf{0} \\ \hline \mathbf{H}^{-1}\mathbf{F}^\top & \mathbf{I} \end{array}\right] \mathbf{y}. \tag{4}$$

## B.1 Proof of Proposition 4.1

*Proof.* For any $n$-dimensional vector $\mathbf{y} = (\mathbf{y}^{a:b}, \mathbf{y}^{-ab})^\top$ partitioned with the first two components corresponding to nodes $a$ and $b$, and the remaining entries to all other nodes, using (4) we take

$$\mathbf{y}^\top \mathbf{C}^X(z,\tau)\,\mathbf{y} = \mathbf{x}^\top \left[\begin{array}{c|c} \mathbf{C}^X(z,\tau)/\mathbf{D}(z,\tau) & \mathbf{0}_{2\times(n-2)} \\ \hline \mathbf{0}_{(n-2)\times 2} & \mathbf{D}(z,\tau) \end{array}\right] \mathbf{x}, \tag{5}$$

where

$$\mathbf{y} = \left[\begin{array}{c|c} \mathbf{I}_2 & \mathbf{0}_{2\times(n-2)} \\ \hline -\mathbf{D}(z,\tau)^{-1}\mathbf{B}(z,\tau)^\top & \mathbf{I}_{n-2} \end{array}\right] \mathbf{x}, \tag{6}$$

so that

$$\mathbf{y}^{a:b} = \mathbf{x}^{a:b} \quad \text{and} \quad \mathbf{y}^{-ab} = \mathbf{x}^{-ab} - \mathbf{D}(z,\tau)^{-1}\mathbf{B}(z,\tau)^\top\,\mathbf{x}^{a:b}. \tag{7}$$

Denoting the spectral domain matrix block $\mathbf{D}_j^S(z) = \left[S_j^{(p,q)}(z)\right]_{p,q \notin \{a,b\}}$, we have

$$\begin{aligned}
\mathbf{y}^\top \mathbf{C}^X(z,\tau)\,\mathbf{y} &= \mathbf{x}^\top \left[\begin{array}{c|c} \mathbf{C}^\epsilon(z,\tau) & \mathbf{0}_{2\times(n-2)} \\ \hline \mathbf{0}_{(n-2)\times 2} & \mathbf{D}(z,\tau) \end{array}\right] \mathbf{x} \quad \text{from (5)} \\
&= \mathbf{x}^\top \left[\begin{array}{c|c} \sum_j \mathbf{S}_j^\epsilon(z)\Psi_j(\tau) & \mathbf{0}_{2\times(n-2)} \\ \hline \mathbf{0}_{(n-2)\times 2} & \sum_j \mathbf{D}_j^S(z)\Psi_j(\tau) \end{array}\right] \mathbf{x} \\
&= \sum_j \left\{ \mathbf{x}^\top \left[\begin{array}{c|c} \mathbf{S}_j^\epsilon(z) & \mathbf{0}_{2\times(n-2)} \\ \hline \mathbf{0}_{(n-2)\times 2} & \mathbf{D}_j^S(z) \end{array}\right] \mathbf{x} \right\} \Psi_j(\tau) \\
&= \sum_j \left\{ (\mathbf{x}^{a:b})^\top \mathbf{S}_j^\epsilon(z) \mathbf{x}^{a:b} + (\mathbf{x}^{-ab})^\top \mathbf{D}_j^S(z) \mathbf{x}^{-ab} \right\} \Psi_j(\tau),
\end{aligned}$$

where in the second line we have used the definition of the cross-covariance and the partition (11), noting that $\mathbf{S}_j^\epsilon(z)$ is a diagonal matrix if $(a,b) \notin E$.

We also have that

$$\mathbf{y}^\top \mathbf{C}^X(z,\tau)\,\mathbf{y} = \mathbf{y}^\top \left\{ \sum_j \mathbf{S}_j^X(z)\Psi_j(\tau) \right\} \mathbf{y} \tag{8}$$

$$= \sum_j \left\{ \mathbf{y}^\top \mathbf{S}_j^X(z)\,\mathbf{y} \right\} \Psi_j(\tau) \tag{9}$$

$$= \sum_j \left\{ (\mathbf{x}_j^S)^\top \left[\begin{array}{c|c} \mathbf{S}_j^X(z)/\mathbf{D}_j^S(z) & \mathbf{0}_{2\times(n-2)} \\ \hline \mathbf{0}_{(n-2)\times 2} & \mathbf{D}_j^S(z) \end{array}\right] \mathbf{x}_j^S \right\} \Psi_j(\tau) \tag{10}$$

$$= \sum_j \left\{ (\mathbf{x}_j^{S;a:b})^\top \left[\mathbf{S}_j^X(z)/\mathbf{D}_j^S(z)\right] \mathbf{x}_j^{S;a:b} + (\mathbf{x}_j^{S;-ab})^\top \mathbf{D}_j^S(z)\,\mathbf{x}_j^{S;-ab} \right\} \Psi_j(\tau), \tag{11}$$

where we recall that we partitioned the spectral matrix as

$$\mathbf{S}_j^X(z) = \left[\begin{array}{c|c} \mathbf{A}_j^S(z) & \mathbf{B}_j^S(z) \\ \hline \mathbf{B}_j^S(z)^\top & \mathbf{D}_j^S(z) \end{array}\right],$$

and

$$\mathbf{y} = \left[\begin{array}{c|c} \mathbf{I}_2 & \mathbf{0}_{2\times(n-2)} \\ \hline -\mathbf{D}_j^S(z)^{-1}\mathbf{B}_j^S(z)^\top & \mathbf{I}_{n-2} \end{array}\right] \mathbf{x}_j^S,$$

so that

$$\mathbf{y}^{a:b} = \mathbf{x}_j^{S;a:b} \quad \text{and} \quad \mathbf{y}^{-ab} = \mathbf{x}_j^{S;-ab} - \mathbf{D}_j^S(z)^{-1}\mathbf{B}_j^S(z)^\top \mathbf{x}_j^{S;a:b}.$$

Equation (11) coupled with the linear independence of $\{\Psi_j(\tau)\}_j$ yields for any $\mathbf{y}$ and for any $j$, $z$,

$$\mathbf{y}^\top \mathbf{S}_j^X(z)\,\mathbf{y} = \mathbf{x}^\top \left[\begin{array}{c|c} \mathbf{S}_j^\epsilon(z) & \mathbf{0}_{2\times(n-2)} \\ \hline \mathbf{0}_{(n-2)\times 2} & \mathbf{D}_j^S(z) \end{array}\right] \mathbf{x} = (\mathbf{x}_j^S)^\top \left[\begin{array}{c|c} \mathbf{S}_j^X(z)/\mathbf{D}_j^S(z) & \mathbf{0}_{2\times(n-2)} \\ \hline \mathbf{0}_{(n-2)\times 2} & \mathbf{D}_j^S(z) \end{array}\right] \mathbf{x}_j^S.$$

We now exploit the first equality above, coupled with (6) that connects the vectors $\mathbf{y}$ and $\mathbf{x}$, whose inverse allows us to write

$$\begin{aligned} \mathbf{y}^\top \mathbf{S}_j^X(z)\,\mathbf{y} &= \mathbf{x}^\top \left[\begin{array}{c|c} \mathbf{S}_j^\epsilon(z) & \mathbf{0}_{2\times(n-2)} \\ \hline \mathbf{0}_{(n-2)\times 2} & \mathbf{D}_j^S(z) \end{array}\right] \mathbf{x} \\ &= \mathbf{y}^\top \left[\begin{array}{c|c} \mathbf{I}_2 & \left(\mathbf{D}(z,\tau)^{-1}\mathbf{B}(z,\tau)^\top\right)^\top \\ \hline \mathbf{0}_{(n-2)\times 2} & \mathbf{I}_{n-2} \end{array}\right] \left[\begin{array}{c|c} \mathbf{S}_j^\epsilon(z) & \mathbf{0}_{2\times(n-2)} \\ \hline \mathbf{0}_{(n-2)\times 2} & \mathbf{D}_j^S(z) \end{array}\right] \mathbf{x} \\ &= \mathbf{y}^\top \left[\begin{array}{c|c} \mathbf{S}_j^\epsilon(z) & \left(\mathbf{D}(z,\tau)^{-1}\mathbf{B}(z,\tau)^\top\right)^\top \mathbf{D}_j^S(z) \\ \hline \mathbf{0}_{(n-2)\times 2} & \mathbf{D}_j^S(z) \end{array}\right] \left[\begin{array}{c|c} \mathbf{I}_2 & \mathbf{0}_{2\times(n-2)} \\ \hline \mathbf{D}(z,\tau)^{-1}\mathbf{B}(z,\tau)^\top & \mathbf{I}_{n-2} \end{array}\right] \mathbf{y}. \end{aligned}$$

Therefore, for any $n$-dimensional vector $\mathbf{y}$, the quadratic form corresponding to the spectral matrix $\mathbf{S}_j^X(z)$ can be re-written as

$$\mathbf{y}^\top \left[\begin{array}{c|c} \mathbf{S}_j^\epsilon(z) + \left(\mathbf{D}(z,\tau)^{-1}\mathbf{B}(z,\tau)^\top\right)^\top \mathbf{D}_j^S(z) \left(\mathbf{D}(z,\tau)^{-1}\mathbf{B}(z,\tau)^\top\right) & \left(\mathbf{D}(z,\tau)^{-1}\mathbf{B}(z,\tau)^\top\right)^\top \mathbf{D}_j^S(z) \\ \hline \left(\left(\mathbf{D}(z,\tau)^{-1}\mathbf{B}(z,\tau)^\top\right)^\top \mathbf{D}_j^S(z)\right)^\top & \mathbf{D}_j^S(z) \end{array}\right] \mathbf{y}, \tag{12}$$

which leads directly to the matrix equalities

$$\mathbf{A}_j^S(z) = \mathbf{S}_j^\epsilon(z) + \left(\mathbf{D}(z,\tau)^{-1}\mathbf{B}(z,\tau)^\top\right)^\top \mathbf{D}_j^S(z) \left(\mathbf{D}(z,\tau)^{-1}\mathbf{B}(z,\tau)^\top\right), \text{ and} \tag{13}$$

$$\mathbf{B}_j^S(z) = \left(\mathbf{D}(z,\tau)^{-1}\mathbf{B}(z,\tau)^\top\right)^\top \mathbf{D}_j^S(z). \tag{14}$$

Using the symmetry and invertibility of $\mathbf{D}_j^S(z)$ and substituting (14) into (13) yields

$$\begin{aligned}\mathbf{S}_j^{\epsilon}(z) &= \mathbf{A}_j^S(z) - \mathbf{B}_j^S(z)\mathbf{D}_j^S(z)^{-1}\mathbf{B}_j^S(z)^{\top}, \\ &= \mathbf{S}_j^X(z)/\mathbf{D}_j^S(z), \, \forall j, \, z,\end{aligned}$$

which concludes the proof. □

## B.2 Proof of Proposition 4.2

*Proof.* In our locally stationary wavelet setup, the fact that two nodes $(a, b) \notin E$ are not connected via an edge means $[\mathbf{C}^{\epsilon}(z, \tau)]_{a,b} = \sum_j [\mathbf{S}_j^{\epsilon}(z)]_{a,b} \Psi_j(\tau) = 0$. Due to the linear independence of $\{\Psi_j(\cdot)\}_j$, this condition translates into a corresponding zero-value in the cross-spectrum of the residual process: $S_j^{\epsilon;(a,b)}(z) = 0$ for any scale $j$ and rescaled time $z$.

Hence, from (13), the Schur complement $[\mathbf{S}_j^X(z)/\mathbf{D}_j^S(z)]$ has a diagonal structure, which means its inverse is also a diagonal matrix, i.e., $[\mathbf{S}_j^X(z)/\mathbf{D}_j^S(z)]_{a,b} = 0$ and $\left[\left(\mathbf{S}_j^X(z)/\mathbf{D}_j^S(z)\right)^{-1}\right]_{a,b} = 0$.

Classical linear algebra inversion of the matrix $\mathbf{S}_j^X(z)$ partitioned as in equation (14) yields its left-upper block to be:

$$\left[\left(\mathbf{S}_j^X(z)\right)^{-1}\right]_{a,b} := \left[\left(\mathbf{S}_j^X(z)/\mathbf{D}_j^S(z)\right)^{-1}\right]_{a,b},$$

which must be part of a diagonal structure, directly implying that $\left[\left(\mathbf{S}_j^X(z)\right)^{-1}\right]_{a,b} = 0$ when $(a, b) \notin E$. This concludes the proof. □

# C Technical proofs for Section 5 (Subprocess-Based Estimation)

## C.1 Proof of Proposition 5.1

*Proof.*

**Expectation**

Since the random increments $\{\boldsymbol{\xi}_{j,m}\}_{j,m}$ are assumed to be zero-mean by Definition 3.1, it follows from the linearity of the wavelet transform and the inherited mean-zero property of the subprocess ($\mathbb{E}[\mathbf{X}_{j,t}] = \mathbf{0}$) that at each scale $j$,

$$\mathbb{E}\left[\mathbf{d}^{j}_{\ell,k}\right] = \sum_{t} \mathbb{E}\left[\mathbf{X}_{j,t}\right] \psi_{\ell,k}(t) = \mathbf{0}, \quad \forall \ell, k.$$

**Covariance**

By definition, the covariance matrix of the subprocess wavelet coefficients is expressed as

$$\begin{aligned}\mathrm{Cov}(\mathbf{d}^{j}_{\ell,k}, \mathbf{d}^{j}_{\ell,k+\eta}) &= \mathbb{E}\left[\left(\sum_{t} \mathbf{X}_{j,t}\psi_{\ell,k}(t)\right)\left(\sum_{t'} \mathbf{X}^{\top}_{j,t'}\psi_{\ell,k+\eta}(t')\right)\right] \\ &= \sum_{t}\sum_{t'} \psi_{\ell,k}(t)\psi_{\ell,k+\eta}(t')\mathbb{E}\left[\mathbf{X}_{j,t}\mathbf{X}^{\top}_{j,t'}\right]. \end{aligned} \tag{15}$$

We evaluate the inner expectation $\mathbb{E}\left[\mathbf{X}_{j,t}\mathbf{X}^{\top}_{j,t'}\right]$ by substituting the Net-LSW subprocess expansion for a fixed scale $j$, $\mathbf{X}_{j,t} = \sum_m \mathbf{W}^{\mathcal{N}}_{j}(m/T)\psi_{j,m}(t)\boldsymbol{\xi}_{j,m}$

$$\begin{aligned}\mathbb{E}\left[\mathbf{X}_{j,t}\mathbf{X}^{\top}_{j,t'}\right] &= \mathbb{E}\left[\left(\sum_{m} \mathbf{W}^{\mathcal{N}}_{j}(m/T)\psi_{j,m}(t)\boldsymbol{\xi}_{j,m}\right)\left(\sum_{m'} \boldsymbol{\xi}^{\top}_{j,m'}\psi_{j,m'}(t')\mathbf{W}^{\mathcal{N}}_{j}(m'/T)\right)\right] \\ &= \sum_{m}\sum_{m'} \mathbf{W}^{\mathcal{N}}_{j}(m/T)\mathbb{E}\left[\boldsymbol{\xi}_{j,m}\boldsymbol{\xi}^{\top}_{j,m'}\right]\mathbf{W}^{\mathcal{N}}_{j}(m'/T)\psi_{j,m}(t)\psi_{j,m'}(t').\end{aligned}$$

Using that $\mathbb{E}[\boldsymbol{\xi}_{j,m}\boldsymbol{\xi}^{\top}_{j,m'}] = \delta_{m,m'}\boldsymbol{\Lambda}^{\mathcal{N}}_{j}(m/T)$ (Property 3.1), the double sum collapses since terms are non-zero only when $m = m'$, yielding

$$\mathbb{E}\left[\mathbf{X}_{j,t}\mathbf{X}^{\top}_{j,t'}\right] = \sum_{m} \mathbf{W}^{\mathcal{N}}_{j}(m/T)\boldsymbol{\Lambda}^{\mathcal{N}}_{j}(m/T)\mathbf{W}^{\mathcal{N}}_{j}(m/T)\psi_{j,m}(t)\psi_{j,m}(t').$$

Recalling the definition of the Net-LSW wavelet spectral matrix (LWS), we have $\mathbf{S}_j(m/T) = \mathbf{W}_j^{\mathcal{N}}(m/T)\mathbf{\Lambda}_j^{\mathcal{N}}(m/T)\mathbf{W}_j^{\mathcal{N}}(m/T)$ from (5) and utilizing the assumption from Definition 3.1 that the transfer functions are slowly varying, we can apply the Taylor expansion approximation and obtain

$$\mathbb{E}\left[\mathbf{X}_{j,t}\mathbf{X}_{j,t'}^{\top}\right] = \mathbf{S}_j(k/T)\sum_m \psi_{j,m}(t)\psi_{j,m}(t') + \mathcal{O}(T^{-1}).$$

Noting that the sum $\sum_m \psi_{j,m}(t)\psi_{j,m}(t')$ is the autocorrelation wavelet $\Psi_j(t-t')$, where $\Psi_j(\tau) = \sum_m \psi_{j,m}(0)\psi_{j,m}(\tau)$ (Nason et al. 2000), the above can be re-written as

$$\mathbb{E}\left[\mathbf{X}_{j,t}\mathbf{X}_{j,t'}^{\top}\right] = \mathbf{S}_j(k/T)\Psi_j(t-t') + \mathcal{O}(T^{-1}).$$

Substituting this result back into equation (15) gives

$$\text{Cov}(\mathbf{d}_{\ell,k}^{j}, \mathbf{d}_{\ell,k+\nu}^{j}) = \mathbf{S}_j(k/T)\sum_t \sum_{t'} \psi_{\ell,k}(t)\psi_{\ell,k+\nu}(t')\Psi_j(t-t') + \mathcal{O}(T^{-1}). \tag{16}$$

To evaluate the scalar double sum $\mathcal{I} = \sum_t \sum_{t'} \psi_{\ell,k}(t)\psi_{\ell,k+\nu}(t')\Psi_j(t-t')$, we change variables by setting $\tau = t - t'$, thus

$$\mathcal{I} = \sum_\tau \Psi_j(\tau)\left(\sum_t \psi_{\ell,k}(t)\psi_{\ell,k+\nu}(t-\tau)\right).$$

By the shift-invariance property of the non-decimated wavelet transform, $\psi_{\ell,k}(t) = \psi_{\ell,0}(t-k)$. Letting $s = t - k$, the inner sum becomes

$$\mathcal{I} = \sum_\tau \Psi_j(\tau)\left(\sum_s \psi_{\ell,0}(s)\psi_{\ell,0}(s-\tau-\nu)\right).$$

Recognizing the inner summation over $s$ as the shifted autocorrelation wavelet $\Psi_\ell(\tau+\nu)$, the scalar term simplifies directly to the inner product matrix element

$$\mathcal{I} = \sum_\tau \Psi_j(\tau)\Psi_\ell(\tau+\nu) = A_{j,\ell}^{(\nu)}.$$

Combining this last expression with (16) yields the final covariance form

$$\text{Cov}(\mathbf{d}_{\ell,k}^{j}, \mathbf{d}_{\ell,k+\nu}^{j}) = A_{j,\ell}^{(\nu)}\mathbf{S}_j(k/T) + \mathcal{O}(T^{-1}).$$

**Asymptotic Distribution**

Under the assumption of joint Gaussianity for the innovation sequence, any linear combination remains Gaussian, hence the subprocesses in Definition 5.1 are Gaussian and their corresponding wavelet coefficients in (18) are also normally distributed. As $T \to \infty$, the $\mathcal{O}(T^{-1})$ error vanishes and evaluating the covariance at matching scales ($\ell = j$) and contemporaneous locations ($\nu = 0$), noting that $A_{j,j}^{(0)} = A_{j,j}$, we obtain (21). □

## C.2 Proof of Proposition 5.2

*Proof.* For notational simplicity, in what follows we drop the index $T$ from the smoothing bandwidth $M_T$. Recalling that the subprocess wavelet coefficients are zero-mean, together with equation (20) and the definition of the smoothed periodogram in (23) , yields the first-order moment

$$\mathbb{E}\Big(\widehat{S}_j^{(p,q)}(k/T)\Big) = S_j^{(p,q)}(k/T) + \mathcal{O}(A_{j,j}^{-1} M T^{-1}).$$

Using the definition of the smoothed periodogram, we have for the second order moment

$$\begin{aligned}\mathbb{E}\Big((\widehat{S}_j^{(p,q)}(k/T))^2\Big) &= \frac{A_{j,j}^{-2}}{(2M+1)^2} \sum_{m=-M}^{M} \sum_{m'=-M}^{M} \mathbb{E}\Big(I_{j,(k+m)(k+m)}^{(p,q)} I_{j,(k+m')(k+m')}^{(p,q)}\Big) \\ &= \frac{A_{j,j}^{-2}}{(2M+1)^2} \sum_{m=-M}^{M} \sum_{\tau=-M-m}^{M-m} \mathbb{E}\Big(d_{j,k+m}^{j,(p)} d_{j,k+m}^{j,(q)} d_{j,k+m+\tau}^{j,(p)} d_{j,k+m+\tau}^{j,(q)}\Big),\end{aligned}$$

where $\tau = m' - m$. With an application of Isserlis' Theorem (Isserlis 1918), the above becomes

$$\begin{aligned}\operatorname{var}\Big(\widehat{S}_j^{(p,q)}(k/T)\Big) = \frac{A_{j,j}^{-2}}{(2M+1)^2} \Bigg\{ &\sum_{m=-M}^{M} \sum_{\tau} \mathbb{E}\Big(d_{j,k+m}^{j,(p)} d_{j,k+m+\tau}^{j,(p)}\Big) \times \mathbb{E}\Big(d_{j,k+m}^{j,(q)} d_{jj,k+m+\tau}^{j,(q)}\Big) \\ &+ \sum_{m=-M}^{M} \sum_{\tau} \mathbb{E}\Big(d_{j,k+m}^{j,(p)} d_{j,k+m+\tau}^{j,(q)}\Big) \times \mathbb{E}\Big(d_{j,k+m+\tau}^{j,(p)} d_{j,k+m}^{j,(q)}\Big) \Bigg\},\end{aligned}$$

and using equation (20) from Proposition 5.1, this further reduces to

$$\frac{A_{j,j}^{-2}}{(2M+1)^2}\sum_{m=-M}^{M}\Bigg\{\sum_{\tau}\left(A_{jj}^{(\tau)}S_j^{(p,p)}(k/T)\right)\times\left(A_{jj}^{(\tau)}S_j^{(q,q)}(k/T)\right)$$
$$+\sum_{\tau}\left(A_{jj}^{(\tau)}S_j^{(p,q)}(k/T)\right)\times\left(A_{jj}^{(-\tau)}S_j^{(p,q)}(k/T)\right)$$
$$+\sum_{\tau}(|m|+1)\mathcal{O}(T^{-1})+\sum_{\tau}(|m|+1)^2\mathcal{O}(T^{-2})\Bigg\}.$$

Using the subprocess absolute covariance summability, $\sup_z \sum_n |c_j^{(p,q)}(z,n)| = \mathcal{O}(1)$ and the autocorrelation wavelet property $\sum_{\tau} |\boldsymbol{\Psi}_j(\tau)| = \mathcal{O}(2^j)$ (Nason et al. 2000), we obtain

$$\sum_{\tau}|A_{jj}^{(\tau)}S_j^{(p,q)}(k/T)| = \sum_{\tau}|\sum_{n}c_j^{(p,q)}(k/T,n)\Psi_j(n+\tau)|,$$
$$\leq \sum_{n}|c_j^{(p,q)}(k/T,n)|\sum_{\tau}|\Psi_j(n+\tau)| = \mathcal{O}(2^j),$$

and further retaining the highest order terms yields

$$\operatorname{var}\left(\widehat{S}_j^{(p,q)}(k/T)\right) = \mathcal{O}(2^{2j}A_{j,j}^{-2}M^{-1}) + \mathcal{O}(A_{j,j}^{-2}MT^{-1}).$$

Since $A_{j,j} = \mathcal{O}(2^j)$, the proposed smoothed wavelet periodogram is asymptotically mean-squared consistent for the true spectrum as $T \to \infty, M \to \infty, MT^{-1} \to 0$.

□

## C.3 Proof of Proposition 5.3

*Proof.* To establish the consistency of our network-constrained localized precision estimator, we adapt the classical foundational results of graphical covariance selection (Dempster 1972, Lauritzen 1996) to the scale-isolated Gaussian wavelet subprocess framework. Recall from equation (21) that isolating resolution level $j$ yields localized subprocess wavelet coefficients following a multivariate normal distribution. This structure aligns the Gaussian likelihood with the target localized spectral matrix $\mathbf{S}_j(k/T)$, allowing us to map the network constraints of $\mathcal{G}$ onto its precision structure, $\boldsymbol{\Theta}_j(k/T)$.

In order to prove consistency, we first establish the following intermediate result for each scale $j$ and rescaled time $z$.

**Proposition C.1.** *Let $\widehat{\mathbf{S}}_j(z)$ be the symmetric, positive-definite estimator defined in (23). Then:*

1. *There exists a unique network-constrained estimator $\widehat{\boldsymbol{\Theta}}^*_j(z)$ that maximizes the objective function in (24) and a corresponding $\widehat{\mathbf{S}}^*_j(z) := \left(\widehat{\boldsymbol{\Theta}}^*_j(z)\right)^{-1}$ satisfying*

$$\widehat{S}_j^{*(p,q)}(z) = \widehat{S}_j^{(p,q)}(z), \quad \forall (p,q) \in \mathcal{E} \text{ or } p = q,$$

   *and the constraint*

$$\widehat{\Theta}_j^{*(p,q)}(z) = 0, \quad \forall (p,q) \notin \mathcal{E}.$$

2. *Among all multivariate normal models meeting the restrictions specified in item 1, the matrix $\widehat{\mathbf{S}}^*_j(z)$ acts as the unique localized maximum likelihood estimator (MLE) of $\mathbf{S}_j(z)$ under the subprocess log-likelihood.*

*Proof of Proposition C.1.*

1. For clarity of the matrix calculus steps, we temporarily suppress the scale subscript $j$ and localized time index $k/T$, writing $\boldsymbol{\Theta} = \boldsymbol{\Theta}_j(k/T)$, $\widehat{\mathbf{S}} = \widehat{\mathbf{S}}_j(k/T)$, and $\Theta^{(p,q)} = \Theta_j^{(p,q)}(k/T)$. The constrained log-likelihood function defined in (24) can be explicitly parametrized over the complement of the network edge set $\mathcal{E}$ using Lagrange multipliers

$$\mathcal{L}(\boldsymbol{\Theta}) = \log\det\boldsymbol{\Theta} - \mathrm{Tr}(\widehat{\mathbf{S}}\boldsymbol{\Theta}) - \sum_{\{(p,q)|(p,q)\notin\mathcal{E}\}} l^{(p,q)}\Theta^{(p,q)}, \tag{17}$$

   where $l^{(p,q)}$ are the Lagrange multipliers restricting $\Theta^{(p,q)} = 0$ for non-adjacent nodes.

   Differentiating the Lagrangian objective (17) with respect to each independent parameter element $\Theta^{(p,q)}$ of the symmetric precision matrix yields the first-order operational conditions for any pair of nodes $p$ and $q$.

*For diagonal entries* where $p = q$:

$$\frac{\partial \mathcal{L}}{\partial \Theta^{(p,p)}} = \frac{1}{2} \frac{\partial \log \det \boldsymbol{\Theta}}{\partial \Theta^{(p,p)}} - \frac{1}{2} \widehat{S}^{(p,p)}, \tag{18}$$

since the structural network configuration enforces no penalty on the individual node variance boundaries.

*For off-diagonal entries* where $p \neq q$, we evaluate based on the structural graph configuration:

- If $(p, q) \in \mathcal{E}$:

$$\frac{\partial \mathcal{L}}{\partial \Theta^{(p,q)}} = \frac{1}{2} \frac{\partial \log \det \boldsymbol{\Theta}}{\partial \Theta^{(p,q)}} - \widehat{S}^{(p,q)}. \tag{19}$$

- If $(p, q) \notin \mathcal{E}$:

$$\frac{\partial \mathcal{L}}{\partial \Theta^{(p,q)}} = \frac{1}{2} \frac{\partial \log \det \boldsymbol{\Theta}}{\partial \Theta^{(p,q)}} - \widehat{S}^{(p,q)} - l^{(p,q)}. \tag{20}$$

Utilizing symmetric matrix calculus derivative properties for log-determinants, we recall that:

$$\frac{\partial \log \det \boldsymbol{\Theta}}{\partial \Theta^{(p,q)}} = \operatorname{Tr}\left( \boldsymbol{\Theta}^{-1} \frac{\partial \boldsymbol{\Theta}}{\partial \Theta^{(p,q)}} \right).$$

Since the partial derivative matrix $\frac{\partial \boldsymbol{\Theta}}{\partial \Theta^{(p,q)}}$ is an indicator matrix containing a value of 1 on the diagonal element $(p, p)$ if $p = q$, or a value of 1 at both symmetric locations $(p, q)$ and $(q, p)$ if $p \neq q$, the trace expression evaluates to

$$\frac{\partial \log \det \boldsymbol{\Theta}}{\partial \Theta^{(p,p)}} = \left[\boldsymbol{\Theta}^{-1}\right]_{p,p}, \quad \text{for diagonal entries,}$$
$$\frac{\partial \log \det \boldsymbol{\Theta}}{\partial \Theta^{(p,q)}} = 2\left[\boldsymbol{\Theta}^{-1}\right]_{p,q}, \quad \text{for off-diagonal entries.}$$

Substituting these matrix algebraic simplifications back into equations (18), (19), and (20), and setting the partial derivatives to zero, we obtain the following.

For $p = q$:

$$\left[\boldsymbol{\Theta}^{-1}\right]_{p,p} = \widehat{S}^{(p,p)}. \tag{21}$$

For $p \neq q$, with $(p,q) \in \mathcal{E}$:

$$\left[\boldsymbol{\Theta}^{-1}\right]_{p,q} = \widehat{S}^{(p,q)}. \tag{22}$$

For $p \neq q$, with $(p,q) \notin \mathcal{E}$:

$$\Theta^{(p,q)} = 0. \tag{23}$$

Restoring the scale index $j$ and rescaled localized time index $k/T$, equations (21), (22), and (23) replicate the structural conditions specified under item 1, defining the optimized matrices $\widehat{\boldsymbol{\Theta}}_j^*(k/T)$ and $\widehat{\mathbf{S}}_j^*(k/T)$.

2. Since the smoothed input matrix $\widehat{\mathbf{S}}_j(k/T)$ is positive-definite, and the objective function $\mathcal{L}(\boldsymbol{\Theta})$ is strictly concave over the network-constrained parameter subspace, a unique global maximum exists and satisfies the zero-sparsity constraints encoded by the network $\mathcal{G}$ for a localized graphical Gaussian MLE (Lauritzen 1996, Theorem 5.3).

□

By application of Proposition C.1, the network-constrained estimators $\widehat{\boldsymbol{\Theta}}_j^*(k/T)$ and $\widehat{\mathbf{S}}_j^*(k/T)$ represent the unique localized maximum likelihood solutions, hence

$$\widehat{\boldsymbol{\Theta}}_j^*(k/T) - \boldsymbol{\Theta}_j(k/T) = o_P(1) \quad \text{and} \quad \widehat{\mathbf{S}}_j^*(k/T) - \mathbf{S}_j(k/T) = o_P(1).$$

□

## C.4 Proof of Proposition 5.4

*Proof.* Following the proof of (Wu et al. 2026, Proposition 4) applied to our setup, it follows that for every scale $j$, time $k$ and node $p$, the wavelet coefficient of the approximated scale $j$-subprocess converges in mean square to its true subprocess counterpart, namely $\mathbb{E}|\widetilde{d}_{j,k}^{j,(p)} - d_{j,k}^{j,(p)}|^2 = \mathcal{O}(T^{-1})$. Coupling the consequent $\widetilde{\mathbf{d}}_{j,k} - \mathbf{d}_{j,k} = o_P(1)$ with the distributional

property of the true wavelet subprocess coefficients in (21),

$$\widetilde{\mathbf{d}}^j_{j,k} \dot\sim N_n\left(\mathbf{0}, A_{j,j}\mathbf{S}_j(k/T)\right),$$

we obtain our desired conclusion. □

## C.5 Proof of Proposition 5.5

*Proof.* For notational simplicity, in what follows we drop the index $T$ from the smoothing bandwidth $M_T$. We denote by $\boldsymbol{\zeta}^j_{j,k} = \widetilde{\mathbf{d}}^j_{j,k} - \mathbf{d}^j_{j,k}$ for each time $k$ and scale $j$, and express the terms in equation (29) using the newly formed $\boldsymbol{\zeta}^j_{j,k}$ terms, as well as the true subprocess wavelet coefficients, $\mathbf{d}^j_{j,k}$. The spectral estimator at $z = k/T$ can then be expressed as

$$\begin{aligned}
\widehat{\widetilde{S}}^{(p,q)}_j(k/T) &= \frac{1}{A_{j,j}(2M+1)}\sum_{m=-M}^{M} \widetilde{d}^{j,(p)}_{j,k+m}\widetilde{d}^{j,(q)}_{j,k+m} \\
&= \frac{1}{A_{j,j}(2M+1)}\sum_{m=-M}^{M} (\zeta^{j,(p)}_{j,k+m} + d^{j,(p)}_{j,k+m})(\zeta^{j,(q)}_{j,k+m} + d^{j,(q)}_{j,k+m}) \\
&= \frac{1}{A_{j,j}(2M+1)}\sum_{m=-M}^{M} (\zeta^{j,(p)}_{j,k+m}\zeta^{j,(q)}_{j,k+m} + d^{j,(p)}_{j,k+m}d^{j,(q)}_{j,k+m} + \zeta^{j,(q)}_{j,k+m}d^{j,(p)}_{j,k+m} + \zeta^{j,(p)}_{j,k+m}d^{j,(q)}_{j,k+m}) \\
&= B1 + B2 + B3 + B4.
\end{aligned}$$

The proof of Proposition 5.4 shows that $\zeta^{j,(p)}_{j,k} = o_P(1)$ at any node $p$, hence the first term is $B1 = o_P(1)$. Observe that the second term term is in fact $B2 = \widehat{S}^{(p,q)}_j(k/T)$, hence by Proposition 5.2 we have $B2 = S^{(p,q)}_j(k/T) + o_P(1)$ as $T \to \infty$.

Let us now turn to the third and fourth terms. We consider

$$\begin{aligned}
|\mathbb{E}(\zeta^{j,(q)}_{j,k}d^{j,(p)}_{j,k})|^2 &\leq \mathbb{E}(\zeta^{j,(q)}_{j,k})^2\mathbb{E}(d^{j,(p)}_{j,k})^2 \text{ from the Cauchy-Schwarz inequality, and there exists } K > 0 \\
&\leq KT^{-1}\mathbb{E}(d^{j,(p)}_{j,k})^2 \text{ since } \mathbb{E}|\zeta^{j,(q)}_{j,k}|^2 = \mathcal{O}(T^{-1}), \\
&= KT^{-1}\left(A_{j,j}S^{(p,p)}_j(k/T)) + \mathcal{O}(T^{-1})\right) \\
&= \mathcal{O}(2^jT^{-1}), \text{ as } A_{j,j} = \mathcal{O}(2^j), \qquad (24)
\end{aligned}$$

where in the penultimate expression we used the result in (20), then collected the maximum order terms. Thus $\zeta_{j,k}^{j,(q)} d_{j,k}^{j,(p)} = o_P(1)$, from which the terms $B3 = o_P(1)$ and $B4 = o_P(1)$. Collecting the results for all four terms, we conclude $\widehat{\widetilde{S}}_j^{(p,q)}(z) = S_j^{(p,q)}(z) + o_P(1)$.

Let us now investigate

$$\mathbb{E}(\widehat{\widetilde{S}}_j^{(p,q)}(k/T)) = \mathbb{E}(B1) + \mathbb{E}(B2) + \mathbb{E}(B3) + \mathbb{E}(B4). \tag{25}$$

For the first term in (25), using the proof of Proposition 5.4, we have seen that $\mathbb{E}\,|\zeta_{j,k}^{j,(p)}|^2 = \mathcal{O}(T^{-1})$ at every node $p$. An application of the Cauchy-Schwarz inequality at nodes $p$, $q$ therefore yields $\mathbb{E}(\zeta_{j,k}^{j,(p)} \zeta_{j,k'}^{j,(q)}) = \mathcal{O}(T^{-1})$ at every time $k$, hence we have $\mathbb{E}(B1) = \mathcal{O}\left(2^{-j}T^{-1}\right)$.

Since $B2 = \widehat{S}_j^{(p,q)}(k/T)$, the second term is $\mathbb{E}(B2) = S_j^{(p,q)}(k/T) + \mathcal{O}\left(2^{-j}MT^{-1}\right)$ from Proposition 5.2.

Let us now turn to the last two terms of (25). From equation (24), we have $\mathbb{E}(\zeta_{j,k}^{j,(q)} d_{j,k}^{j,(p)}) = \mathcal{O}(2^{j/2}T^{-1/2})$ hence $\mathbb{E}(B3) = \mathcal{O}(2^{-j/2}T^{-1/2})$ and similar for $\mathbb{E}(B4)$. Substituting these terms back into (25), we obtain

$$\mathbb{E}(\widehat{\widetilde{S}}_j^{(p,q)}(k/T)) = S_j^{(p,q)}(k/T) + \mathcal{O}(M_T'), \tag{26}$$

where $M_T' = 2^{-j}MT^{-1} \vee 2^{-j/2}T^{-1/2} \to 0$ when $M$, $T \to \infty$ with $MT^{-1} \to 0$.

Moving onto second order quantities, the general terms that contribute to the variance of $\widehat{\widetilde{S}}_j^{(p,q)}(k/T)$ are of the form

$$\mathbb{E}\Big((\widehat{S}_j^{(p,q)}(k/T))^2\Big) = \frac{A_{j,j}^{-2}}{(2M+1)^2} \sum_{m=-M}^{M} \sum_{\tau=-M-m}^{M-m} \mathbb{E}\Big(\widetilde{d}_{j,k+m}^{j,(p)} \widetilde{d}_{j,k+m}^{j,(q)} \widetilde{d}_{j,k+m+\tau}^{j,(q)} \widetilde{d}_{j,k+m+\tau}^{j,(q)}\Big).$$

Following Isserlis (1918), the above becomes

$$\begin{aligned}
\text{var}\left(\widehat{\widetilde{S}}_j^{(p,q)}(k/T)\right) &= \frac{A_{j,j}^{-2}}{(2M+1)^2}\Bigg\{\sum_{m=-M}^{M}\sum_{\tau}\mathbb{E}\left(\widetilde{d}_{j,k+m}^{j,(p)}\widetilde{d}_{j,k+m+\tau}^{j,(p)}\right)\times\mathbb{E}\left(\widetilde{d}_{j,k+m}^{j,(q)}\widetilde{d}_{jj,k+m+\tau}^{j,(q)}\right) \\
&+\sum_{m=-M}^{M}\sum_{\tau}\mathbb{E}\left(\widetilde{d}_{j,k+m}^{j,(p)}\widetilde{d}_{j,k+m+\tau}^{j,(q)}\right)\times\ \mathbb{E}\left(\widetilde{d}_{j,k+m+\tau}^{j,(p)}\widetilde{d}_{j,k+m}^{j,(q)}\right)\Bigg\}, \text{ then akin to (26)} \\
&= \frac{A_{j,j}^{-2}}{(2M+1)^2}\sum_{m=-M}^{M}\Bigg\{\sum_{\tau}\left(A_{jj}^{(\tau)}S_j^{(p,p)}(k/T)\right)\times\left(A_{jj}^{(\tau)}S_j^{(q,q)}(k/T)\right) \\
&+\sum_{\tau}\left(A_{jj}^{(\tau)}S_j^{(p,q)}(k/T)\right)\times\left(A_{jj}^{(-\tau)}S_j^{(p,q)}(k/T)\right) \\
&+\sum_{\tau}(|m|+1)\mathcal{O}(M_T'')+\sum_{\tau}(|m|+1)^2\mathcal{O}((M_T'')^2)\Bigg\},
\end{aligned}$$

where the last equality followed using $\widetilde{d}_{j,k}^{j,(p)} = \zeta_{j,k}^{j,(p)} + d_{j,k}^{j,(p)}$ coupled with the properties $\mathbb{E}\left(d_{j,k}^{j,(p)}d_{j,k+\tau}^{j,(q)}\right) = A_{j,j}^{(\tau)}S_j^{(p,q)}(k/T)+\mathcal{O}(T^{-1})$ in (20), $\mathbb{E}(\zeta_{j,k}^{j,(q)}d_{j,k+\tau}^{j,(p)}) = \mathcal{O}(A_{j,j}^{1/2}T^{-1/2})$ and $\mathbb{E}(\zeta_{j,k}^{j,(p)}\zeta_{j,k+\tau}^{j,(q)}) = \mathcal{O}(T^{-1})$ above, which lead to $\mathbb{E}\left(\widetilde{d}_{j,k}^{j,(p)}\widetilde{d}_{j,k+\tau}^{j,(q)}\right) = A_{j,j}^{(\tau)}S_j^{(p,q)}(k/T)+\mathcal{O}(M_T'')$, with $M_T'' = 2^{j/2}T^{-1/2}$.

Similar arguments to those in the proof of Proposition 5.2 and retaining highest order terms yield

$$\text{var}\left(\widehat{\widetilde{S}}_j^{(p,q)}(k/T)\right) = \mathcal{O}(M^{-1}\vee 2^{-3j/2}MT^{-1/2}\vee 2^{-j}M^2T^{-1}),$$

thus the proposed smoothed wavelet periodogram $\widehat{\widetilde{S}}_j^{(p,q)}(k/T)$ is asymptotically mean-squared consistent for the true spectrum as $T\to\infty, M\to\infty, MT^{-1/2}\to 0$.

□

## C.6 Proof of Corollary 5.1

*Proof.* Coupling the consistency of the spectral estimator that employs the approximated subprocess construction from Proposition 5.5 with the result in Proposition 5.3, the resulting spectral and precision estimators are consistent for their corresponding true quantities, namely $\widetilde{\mathbf{S}}_j^*(z) - \mathbf{S}_j(z) = o_P(1)$ and $\widetilde{\boldsymbol{\Theta}}_j^*(z) - \boldsymbol{\Theta}_j(z) = o_P(1)$. Then the desired result follows by a direct application of the continuous mapping theorem (Billingsley 1999). □

# D Simulation Study

To evaluate the finite-sample performance of our framework, we conduct a comprehensive simulation study. The objective is twofold: (1) to assess the model's ability to recover multiscale, nonstationary dependence structures constrained by a network topology, and (2) to evaluate the scalability and robustness of our framework under several configurations.

## D.1 Data Generating Process

To simulate a Net-LSW process, a time-series realization $\widehat{\mathbf{X}}_t^{\mathcal{N}}$ is synthesized through a hierarchical multiscale procedure. Here, $\widehat{\mathbf{X}}_t^{\mathcal{N}}$ denotes a finite realization of a true underlying Net-LSW process $\mathbf{X}_t^{\mathcal{N}}$ as specified by the properties in Section 3.1. Using the subprocess-based representation described in Section 5.1, the simulated Net-LSW process $\widehat{\mathbf{X}}_t^{\mathcal{N}}$ is a superposition of independent subprocesses across a set of active scales $\mathcal{J}$ (Remark 6),

$$\widehat{\mathbf{X}}_t^{\mathcal{N}} = \sum_{j \in \mathcal{J}} \widehat{\mathbf{X}}_{j,t}.$$

To thoroughly examine the flexibility of our setup, we consider different numbers of active resolution levels and nonstationary parameter modulations. The generation of each scale-specific component $\widehat{\mathbf{X}}_{j,t}$ follows a four-stage simulation pipeline:

1. **Network-constrained precision dynamics:** Let $\mathcal{G} = (\mathcal{N}, \mathcal{E})$ be a network topology under consideration. For each scale $j$ and rescaled time $z = k/T$, we define a time-varying spectral precision matrix $\widehat{\mathbf{\Theta}}_j(z)$. The sparsity pattern of $\widehat{\mathbf{\Theta}}_j(z)$ is strictly constrained by the adjacency, ensuring that $[\widehat{\mathbf{\Theta}}_j(z)]_{p,q} = 0$ if $(p, q) \notin \mathcal{E}$ for $p \neq q$.

   We allocate different nonstationary activity types to control the temporal evolution of the off-diagonal strength parameter $\theta_j(z)$ at each scale $j \in \mathcal{J}$:

- *Transient burst (Type 1):*

$$\theta_j(z) = \begin{cases} 0.8 & \text{if } 0.3 \le z \le 0.4 \\ 0.1 & \text{otherwise} \end{cases}$$

- *Changepoint (Type 2):*

$$\theta_j(z) = \begin{cases} 0.1 & \text{if } z < 0.5 \\ 0.6 & \text{if } z \ge 0.5 \end{cases}$$

- *Linear trend (Type 3):*

$$\theta_j(z) = 0.2 + 0.3z \quad z \in [0,1]$$

- *Piecewise constant step function (Type 4):*

$$\theta_j(z) = \begin{cases} -0.7 & \text{if } z \le 0.5 \\ 0.4 & \text{if } z > 0.5 \end{cases}$$

- *Uniform CDF (linear transition) (Type 5):*

$$\theta_j(z) = \begin{cases} 0 & \text{if } z < 0.25 \\ 2z - 0.5 & 0.25 \le z \le 0.75 \\ 1 & \text{if } z > 0.75 \end{cases}$$

- *Cosine wave (Type 6):*

$$\theta_j(z) = 0.2\left[1 + 0.5\cos(2\pi z)\right]$$

The functional forms of these time-varying modulations are illustrated in Figure 1.

In multiple-scale configurations, we mix these behaviours across active scales to test the model's capacity to decouple distinct, overlapping dynamics.

2. **Generation of correlated innovations:** The local spectral density matrix is derived as the inverse of the simulated precision matrix, $\widehat{\mathbf{S}}_j(z) = \left(\widehat{\boldsymbol{\Theta}}_j(z)\right)^{-1}$. From this matrix, we extract the innovation correlation structure,

$$\widehat{\boldsymbol{\Lambda}}_j(z) = \operatorname{diag}\left(\widehat{\mathbf{S}}_j(z)\right)^{-1/2} \widehat{\mathbf{S}}_j(z) \operatorname{diag}\left(\widehat{\mathbf{S}}_j(z)\right)^{-1/2}. \tag{27}$$

   At each scale $j$ and time point $k$, we sample a multivariate Gaussian innovation vector

$$\widehat{\boldsymbol{\xi}}_{j,k} \sim N_n\left(\mathbf{0}, \widehat{\boldsymbol{\Lambda}}_j(z)\right).$$

3. **Latent subprocess construction:** The innovations are modulated by both the local spectral amplitude and a dyadic scaling factor. Let $\widehat{\mathbf{W}}_j(z) = \operatorname{diag}\left(\widehat{\mathbf{S}}_j(z)\right)^{1/2}$. For each network node $p \in \mathcal{N}$, the scaled innovations serve as the inputs to the synthesis step.

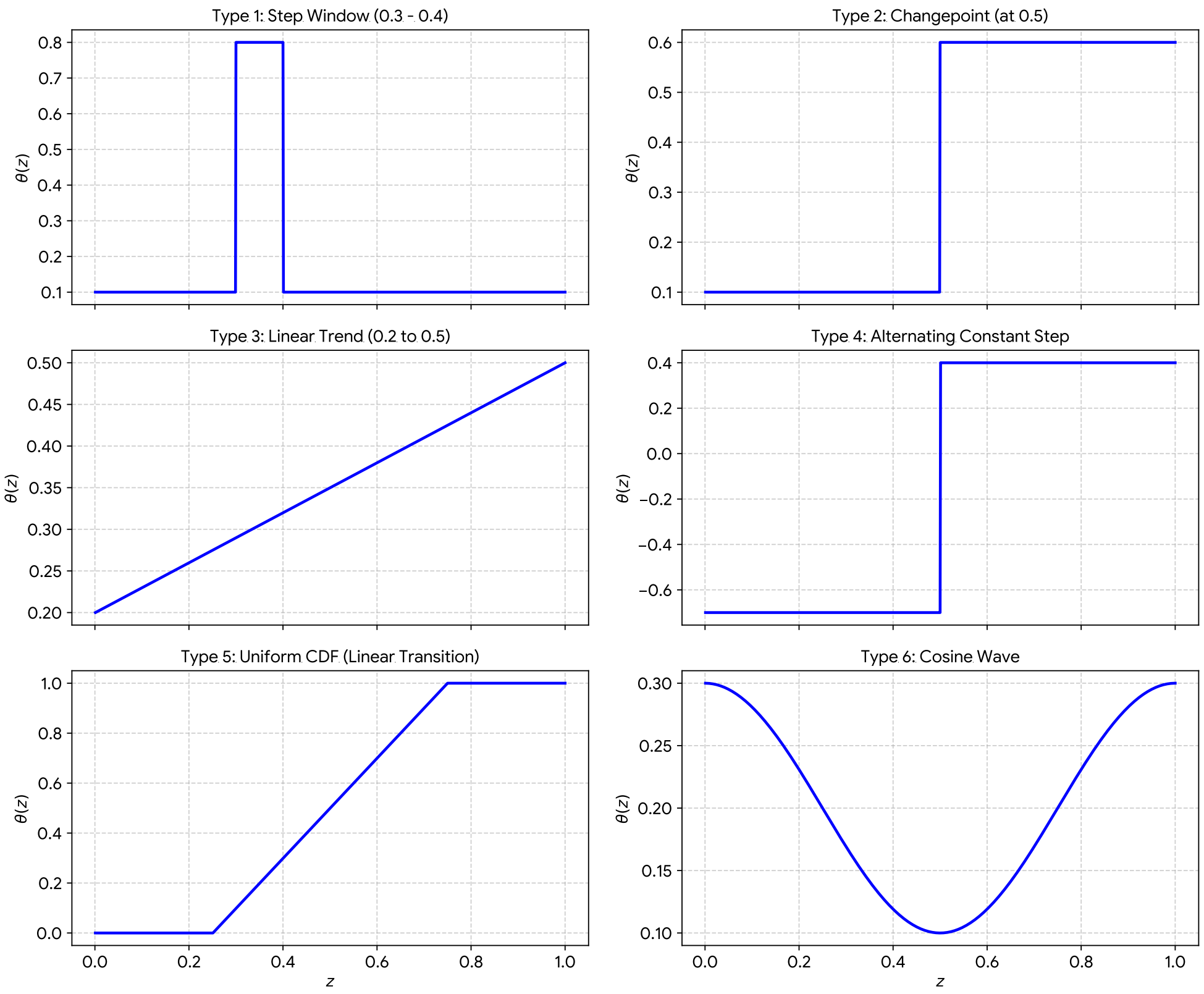

Figure 1: Time-varying profiles for the off-diagonal strength parameter $\theta_j(z)$ across different nonstationary modulation types.

The true subprocess $\widehat{X}_{j,t}^{(p)}$ is reconstructed using the average basis inverse transform (Nason & Silverman 1995)

$$\widehat{X}_{j,t}^{(p)} = \sum_{k=0}^{T-1} \widehat{\mathbf{W}}_j^{(p,\cdot)}(k/T)\widehat{\boldsymbol{\xi}}_{j,k}\psi_{j,k}(t).$$

The simulated Net-LSW process $\widehat{\mathbf{X}}_t^{\mathcal{N}}$ is then constructed by subprocess aggregation

$$\widehat{\mathbf{X}}_t^{\mathcal{N}} = \sum_{j\in\mathcal{J}} \widehat{\mathbf{X}}_{j,t}.$$

This formulation guarantees that the simulated power at each resolution level matches the theoretical target spectrum while respecting the structural zeros imposed by $\mathcal{E}$.

**Remark 1** (**Notational conventions).** *Throughout this section,* $\widehat{\mathbf{X}}_t^{\mathcal{N}}$ *refers exclusively to a simulated realization of the Net-LSW process. This must be conceptually distinguished* *from the approximated subprocesses $\widetilde{\mathbf{X}}_{j,t}$ in Section 5.5. This distinction forms the basis of our assessment framework: in practice, we treat the realized process $\widehat{\mathbf{X}}_t^{\mathcal{N}}$ as the sole observed quantity. The estimation task consists of using $\widehat{\mathbf{X}}_t^{\mathcal{N}}$ to isolate $\widetilde{\mathbf{X}}_{j,t}$ and recover spectral precision estimates, $\widetilde{\mathbf{\Theta}}_j^*(z)$, under the structural zeros of the fixed network $\mathcal{G}$. This is subsequently validated against the generative truth, $\widehat{\mathbf{\Theta}}_j(z)$.*

## D.2 Performance Evaluation Framework

To evaluate the finite-sample performance of the proposed estimation procedure (summarized in Algorithm 1), we track its estimation fidelity against the truth.

For a given active scale $j$ and time-series length $T$, let $R$ denote the total number of independent simulation replicates. Let $\varphi_j^{(p,q)}(k/T)$ represent a localized target dependence structure at scale $j$ and time $k$—denoting either the true local coherence $\rho_j^{(p,q)}(k/T)$ or the true local partial coherence $\eta_j^{(p,q)}(k/T)$ between node pair $(p,q)$. The corresponding estimator obtained from the $r$-th simulation replicate is denoted by $\widetilde{\varphi}_j^{(r);(p,q)}(k/T)$.

We assess the quality of the localized dependency measures using three statistical criteria calculated over the network topology: Mean Squared Error (MSE), Squared Bias (Bias$^2$), and Variance (Var) at each scale $j \in \mathcal{J}$. To evaluate the accuracy across the entire network topology and time grid, the summary metrics are defined in a multiscale manner, using the spatial-temporal expectation over all node paths $p, q \in \mathcal{N}$ and discrete time points $k = 0, \ldots, T-1$, namely,

1. **Mean Squared Error** (MSE)**:**
$$\mathrm{MSE}_j = \frac{1}{n^2 \cdot T} \sum_{p=1}^{n} \sum_{q=1}^{n} \sum_{k=0}^{T-1} \left[ \frac{1}{R} \sum_{r=1}^{R} \left( \widetilde{\varphi}_j^{(r);(p,q)}(k/T) - \varphi_j^{(p,q)}(k/T) \right)^2 \right].$$

2. **Squared Bias** (Bias$^2$)**:**
$$\mathrm{Bias}_j^2 = \frac{1}{n^2 \cdot T} \sum_{p=1}^{n} \sum_{q=1}^{n} \sum_{k=0}^{T-1} \left[ \left( \frac{1}{R} \sum_{r=1}^{R} \widetilde{\varphi}_j^{(r);(p,q)}(k/T) \right) - \varphi_j^{(p,q)}(k/T) \right]^2.$$

3. **Variance** (Var):

$$\mathrm{Var}_j = \mathrm{MSE}_j - \mathrm{Bias}_j^2.$$

Across all settings, we evaluate the proposed Net-LSW framework against standard unconstrained multivariate LSW (**MvLSW**) estimation (Park et al. 2014), as well as benchmark performance against the **Oracle** that uses true subprocess information, $\{\widehat{\mathbf{X}}_{j,t}\}$.

**Remark 2 (Implementation details and limitations).** *The MvLSW estimates are obtained using the* `MvLSW` `R` *package (Taylor et al. 2019), while the constrained optimization step of the Net-LSW framework (described in Section 5.4) is performed via the* `glasso` *package in* `R` *(Friedman et al. 2007, Hastie et al. 2009). A naive approach might also directly apply the constrained optimization step to the MvLSW network-agnostic estimates. However, due to the cross-scale bias correction inherent to the MvLSW framework, their estimates in practice often suffer from non-positive definiteness (although theoretically this should hold). Consequently, the* `glasso` *function cannot accept these matrices as valid empirical covariance inputs (recall the underlying* `glasso` *algorithm relies on recursively partitioning the target matrix by iteratively isolating rows and columns to solve regression problems, as detailed in Algorithm 1). Hence a non-positive-definite input matrix breaks the optimization routine, making direct constrained optimization on the MvLSW estimator impossible in practice. In contrast, the simple form of our proposed estimator that does not require correction renders it always positively definite, both theoretically and practically.*

## D.3 Estimation Accuracy under Fixed Network Topologies

We first investigate estimation performance under two explicit network dimensions represented in Figure 2: the 5-node `fiveNet` configuration from the `GNAR` `R` package (Knight et al. 2020) (left panel) and the 10-node randomly generated Erdös-Renyi network (central panel). We evaluate these graphs across three distinct scenarios corresponding to different

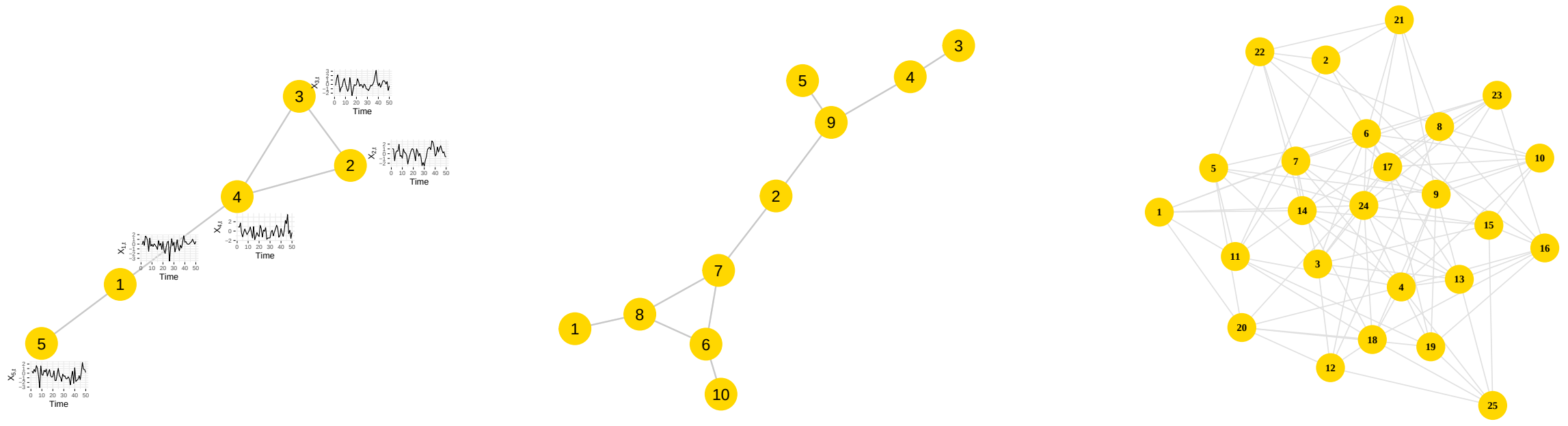

Figure 2: Example network topologies used within the simulation setup: the 5-node `fiveNet` configuration from the `GNAR` package (Knight et al. 2020) in `R` (left), alongside randomly generated Erdös-Renyi network structures with 10 nodes (center) and 25 nodes (right).

activity types as represented in Figure 1:

- **Scenario 1 (single scale):** $\mathcal{J} = \{1\}$, where scale 1 follows Type 1 dynamics.
- **Scenario 2 (two scales):** $\mathcal{J} = \{1, 4\}$, where scale 1 follows Type 1 dynamics and scale 4 follows Type 6 dynamics.
- **Scenario 3 (multiple-scales):** $\mathcal{J} = \{1, 2, 3, 4, 5\}$, mapping respectively to parameter activity Types 1, 2, 3, 4, and 6.

For each scenario, data are generated over $R = 50$ independent replicates for sample sizes $T \in \{512, 1024, 2048\}$. In what follows, we provide a detailed comparative analysis highlighting the relative strengths and asymptotic properties of the competing frameworks. The full empirical results supporting these observations are consolidated in Table 1 (coherence) and Table 2 (partial coherence), which present MSE and Bias$^2$ across all configurations (with variance inferred as their difference).

Across both tables, our proposed estimation procedure consistently outperforms the baseline

MvLSW approach. For instance, in Table 1 (Scenario 1, $n = 5, j = 1, T = 512$), the proposed method records an MSE improvement of 59% when compared to MvLSW. This efficiency gap widens further in the partial coherence setting of Table 2 ($n = 10, j = 1, T = 2048$), where the proposed framework achieves a highly precise MSE of .00272 compared to .03681 for that of Park et al. (2014), representing an error reduction of over 92%.

When handling structural scale-localized behaviour at scale $j = 4$ (Scenario 2), MvLSW exhibits severe structural deficiencies; as seen in Table 1 ($n = 5, j = 4, T = 512$), MvLSW has an MSE of .12697 driven heavily by a Bias$^2$ of .07963, while our proposed method maintains an MSE of just .04832. A similar trend occurs in Table 2 ($n = 10, j = 4, T = 1024$), where our estimation technique achieves a substantial reduction in MSE compared to MvLSW (.01230 versus .02444).

As the series length $T$ increases, our technique demonstrates strong consistency across both tables, as seen in Table 1 (Scenario 1, $n = 5, j = 1$) where its MSE drops cleanly from .01862 to .01261 to .00964 for $T \in \{512, 1024, 2048\}$. Conversely, the asymptotics completely break down for MvLSW in complex environments; in Table 2 (Scenario 3, $n = 10, j = 1$), its MSE actually degrades from .04066 to .09408 as $T$ increases.

Furthermore, doubling the network size highlights the scaling power of the proposed framework; in Table 2 (Scenario 1, $j = 1, T = 512$), expanding the network drops the MSE for our framework from .01117 ($n = 5$) down to .00448 ($n = 10$), whereas MvLSW has its error increase with larger network dimensions, with its MSE climbing from .06262 to .11981.

While MvLSW never beats our proposed method in both metrics simultaneously, there are settings in which it can achieve a lower bias under large samples at fine scales (e.g., Table 1, $n = 5$, Scenario 1, $j = 1, T = 2048$, where MvLSW achieves a lower Bias$^2$ compared to our estimation scheme, though its variance renders its MSE uncompetitive). Our method is most challenged by Scenario 3 at coarse scales ($j = 5, T = 512$), leading to its highest overall MSE of .10784 in Table 1, which is, however, stabilized to .07125 under the partial coherence framework of Table 2.

Table 1: Coherence accuracy metrics (MSE and Bias$^2$) for the $n = 5$ and $n = 10$ node networks across Scenarios 1, 2, and 3.

| | | | MSE | | | Bias$^2$ | | |
|---|---|---|---|---|---|---|---|---|
| **Scenario** | **Scale** | **Method** | $T = 512$ | $T = 1024$ | $T = 2048$ | $T = 512$ | $T = 1024$ | $T = 2048$ |
| **n=5** | | | | | | | | |
| **Scenario 1** | $j = 1$ | Oracle | .01524 | .01026 | .00765 | .00628 | .00350 | .00220 |
| | | Proposed | .01862 | .01261 | .00964 | .00635 | .00352 | .00228 |
| | | MvLSW | .04537 | .03049 | .02177 | .00336 | .00248 | .00183 |
| **Scenario 2** | $j = 1$ | Oracle | .01524 | .01026 | .00765 | .00628 | .00350 | .00220 |
| | | Proposed | .01854 | .01287 | .00987 | .00655 | .00398 | .00258 |
| | | MvLSW | .05341 | .03473 | .02450 | .00474 | .00274 | .00188 |
| | $j = 4$ | Oracle | .03483 | .02751 | .02396 | .00125 | .00075 | .00067 |
| | | Proposed | .04832 | .03796 | .03335 | .00186 | .00117 | .00099 |
| | | MvLSW | .12697 | .11069 | .09686 | .07963 | .05205 | .02340 |
| **Scenario 3** | $j = 1$ | Oracle | .01524 | .01026 | .00765 | .00628 | .00350 | .00220 |
| | | Proposed | .01839 | .01425 | .01188 | .00890 | .00671 | .00597 |
| | | MvLSW | .08356 | .06638 | .04955 | .02471 | .01627 | .01221 |
| | $j = 2$ | Oracle | .01043 | .00724 | .00580 | .00217 | .00127 | .00085 |
| | | Proposed | .02067 | .01841 | .01647 | .01056 | .00999 | .00987 |
| | | MvLSW | .18876 | .16512 | .14514 | .12491 | .11061 | .09826 |
| | $j = 3$ | Oracle | .01419 | .01192 | .00940 | .00036 | .00034 | .00023 |
| | | Proposed | .03903 | .03627 | .03111 | .01890 | .01937 | .01812 |
| | | MvLSW | .24416 | .23151 | .19657 | .15834 | .15271 | .13163 |
| | $j = 4$ | Oracle | .03184 | .02567 | .01956 | .00335 | .00209 | .00084 |
| | | Proposed | .06442 | .05537 | .04499 | .02102 | .01965 | .01695 |
| | | MvLSW | .28555 | .27019 | .24264 | .18920 | .18067 | .16828 |
| | $j = 5$ | Oracle | .07373 | .05746 | .04564 | .00349 | .00150 | .00151 |
| | | Proposed | .10784 | .08670 | .07847 | .01718 | .01430 | .01877 |
| | | MvLSW | .22199 | .19317 | .18302 | .12341 | .11223 | .10626 |
| **n = 10** | | | | | | | | |
| **Scenario 1** | $j = 1$ | Oracle | .01024 | .00708 | .00513 | .00492 | .00285 | .00175 |

Continued on next page

Table 1: Coherence accuracy metrics (MSE and Bias$^2$) for the $n = 5$ and $n = 10$ node networks across Scenarios 1, 2, and 3. – Continued from previous page

| | | | MSE | | | Bias$^2$ | | |
|---|---|---|---|---|---|---|---|---|
| **Scenario** | **Scale** | **Method** | $T = 512$ | $T = 1024$ | $T = 2048$ | $T = 512$ | $T = 1024$ | $T = 2048$ |
| | | Proposed | .01220 | .00861 | .00634 | .00499 | .00292 | .00178 |
| | | MvLSW | .05388 | .03811 | .02674 | .00485 | .00229 | .00145 |
| **Scenario 2** | $j = 1$ | Oracle | .01024 | .00708 | .00513 | .00492 | .00285 | .00175 |
| | | Proposed | .01225 | .00877 | .00657 | .00514 | .00320 | .00205 |
| | | MvLSW | .05884 | .04891 | .03331 | .01737 | .00919 | .00314 |
| | $j = 4$ | Oracle | .02613 | .02033 | .01652 | .00072 | .00047 | .00040 |
| | | Proposed | .03675 | .02840 | .02312 | .00104 | .00070 | .00067 |
| | | MvLSW | .06061 | .05895 | .05743 | .05748 | .05600 | .05365 |
| **Scenario 3** | $j = 1$ | Oracle | .01024 | .00708 | .00513 | .00492 | .00285 | .00175 |
| | | Proposed | .01239 | .00977 | .00775 | .00645 | .00512 | .00396 |
| | | MvLSW | .04765 | .05108 | .04878 | .03558 | .02731 | .02133 |
| | $j = 2$ | Oracle | .00798 | .00577 | .00464 | .00157 | .00098 | .00057 |
| | | Proposed | .01522 | .01355 | .01141 | .00791 | .00757 | .00647 |
| | | MvLSW | .09494 | .09127 | .08811 | .08854 | .08625 | .08343 |
| | $j = 3$ | Oracle | .01272 | .01034 | .00845 | .00022 | .00025 | .00022 |
| | | Proposed | .02672 | .02295 | .02037 | .01096 | .01038 | .01022 |
| | | MvLSW | .11151 | .10625 | .10125 | .09875 | .09664 | .09395 |
| | $j = 4$ | Oracle | .02867 | .02064 | .01694 | .00353 | .00169 | .00095 |
| | | Proposed | .04433 | .04086 | .03521 | .01413 | .01575 | .01417 |
| | | MvLSW | .13524 | .13212 | .12670 | .11931 | .11895 | .11723 |
| | $j = 5$ | Oracle | .04979 | .03957 | .03321 | .00166 | .00116 | .00096 |
| | | Proposed | .07027 | .06056 | .05195 | .00982 | .01130 | .00910 |
| | | MvLSW | .07334 | .07008 | .06654 | .06222 | .06163 | .05973 |

Table 2: Partial coherence accuracy metrics (MSE and $\text{Bias}^2$) for the $n = 5$ and $n = 10$ node networks across Scenarios 1, 2, and 3.

| | | | **MSE** | | | **Bias²** | | |
|---|---|---|---|---|---|---|---|---|
| **Scenario** | **Scale** | **Method** | $T = 512$ | $T = 1024$ | $T = 2048$ | $T = 512$ | $T = 1024$ | $T = 2048$ |
| **n=5** | | | | | | | | |
| **Scenario 1** | $j = 1$ | Oracle | .00838 | .00617 | .00485 | .00044 | .00029 | .00018 |
| | | Proposed | .01117 | .00813 | .00648 | .00050 | .00033 | .00021 |
| | | MvLSW | .06262 | .03856 | .02564 | .00134 | .00088 | .00052 |
| **Scenario 2** | $j = 1$ | Oracle | .00838 | .00617 | .00485 | .00044 | .00029 | .00018 |
| | | Proposed | .01076 | .00790 | .00625 | .00049 | .00032 | .00020 |
| | | MvLSW | .07889 | .04885 | .03183 | .00189 | .00119 | .00068 |
| | $j = 4$ | Oracle | .02777 | .02188 | .01774 | .00070 | .00058 | .00047 |
| | | Proposed | .03886 | .02995 | .02465 | .00102 | .00090 | .00072 |
| | | MvLSW | .10730 | .12187 | .14376 | .03249 | .02196 | .01103 |
| **Scenario 3** | $j = 1$ | Oracle | .00838 | .00617 | .00485 | .00044 | .00029 | .00018 |
| | | Proposed | .00864 | .00670 | .00527 | .00057 | .00052 | .00046 |
| | | MvLSW | .13022 | .10332 | .06877 | .00963 | .00546 | .00331 |
| | $j = 2$ | Oracle | .00689 | .00533 | .00461 | .00022 | .00020 | .00012 |
| | | Proposed | .00891 | .00728 | .00605 | .00068 | .00070 | .00061 |
| | | MvLSW | .17873 | .14960 | .13389 | .04532 | .03560 | .02933 |
| | $j = 3$ | Oracle | .01164 | .00945 | .00766 | .00028 | .00028 | .00018 |
| | | Proposed | .02081 | .01727 | .01403 | .00347 | .00335 | .00323 |
| | | MvLSW | .20431 | .19149 | .16983 | .06401 | .05803 | .04891 |
| | $j = 4$ | Oracle | .02823 | .02212 | .01826 | .00117 | .00093 | .00045 |
| | | Proposed | .03792 | .03165 | .02558 | .00386 | .00381 | .00266 |
| | | MvLSW | .20493 | .19982 | .18574 | .06378 | .06018 | .05288 |
| | $j = 5$ | Oracle | .05149 | .04235 | .03467 | .00263 | .00132 | .00111 |
| | | Proposed | .07125 | .05794 | .05043 | .00550 | .00347 | .00491 |
| | | MvLSW | .19325 | .16852 | .17373 | .06635 | .05645 | .05458 |
| **n = 10** | | | | | | | | |
| **Scenario 1** | $j = 1$ | Oracle | .00337 | .00260 | .00204 | .00022 | .00013 | .00009 |
| | | Proposed | .00448 | .00345 | .00272 | .00025 | .00016 | .00012 |
| | | MvLSW | .11981 | .06296 | .03681 | .00334 | .00140 | .00076 |

Continued on next page

Table 2: Partial coherence accuracy metrics (MSE and Bias$^2$) for the $n = 5$ and $n = 10$ node networks across Scenarios 1, 2, and 3. – Continued from previous page

| | | | **MSE** | | | **Bias$^2$** | | |
|---|---|---|---|---|---|---|---|---|
| **Scenario** | **Scale** | **Method** | $T = 512$ | $T = 1024$ | $T = 2048$ | $T = 512$ | $T = 1024$ | $T = 2048$ |
| **Scenario 2** | $j = 1$ | Oracle | .00337 | .00260 | .00204 | .00022 | .00013 | .00009 |
| | | Proposed | .00433 | .00335 | .00265 | .00025 | .00016 | .00011 |
| | | MvLSW | .12318 | .08912 | .05112 | .00658 | .00337 | .00136 |
| | $j = 4$ | Oracle | .01114 | .00905 | .00752 | .00045 | .00032 | .00025 |
| | | Proposed | .01494 | .01230 | .01036 | .00067 | .00052 | .00040 |
| | | MvLSW | .02541 | .02444 | .02419 | .02312 | .02234 | .02130 |
| **Scenario 3** | $j = 1$ | Oracle | .00337 | .00260 | .00204 | .00022 | .00013 | .00009 |
| | | Proposed | .00378 | .00284 | .00225 | .00033 | .00024 | .00020 |
| | | MvLSW | .04066 | .08374 | .09408 | .01437 | .01087 | .00759 |
| | $j = 2$ | Oracle | .00309 | .00236 | .00193 | .00013 | .00008 | .00006 |
| | | Proposed | .00385 | .00322 | .00267 | .00036 | .00037 | .00031 |
| | | MvLSW | .03568 | .03120 | .02984 | .02632 | .02447 | .02318 |
| | $j = 3$ | Oracle | .00467 | .00383 | .00311 | .00012 | .00012 | .00009 |
| | | Proposed | .00962 | .00776 | .00657 | .00233 | .00201 | .00191 |
| | | MvLSW | .05648 | .04835 | .04173 | .03435 | .03224 | .03012 |
| | $j = 4$ | Oracle | .01198 | .00904 | .00753 | .00067 | .00030 | .00027 |
| | | Proposed | .01562 | .01298 | .01102 | .00153 | .00128 | .00121 |
| | | MvLSW | .05162 | .04698 | .04078 | .02961 | .02859 | .02699 |
| | $j = 5$ | Oracle | .01994 | .01668 | .01394 | .00110 | .00078 | .00062 |
| | | Proposed | .02716 | .02382 | .02069 | .00307 | .00304 | .00250 |
| | | MvLSW | .04055 | .03742 | .03323 | .02774 | .02685 | .02536 |

## D.4 Effect of Network Sparsity

In this section, we study how varying levels of edge density in the underlying network affect estimation stability. We isolate this structural effect by fixing the dimension of the network to $n = 25$ nodes (as exemplified in Figure 2, right panel) and varying the underlying edge generation probability $\wp \in \{10\%, 30\%, 50\%\}$. To achieve this, the underlying topologies are generated as Erdös-Renyi random graphs using the `igraph` `R` package (Csárdi et al. 2026).

The experimental design for this simulation is organized as follows:

1. For each sparsity level $\wp$, we randomly generate $K = 30$ independent Erdös-Renyi network topologies.

2. For each network, we simulate $R = 50$ independent time-series replicates according to the data generating process detailed in Section D.1.

3. For each of the 30 networks, we compute its MSE and Bias$^2$ averaged over its $R = 50$ runs.

The impact of network sparsity is evaluated across the same series of scenarios (Scenarios 1, 2, and 3) and sample lengths ($T \in \{512, 1024, 2048\}$) used in the previous section. Below, we examine the distributional performance and sensitivity under changing edge configurations sequentially across all three scenarios.

### D.4.1 Scenario 1 Results

Figures 3 through 6 present the empirical tracking accuracy metrics under a single active scale environment ($j$).

Under Scenario 1, the Oracle (Subprocess (True)) and proposed Net-LSW (Subprocess (Approx)) estimators demonstrate a clear performance advantage over the MvLSW framework, producing narrower error distributions that sit consistently lower on the vertical axis across all three edge probability levels ($\wp$). The MvLSW framework exhibits noticeably wider boxplots and higher median errors in comparison. Furthermore, while the proposed method and the Oracle follow a shared pattern of steady reduction in error magnitude as $T$ increases, this asymptotic behaviour is markedly less evident for MvLSW. In fact, under partial coherence at $\wp = 50\%$, the MvLSW framework fails to show any systematic error reduction and instead exhibits wider distributional spreads and higher median errors as sample size increases.

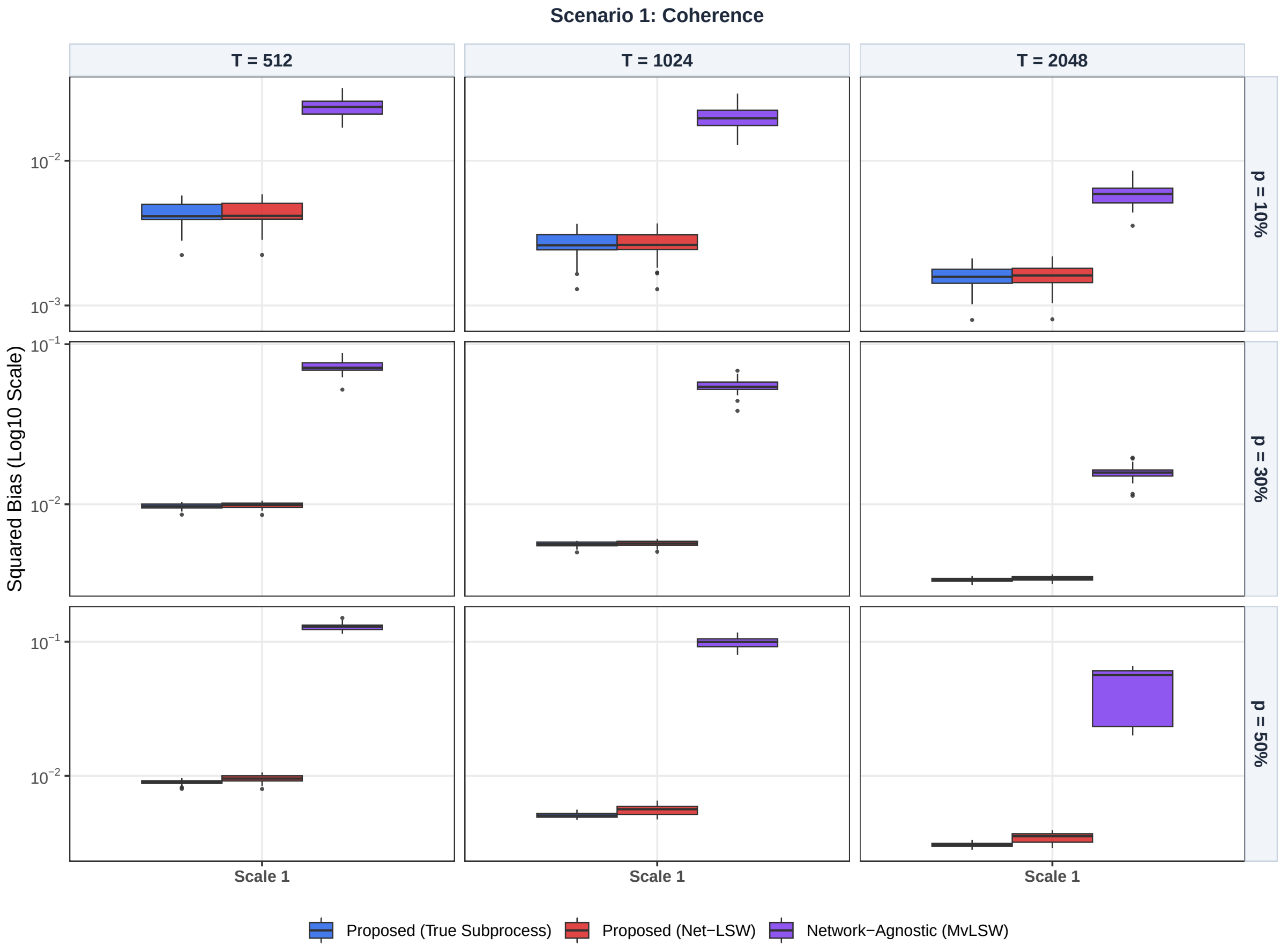

Figure 3: Empirical **Bias**$^2$ distributions for coherence under Scenario 1 over $K = 30$ random network topologies for edge probabilities $\wp \in \{10\%, 30\%, 50\%\}$ (rows) across varying sample sizes $T$. Note our proposed method (red) displays only a very slight departure from the Oracle (blue), and outperforms MvLSW (purple). The asymptotic regime is more pronounced for our method.

### D.4.2 Scenario 2 Results

To visually analyse the distributional performance and sensitivity under changing edge configurations in a two-scale setting, the results across the 30 random graphs for Scenario 2

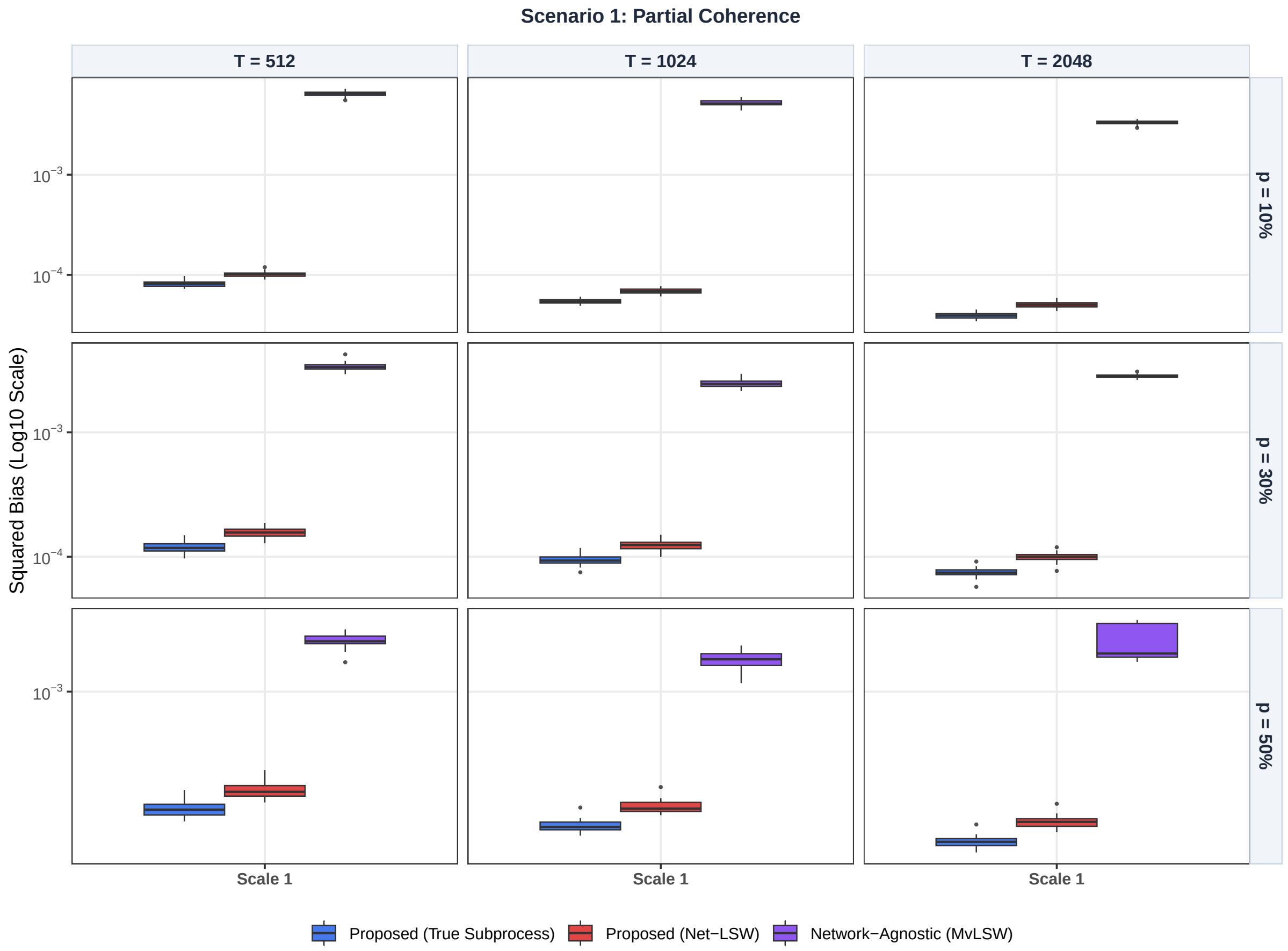

Figure 4: Empirical $\mathbf{Bias}^2$ distributions for partial coherence under Scenario 1 over $K = 30$ random network topologies for edge probabilities $\wp \in \{10\%, 30\%, 50\%\}$ (rows) across varying sample sizes $T$. Note our proposed method (red) displays only slight departures from the Oracle (blue), and outperforms MvLSW (purple). The asymptotic regime is clearly visible.

are presented as boxplots in Figures 7 and 8.

For both the squared bias ($\text{Bias}^2$) and mean squared error (MSE), the proposed Net-LSW estimator (red) consistently achieves lower median errors and narrower interquartile ranges than the network-agnostic MvLSW of Park et al. (2014) (purple) under sparse network conditions ($\wp = 10\%$), where it tracks the true subprocess benchmark (blue) across all scales and sample sizes. As the sample size $T$ increases from 512 to 2048 in this sparse setting, both the proposed estimator and the blue benchmark exhibit a clear downward shift in their error distributions alongside a visible tightening of the boxplots, highlighting empirical consistency. Conversely, the network-agnostic MvLSW largely fails to show a clear systematic error reduction and maintains wide distributional spreads.

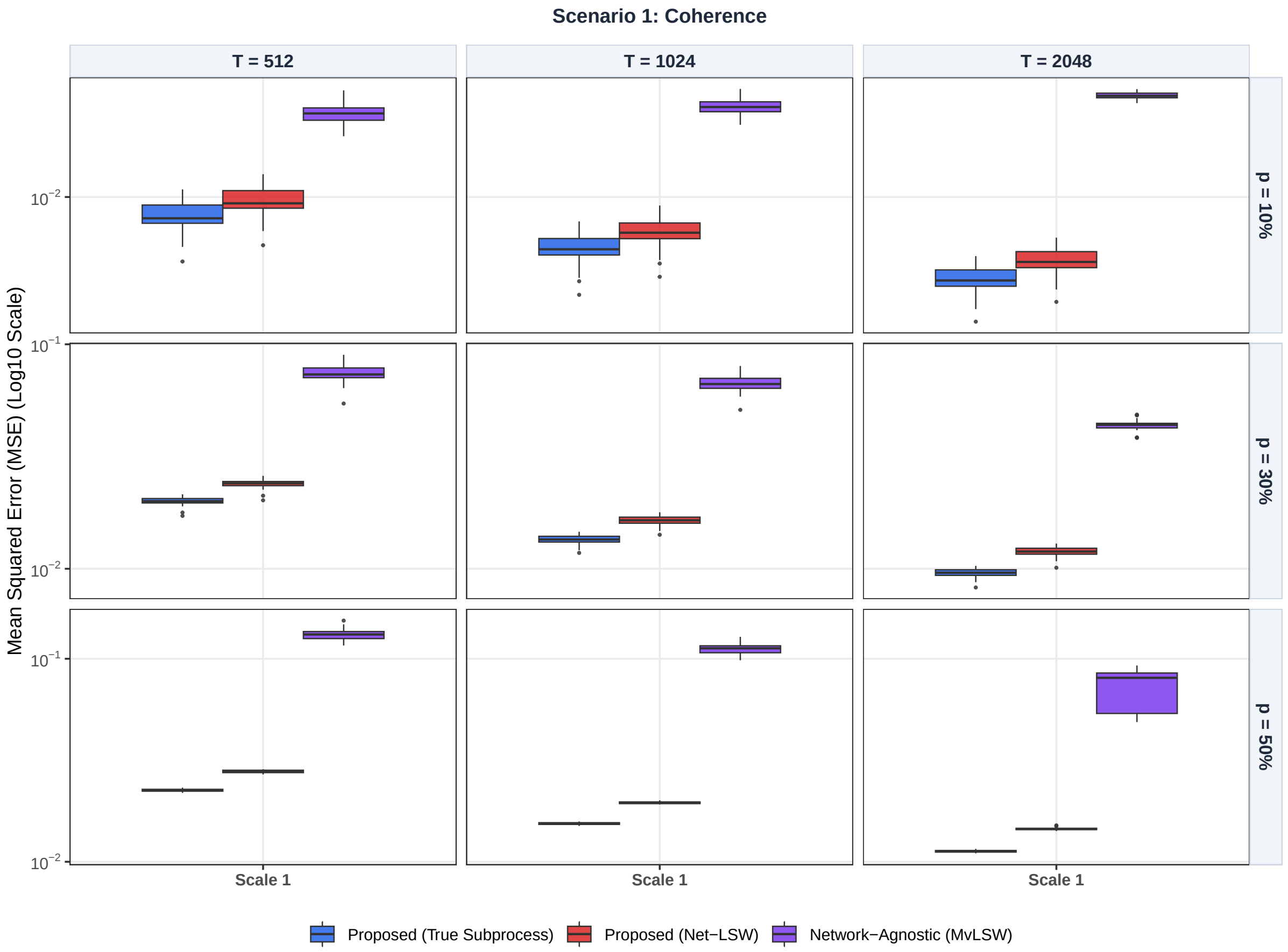

Figure 5: Empirical MSE distributions for coherence under Scenario 1 over $K = 30$ random network topologies for edge probabilities $\wp \in \{10\%, 30\%, 50\%\}$ (rows) across varying sample sizes $T$. Note our proposed method (red) displays slight departures from the Oracle (blue), and outperforms MvLSW (purple). The MvLSW asymptotic regime does not function for sparser graphs (top row).

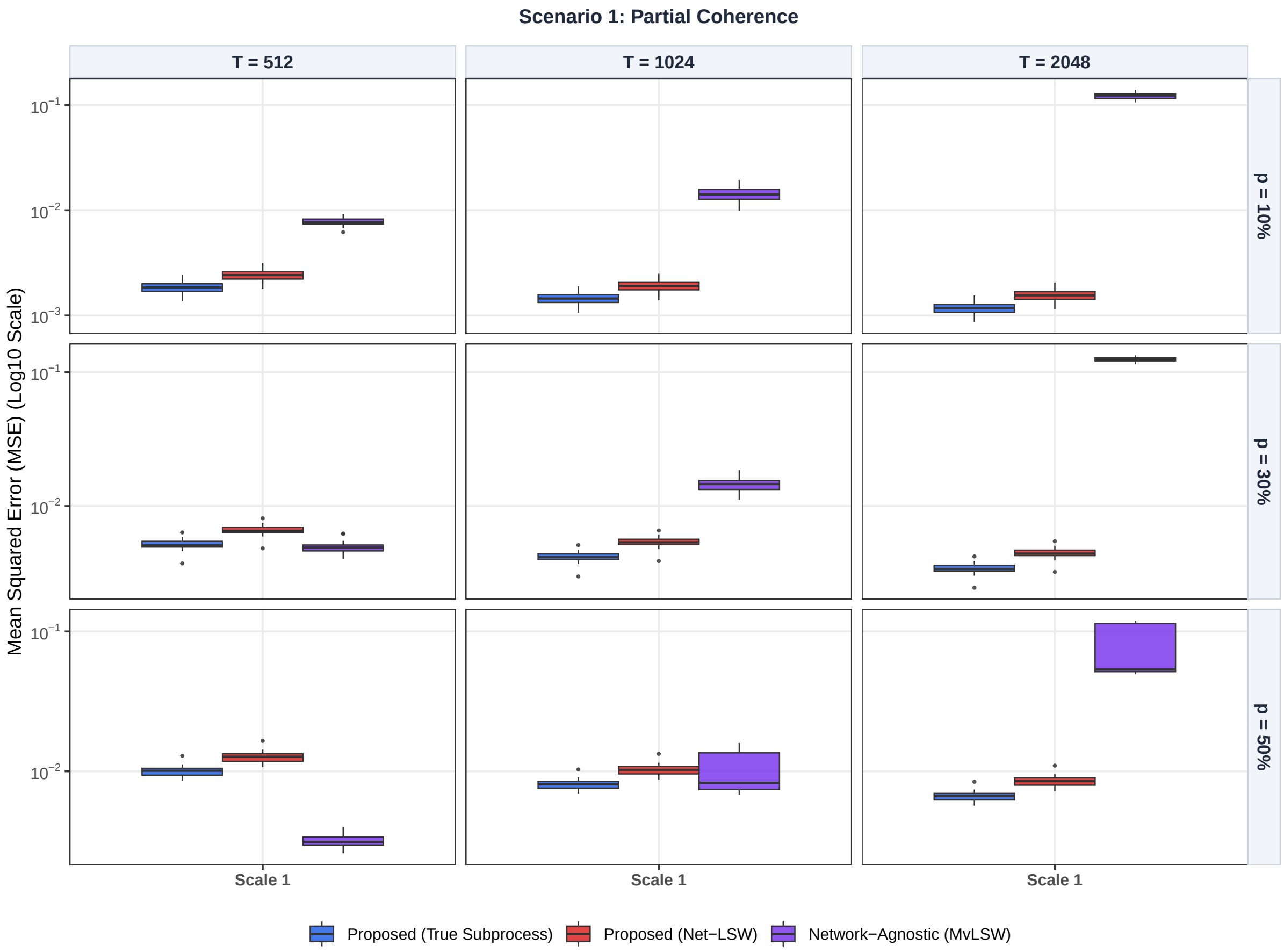

Figure 6: Empirical MSE distributions for partial coherence under Scenario 1 over $K = 30$ random network topologies for edge probabilities $\wp \in \{10\%, 30\%, 50\%\}$ (rows) across varying sample sizes $T$. Note our proposed method (red) displays slight departures from the Oracle (blue), and outperforms MvLSW (purple) for large $T$. The MvLSW asymptotic regime does not function regardless of the graph density.

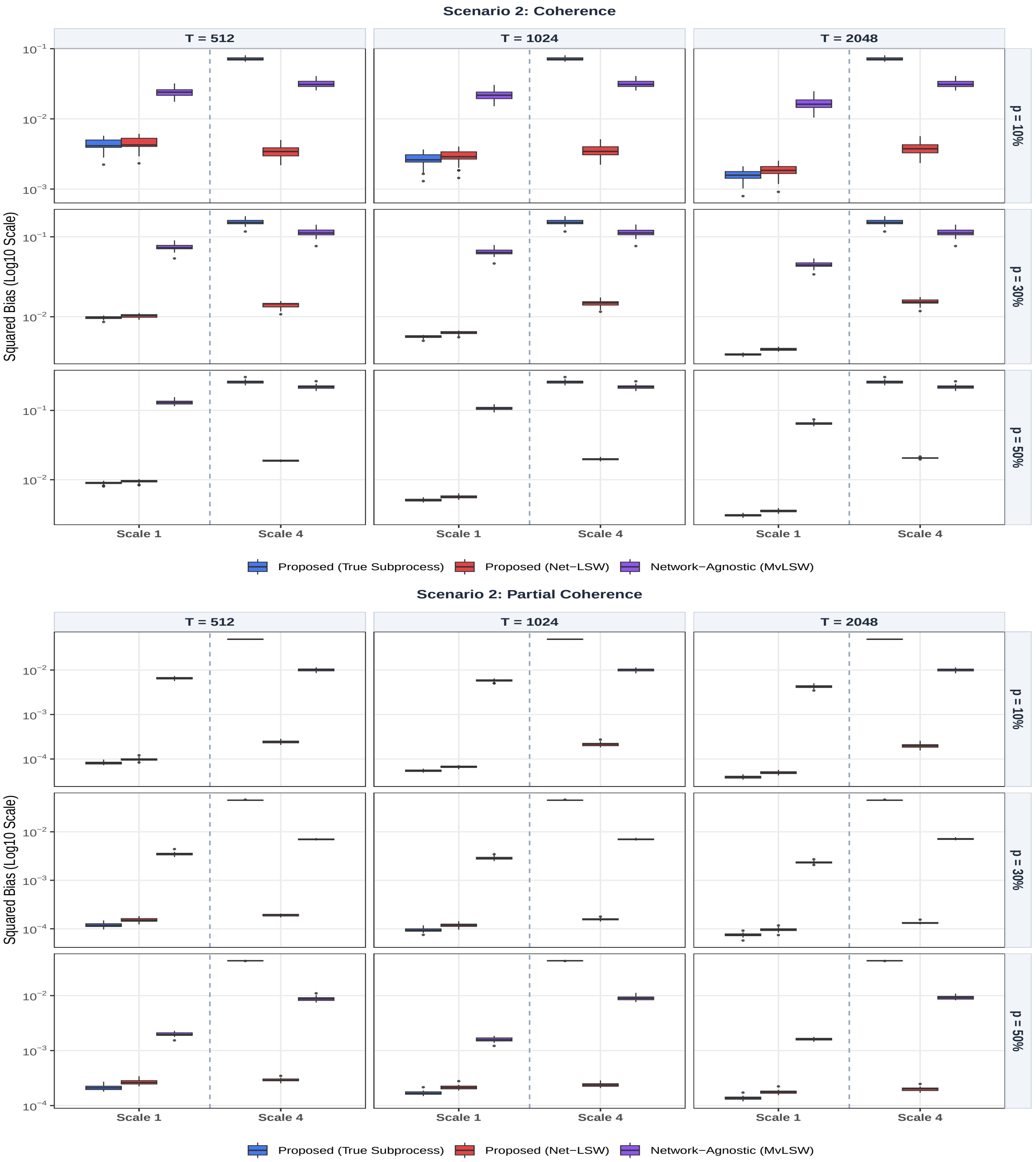

Figure 7: Empirical Bias$^2$ distributions under Scenario 2 over $K = 30$ random network topologies for edge densities $\wp \in \{10\%, 30\%, 50\%\}$ (rows) under varying sample sizes $T$. Top panel: the $3 \times 3$ evaluation grid for coherence; Bottom panel: the corresponding grid for partial coherence.

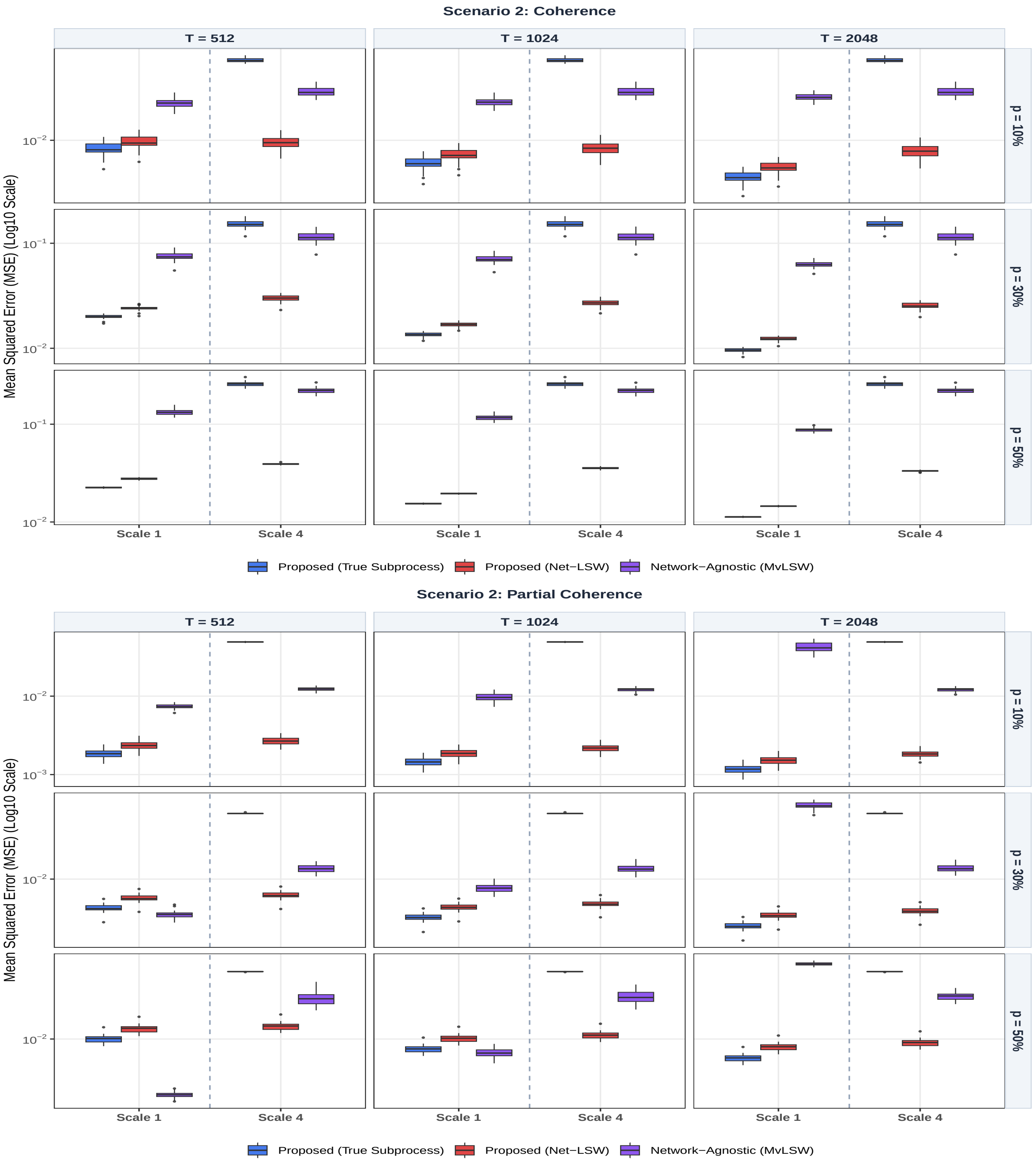

Figure 8: Empirical MSE distributions under Scenario 2 over $K = 30$ random network topologies for edge densities $\wp \in \{10\%, 30\%, 50\%\}$ (rows) under varying sample sizes $T$. Top panel: the $3 \times 3$ evaluation grid for coherence; Bottom panel: the corresponding grid for partial coherence.

#### D.4.3 Scenario 3 Results

Figures 9 through 12 display the empirical tracking accuracy metrics under the full multiple-scale setting.

Under the complex, multiple-scale environment of Scenario 3, the performance gap widens significantly across the finer scales $j = 1$ through 3, where both the true and approximate subprocess estimators maintain much lower medians and tighter error bounds than the standard MvLSW framework. This distinction is particularly pronounced in the partial coherence metrics, where the MvLSW boxplots remain highly elevated on the vertical axis across $\wp = 10\%$, $\wp = 30\%$, and $\wp = 50\%$.

However, distinct exceptions to these general asymptotic trends emerge as network density increases. At moderate density ($\wp = 30\%$) and the largest sample size ($T = 2048$), intermediate resolutions (Scales 3 and 4) experience a severe inflation in both interquartile ranges and median errors for both the Net-LSW estimator and the true subprocess benchmark. As the density increases further to $\wp = 50\%$, this structural behaviour shifts again: the error distributions for both the proposed estimator and the true subprocess benchmark exhibit extremely compressed, tight variations across all sample sizes, settling at a highly stable, low-error baseline. In these dense settings ($\wp = 50\%$), the proposed Net-LSW estimator maintains its superior performance across the finer scales (Scales 1–3), while the standard MvLSW framework produces lower median errors at scale 4. Because the proposed Net-LSW estimator tracks the true subprocess benchmark identically through both the temporary variance inflation at $\wp = 30\%$ and the highly compressed distributions at $\wp = 50\%$, this behaviour reinforces the robustness of the network regularization framework even under dense network topologies.

Finally, at the coarsest scale ($j = 5$), the standard MvLSW framework produces lower MSE values compared to the subprocess estimators. Crucially, as reported in Figure 10, MvLSW displays a very high squared bias at this same scale. Because its MSE remains low despite this large bias, it implies that the MvLSW estimator operates with an artificially low variance at scale 5, tightly tracking an incorrect signal trajectory.

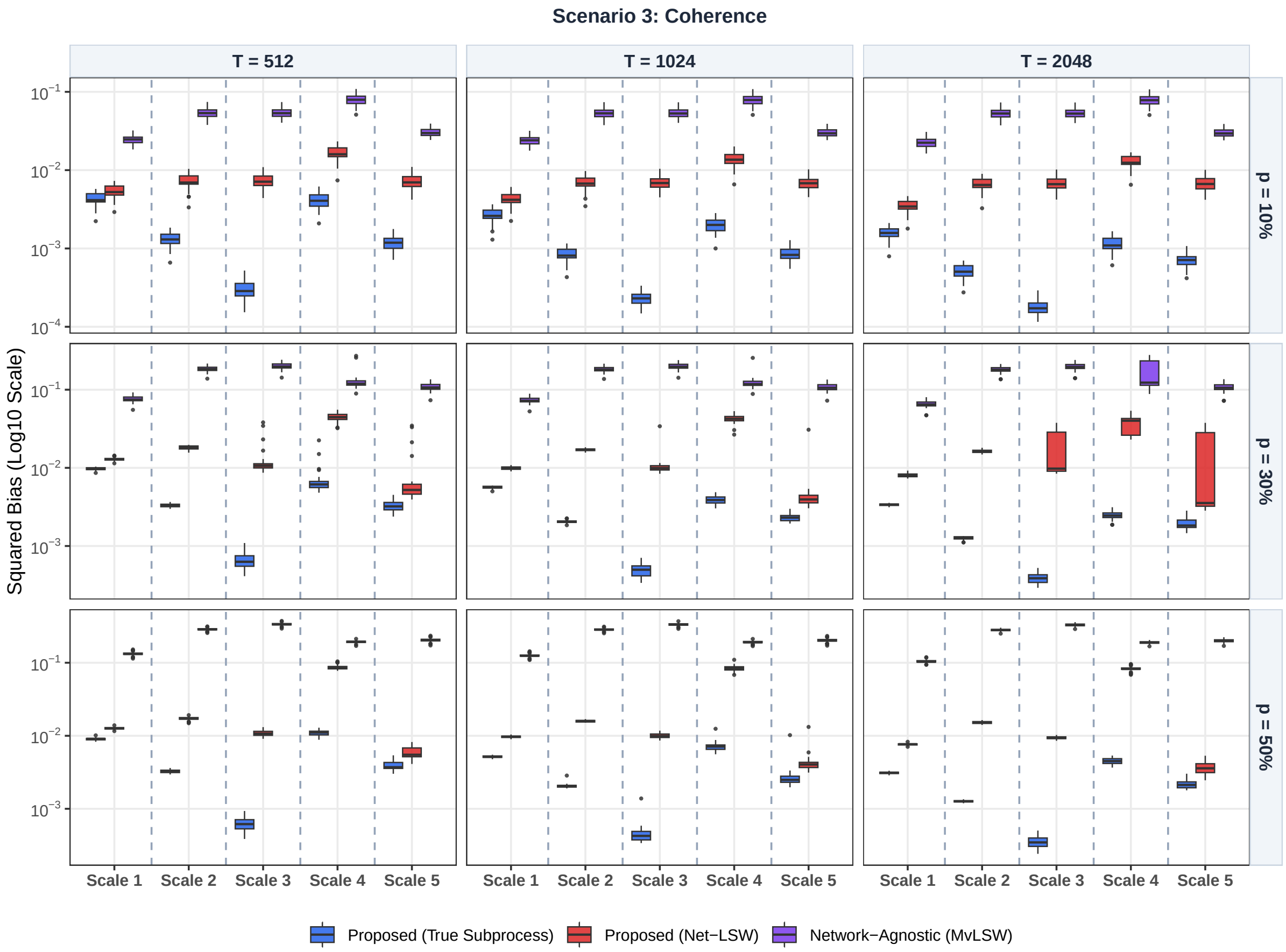

Figure 9: Empirical $\mathbf{Bias}^2$ distributions for coherence under Scenario 3 over $K = 30$ random network topologies for edge probabilities $\wp \in \{10\%, 30\%, 50\%\}$ (rows) across varying sample sizes $T$.

As a specific example, Figure 13 displays the partial coherence trajectory averaged across the simulations for a representative node pair. Here, it is evident that the MvLSW framework fails to accurately capture the true underlying energy profile at this scale, flatlining along an incorrect path despite its low variance.

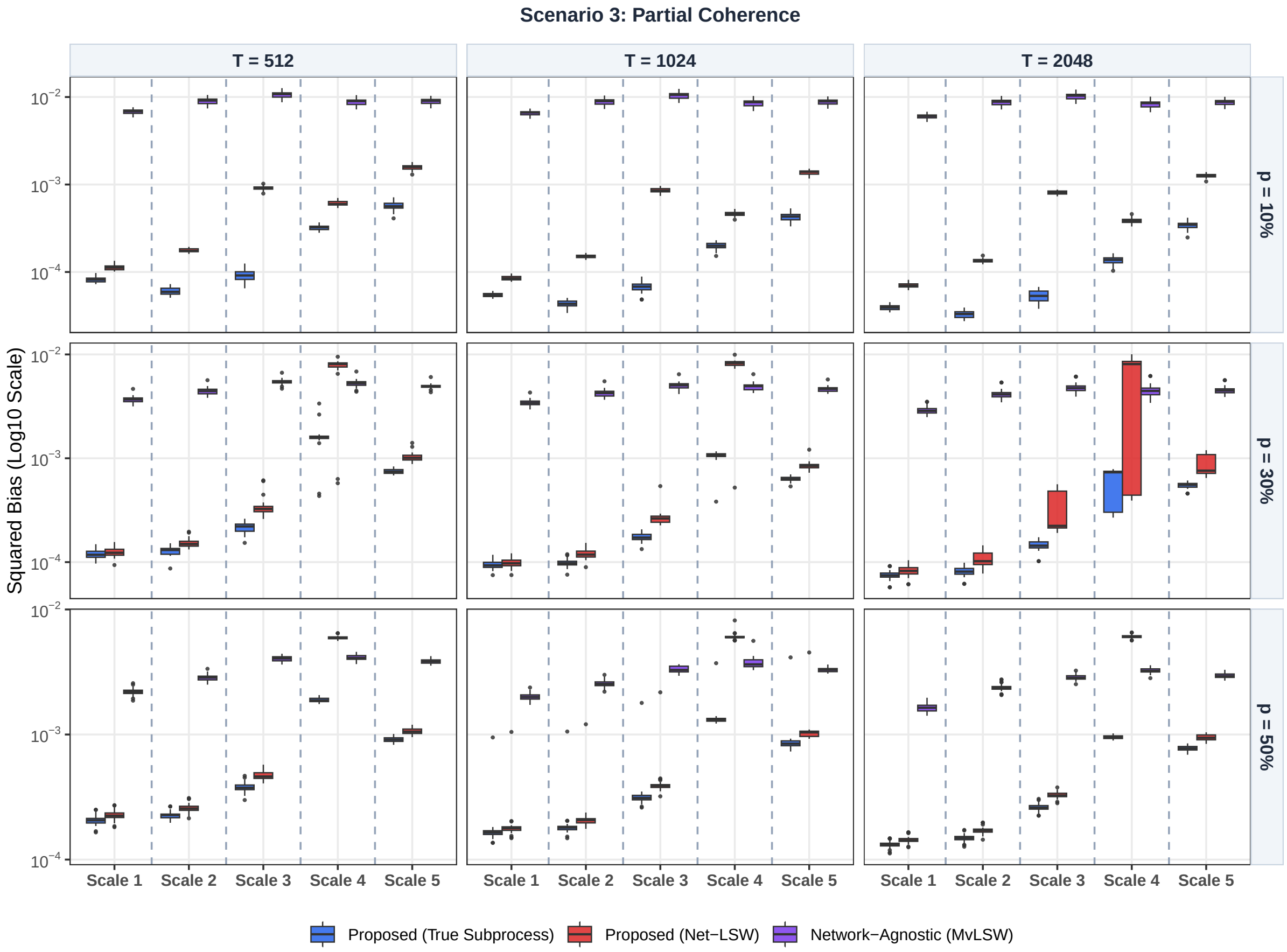

Figure 10: Empirical $\mathbf{Bias}^2$ distributions for partial coherence under Scenario 3 over $K = 30$ random network topologies for edge probabilities $\wp \in \{10\%, 30\%, 50\%\}$ (rows) across varying sample sizes $T$.

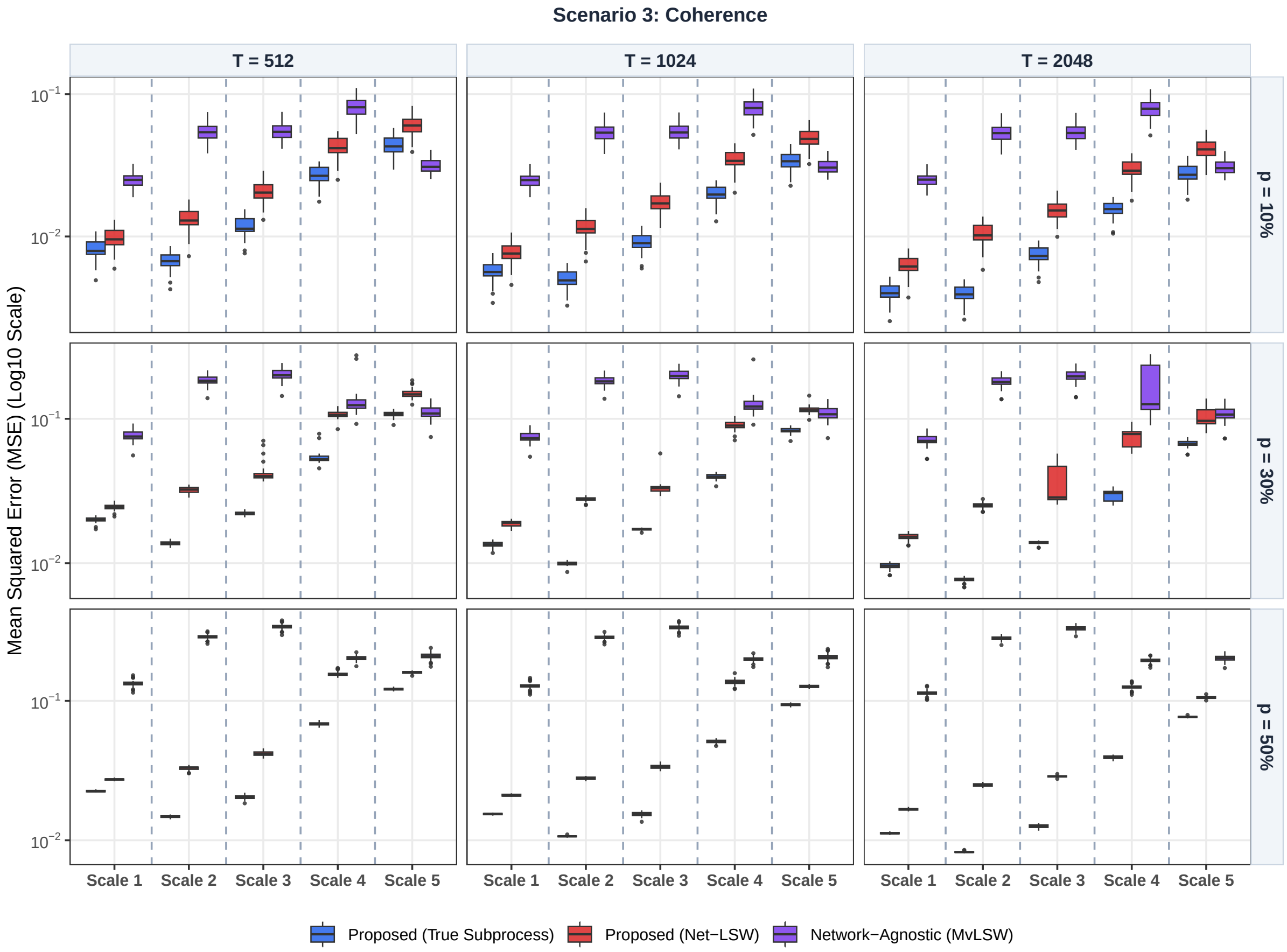

Figure 11: Empirical MSE distributions for coherence under Scenario 3 over $K = 30$ random network topologies for edge probabilities $\wp \in \{10\%, 30\%, 50\%\}$ (rows) across varying sample sizes $T$.

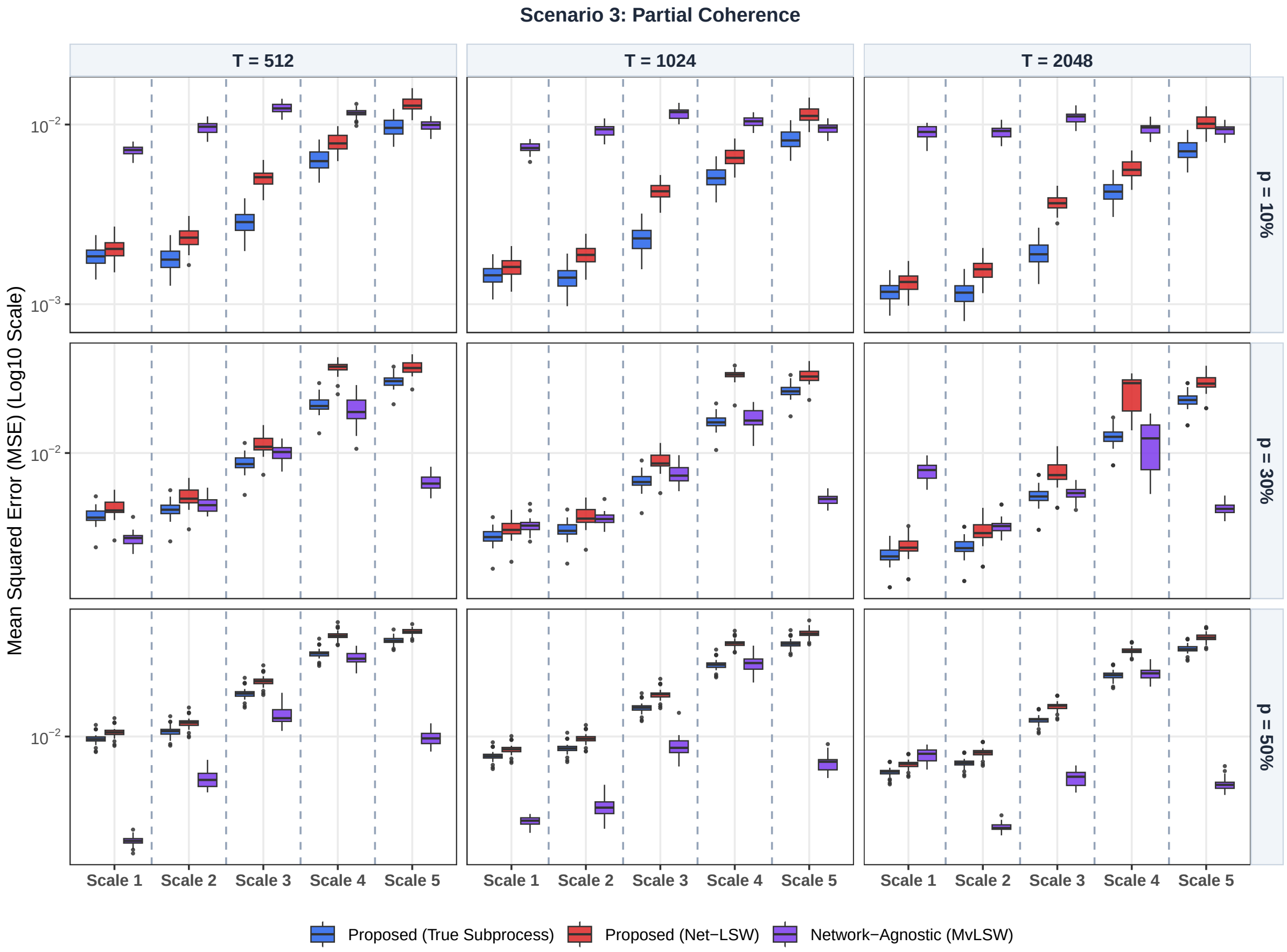

Figure 12: Empirical MSE distributions for partial coherence under Scenario 3 over $K = 30$ random network topologies for edge probabilities $\wp \in \{10\%, 30\%, 50\%\}$ (rows) across varying sample sizes $T$.

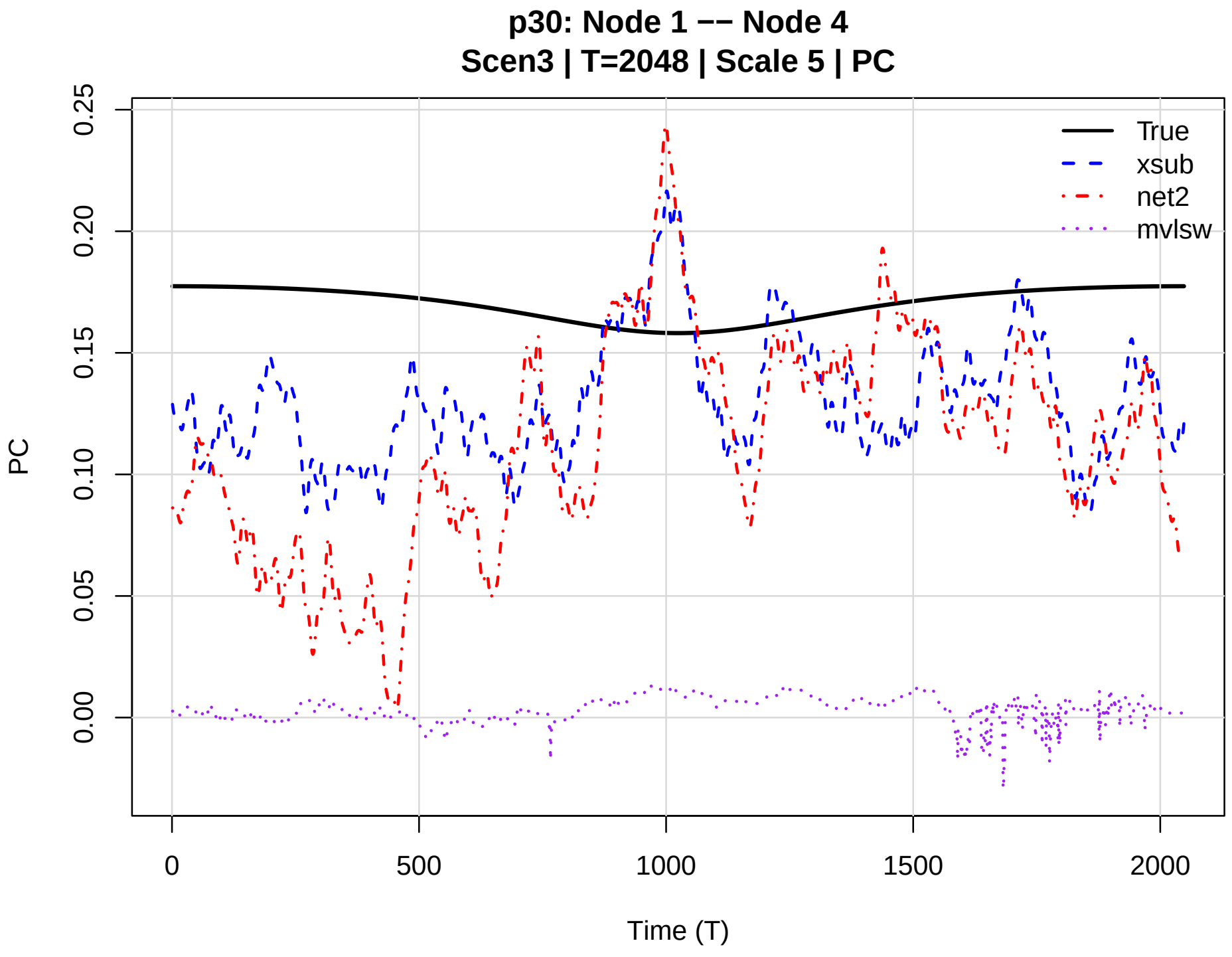

Figure 13: Time-varying partial coherence between Node 1 and Node 4 under Scenario 3 ($\wp = 30\%$, $T = 2048$) at scale $j = 5$. The averaged trajectories highlight that the MvLSW (purple dotted curve) framework fails to follow the true signal (black curve) path despite exhibiting low variability.